\documentclass[manuscript]{aastex}

\usepackage{hyperref}
\usepackage{ulem}

\begin{document}
\title{
A Precision Multi-Band Two-Epoch Photometric Catalog of 44 Million Sources in the Northern Sky from Combination of the USNO-B and Sloan Digital Sky Survey Catalogs
} 

\author{G.\ J.\ Madsen\altaffilmark{1,2,3} \&  B.\ M.\ Gaensler\altaffilmark{1,2}}
\altaffiltext{1}{Sydney Institute for Astronomy, School of Physics, The University of Sydney, NSW 2006, Australia}
\altaffiltext{2}{ARC Centre of Excellence for All-sky Astrophysics (CAASTRO)}
\altaffiltext{3}{Present address: Institute of Astronomy, University of Cambridge, Madingley Road, Cambridge CB3 0HA, UK}
\email{gmadsen@ast.cam.ac.uk}


\begin{abstract}

A key science driver for the next generation of wide-field optical and radio surveys is the exploration of the time variable sky.  These surveys will have unprecedented sensitivity and areal coverage, but will be limited in their ability to detect variability on time scales longer than the lifetime of the surveys. 
We present a new precision, multi-epoch photometric catalog that spans 60 years by combining the USNO-B and SDSS Data Release  {{9}}  catalogs.  We recalibrate the photometry of the original USNO-B catalog and create a catalog with two epochs of photometry in up to five different bands for {{43,647,887}} optical point sources  {{that lie in the DR9 footprint of the northern sky. The recalibrated objects  span a magnitude range $14 \lesssim m \lesssim 20$ and are accurate to $\approx$ 0.1 mag.}}
We minimize the presence of spurious objects and those with inaccurate magnitudes by identifying and removing several sources of systematic errors in the two originating catalogs, with a focus on spurious objects that exhibit large apparent magnitude variations.  
After accounting for these effects, we find $\approx$ {{250,000}}  stars and quasars that show significant ($\ge$ {{4}}$\sigma$) changes in brightness between the USNO-B and SDSS  {{DR9}} epochs.  We discuss the historical value of the catalog and its application to the study of long time-scale, large amplitude variable stars and quasars.

\end{abstract}

\section{INTRODUCTION}

For more than one hundred years, formidable work has gone into recording and preserving optical images of the celestial sphere.  Coordinated large-scale efforts began with the Carte du Ciel project \citep[see review by][]{Bigg00} and continued with dedicated, deep surveys such as those carried out with the Palomar Oschin Schmidt and the UK Schmidt telescopes in the northern and southern hemispheres, respectively. 
The photographic plates from these surveys have been digitized and processed by several groups to create the Digitised Sky Survey \citep[DSS;][]{McLean+00} and all-sky astrometric and photometric catalogs such as the US Naval Observatory-B catalog \citep[USNO-B;][]{usnob}, the Second Generation Guide Star Catalog \citep[GSC-II;][]{Lasker+08}, the Digitized Palomar Observatory Sky Survey \citep[DPOSS;][]{Djorgovski+98}, and the SuperCosmos Sky Survey \citep[SSS;][]{SSS-Overview}.
Ongoing efforts such as the Digital Access to a Sky Century at Harvard \citep[DASCH;][]{Grindlay+12} are processing hundreds of thousands of plates of shallower depth to provide hundreds of epochs of photometry over the entire sky. 

These data, when combined with modern very wide-field optical surveys such as the Sloan Digital Sky Survey \citep[SDSS;][]{York+00}, provide more than fifty years of astrometry from which very high quality proper motions can be derived over a large part of the sky \citep[e.g.,][]{Munn+04}.  
However, the long time baselines between observations also {{have}} a high legacy value for the study of optical variability.  The investigation of long period variable stars, cataclysmic binaries, quasars and the search for rare objects such {{as}} novae and R CrB stars all benefit from observations that span decades or more \citep[e.g.,][]{Zijlstra+02, Sesar+06, Wils+10, Hudec11, MMS11, Macleod+12}. {{For example, the timing of outbursts from recurrent novae, changes in the periods of Mira variables, and observations of late thermal pulses such as those from FG Sge \citep{HB68} or WISE J1810-3305 \citep{GYT12} provide insight into the late stages of stellar evolution and mass loss.}}
However, the photometric accuracy of historical catalogs is typically much worse than modern optical photometric surveys.  {{Indeed}}, the authors of USNO-B state that the photometry is probably the weakest aspect of the catalog \citep{usnob}.  The lack of accurate photometry and the presence of spurious objects in such catalogs diminish their legacy value.

In this paper, we present an effort to improve the photometry of tens of millions of point sources over time scales spanning 60 years.  We take advantage of the high photometric accuracy of SDSS to recalibrate a large fraction of USNO-B and to create a precision multi-band two-epoch photometric catalog.  We develop a careful recalibration process that maximizes the size and reliability of the catalog while minimizing the presence of spurious objects or unreliable photometric measurements. We make the full catalog publicly available and include metadata that describe the quality of the data.  In \S\ref{sec: source}, we describe our source catalogs, the cross-matching procedure, spurious objects, and the common photometric system.  In \S\ref{sec: pre-flag}, we describe a process for identifying blended objects. We outline the photometric recalibration in \S\ref{sec: recal}, followed by the identification of anomalous magnitudes in \S\ref{sec: consistent}. We present our results in \S\ref{sec: results} and include a comparison of our results with previous work by others.  The limitations of the catalog are given in \S\ref{sec: limitations}. We provide  suggested applications of the catalog in \S\ref{sec: applications} and summarize our work in \S\ref{sec: summary}.  A flowchart summarizing our work is shown in Figure \ref{fig: flowchart}; the reader is advised to refer to this Figure throughout the paper. 

\section{SOURCE CATALOGS}
\label{sec: source}

\subsection{USNO-B}
\label{sec: usno}

The USNO-B catalog is a compilation of optical astrometric and photometric measurements for more than a billion objects over the entire sky \citep{usnob}.  It is derived from digital scans of original or glass copies of 7,435 Schmidt plates that were taken between 1949 and 2002 as part of the Palomar Observatory Sky Survey \citep[POSS;][]{MA63,Reid+91}, ESO/SERC, and Anglo-Australian Observatory UK Schmidt surveys \citep{HD81, Cannon84, West84}.  The catalog provides up to five photometric measurements for each object in five bands ($O, E, J, F,$ and $N$; named after the plate emulsions).  The five bands are alternatively referred to as one of three colours: blue ($O$ and $J$), red ($E$ and $F$), and infrared ($N$).   
The effective wavelengths of bands $O, E, J, F,$ and $N$ are $\approx$ 4100\AA, 6500\AA, 4700\AA, 6600\AA, and 8400\AA\ with full-width at half-maxima of $\approx$\ 1500\AA, 450\AA, 1500\AA, 800\AA, and 1700\AA, respectively\footnote{These values are derived from data available at \href{http://gsss.stsci.edu/SkySurveys/Surveys.htm}{http://gsss.stsci.edu/SkySurveys/Surveys.htm}}.
\citet{usnob} report that the photometry is complete to $\approx21$ mag, although this varies by up to $\approx$ 2 magnitudes from plate to plate. The five filter measurements were taken at different epochs; the time between epochs varies from hours to decades. The  astrometric accuracy of the catalog is $\approx$ 0.2\arcsec; this is the median dispersion of the positions of a given object imaged on multiple overlapping plates.  The catalog matches objects on plates from different epochs to calculate proper motions. However, these proper motions are normalized to another astrometric catalog, and hence they are relative and not absolute. 

The photometric accuracy for point sources in USNO-B is difficult to estimate, but is reported by \citet{usnob} as 0.3 mag. This number is the standard deviation of the  photometric solution among a large set of calibration stars. However, as noted by \citet{usnob} and demonstrated by \citet{Sesar+06}, the systematic uncertainty in the photometry can often exceed 0.3 mag by several magnitudes.

For the purposes of this work, there are a few advantages of using the USNO-B catalog over other historical catalogs that are based on many of the same photographic plates, such as the Guide Star Catalog \citep{Lasker+08}, the SuperCOSMOS Sky Survey catalog \citep{SSS-Overview}, and the Digitized Second Palomar Observatory Sky Survey \citep[][]{Djorgovski+98}.  The USNO-B catalog has a wider range of epochs and/or sky coverage compared to other catalogs.  As noted by \citet{Munn+04}, USNO-B has a higher threshold for rejecting possible false detections (e.g., plate artifacts/defects, diffraction spikes) and therefore includes many real objects that are not reported in other catalogs.  Most importantly, the original survey and plate identification are reported for every object, enabling a robust and accurate photometric recalibration (see \S\ref{sec: recal} below).  
Several of the POSS, ESO/SERC, and AAO surveys used plates that overlap, i.e. a small fraction of the plates cover the same part of the sky.  These overlap regions were used to aid in the photometric calibration. However, the photometry of each cataloged object in USNO-B is from a single plate. We also note that within overlap regions, objects grouped together by plate identification do not always form a distinct boundary on the sky.

We selected all objects in USNO-B that lie within $1^\circ$ of the areal footprint of the  {{DR9}} catalog of SDSS-III ($\approx$ 14,500 deg$^2$).  The extra $1^\circ$ boundary is included to allow for the cross-matching of objects with large or inaccurate proper motions.  A total of $\approx$ {{260}} million objects satisfy this criterion (see Figure \ref{fig: flowchart}).  {{Approximately 77\%  of these USNO-B objects were observed as part of one of the POSS surveys for all five bands; the remaining 23\% were observed as part of the southern ESO/SERC or AAO UK Schmidt surveys in at least one band.}}

The USNO-B catalog does not provide directly the epoch of the photometry for every object. 
We derived the UT date and time (at mid-exposure) for most entries in the catalog by using the plate identifications and data available through the Image and Catalog Archive hosted by USNO\footnote{\href{http://www.usno.navy.mil/USNO/astrometry/optical-IR-prod/icas/icas-guide}{http://www.usno.navy.mil/USNO/astrometry/optical-IR-prod/icas/icas-guide}}. This resource provides information for all of the `blue' and `red', i.e.\ the $O, E, J, $ and $F$ bands. 

For most of the {{`infrared'}} $N$ band plate epochs, we used data available through the German Astrophysical Virtual Observatory\footnote{\href{http://dc.zah.uni-heidelberg.de/usnob/res/plates/pq/info}{http://dc.zah.uni-heidelberg.de/usnob/res/plates/pq/info}}. A small number of $N$ band plates (25) were taken as part of the SERC-I survey but were assigned plate identifications from POSS-II infrared (IR) survey; these plates filled in gaps in the northern POSS-II IR survey.  We identified the epochs for 14 of these 25 plates in the Mikulski Archive for Space Telescopes\footnote{\href{http://gsss.stsci.edu/SkySurveys/Surveys.htm}{http://gsss.stsci.edu/SkySurveys/Surveys.htm}}. The epochs for the remaining 11 $N$-band plates ($<$ 1\% of the sky) are unidentified.  Objects that are on plates with unidentifiable epochs remain in the list of $\approx$ {{260}} million objects, but are noted as having missing epochs. 
All of the epochs ($t_{USNO}$) are recorded in units of decimal years with a precision of 10$^{-4}$ years, for a temporal resolution of $\approx$ 1 hour (the exposure time for individual plates varies from $\approx$ 5 minutes up to  {{3}} hours).

\subsection{SDSS-III DR9}
\label{sec: dr8}

The Sloan Digital Sky Survey (SDSS) is an ongoing effort to obtain high precision photometry and spectroscopy of a large fraction of the northern sky \citep{York+00}.  The survey commenced in {{the year}} 2000 and provides nearly simultaneous imaging in five optical and near infrared bands ($u,g,r,i,z$).  The survey is now in its third phase (SDSS-III) and has made data available through Data Release  {{9}} \citep[DR9;][]{DR9}.   The  {{DR9}} imaging data cover more than 14,500 deg$^2$ with a 50\% completeness limit for point sources of $r = 22.5$ mag (about one magnitude fainter than USNO-B), and a relative photometric accuracy of $<2\%$.   
{{The global absolute astrometric precision of the survey is 0\farcs1 (rms); the reported proper motions are based on the method described in \citet{Munn+04, Munn+08}.}}

{{DR9}} provides five band photometry for $\sim$ 470 million unique stars, quasars and galaxies. It provides additional five band photometry for $\sim$ 320 million objects derived from repeat observations of the same part of the sky.  To avoid ambiguity, we only consider `primary' measurements, i.e.\  {{those that are classified as the `best' observation for each object. We also exclude objects that have moved during the $\approx$ 70 seconds that elapse between observation in different filters (e.g., main belt asteroids). }}
We further refine our selection in order to choose appropriate objects that can be robustly matched to those in USNO-B. The full list of our selection criteria applied to  {{DR9}} are:

\begin{enumerate}
\item{the object is classified by {{DR9}} as an unresolved point source, e.g. a star or quasar. The classification is accurate for objects brighter than $\approx$ 22 mag \citep{Scranton+02}.
This criterion applies to about half of all unique objects ($\approx$ 260 million). We exclude galaxies because the photometric calibration in USNO-B is not appropriate for extended sources.}

\item{
the object's  magnitude is brighter than that at which the completeness of USNO-B is less than  {{10\%}} \citep[$g,r,$ or $i < $~{{22}};][]{Munn+04}.  {{This enables the cross identification of objects with extreme colours or those that are potentially variable objects that were brighter than the USNO-B sensitivity limit at the USNO-B epoch.}}
This criterion reduces the sample to $\approx$  {{99}} million objects.
}

\item{the object was observed under photometric sky conditions or was calibrated based on overlaps with calibrated data; this reduces the sample size to $\approx$  {{90}} million objects.}

\item{the photometric pipeline reports that the object has reliable photometry in each of the $g$, $r$, and $i$ bands. This criterion excludes objects that are saturated ($m \lesssim$  {{12}} mag), are near the chip edge, have a bad point spread function (PSF), are not deblended accurately, etc.  This reduces the sample to $\approx$  {{77}} million objects.}

\item{the proper motion is reported as reliable, following the recommendations of \citet{Munn+04}. This criterion uses the results of \citet{Munn+04} and requires that an object is a) matched to exactly one USNO-B object within 1$\arcsec$ (after a correction for proper motion); b) is identified on four or more USNO-B plates, and c) has an uncertainty in each celestial coordinate of $< $ 0.35\arcsec. We note, however, that \citet{Munn+04} {{do}} not report the identity of the match in USNO-B (see \S\ref{sec: xmatch} below).  This criterion reduces the sample from  {{DR9}} to $\approx$  {{44}} million objects.}
\end{enumerate}

We convert the mean time of the observations in the $g,r$, and $i$ bands for each object to decimal years with a precision of 10$^{-4}$ years, the same precision as for the USNO-B epochs.
The full details of the selection criteria, including the constraints on various photometric flags, are given in Appendix \ref{sec: app}.

\subsection{Cross Matching }
\label{sec: xmatch}

An accurate comparison between objects in USNO-B and  {{DR9}} requires a careful cross matching between the two catalogs, incorporating proper motion measurements. For example, the longest interval between the photographic plate and  {{DR9}} observations is $\approx$ 60 years; the distribution of time intervals between observations is shown in Figure \ref{fig: histo epochs}.
(Note that some $N$ band observations were taken more than 2 years after the SDSS data.)
Over the course of 60 years, 10\% of the  {{DR9}} objects have moved by more than 1\arcsec.

The proper motions reported by DR9  are calculated using the procedure described by \citet{Munn+04}. However, \citet{Munn+04} do not report the identities of the cross matched objects. In addition, while their proper motions are placed in an absolute reference frame, they are based on the less accurate proper motions reported by USNO-B itself.  Consequently, we performed an independent cross matching procedure described below.

The USNO-B catalog reports celestial coordinates at the equinox J2000 and at epoch 2000.0. The USNO-B coordinates were converted by \citet{usnob} to epoch 2000.0 using the proper motions reported by USNO-B itself. These proper motions are only relative measurements, not absolute, and suffer from a larger uncertainty compared to those of \citet{Munn+04, Munn+08}.  For example, about 500,000 objects in USNO-B have proper motion values that differ by more than 10 milli-arcseconds per year from their proper motions in  {{DR9}}.

The {{DR9}} catalog reports celestial coordinates at the equinox J2000 but at the epoch of each individual {{DR9}} observation. In order to account for these discrepancies in the coordinate systems in the two catalogs, we cross match our sample of USNO-B and  {{DR9}} objects in two stages.  

In the first stage, we convert all  {{DR9}} objects to the epoch 2000.0 using the proper motions reported in DR9.  We then match all USNO-B objects with valid coordinate uncertainties to  {{DR9}} objects {{(at epoch 2000.0)}} using the quad-tree cube algorithm of \citet{KB06}.   We allow for multiple objects in each catalog to be matched to multiple objects (many-to-many) with a large matching radius of 30\arcsec. The large matching radius was chosen to correct for potentially highly incorrect proper motions reported by USNO-B.

In the second stage, we convert the coordinates of each preliminarily matched USNO-B object back to the mean epoch of the corresponding USNO-B observation using the proper motions reported by USNO-B.  (This mitigates the potentially adverse effects of incorrect USNO-B proper motions.) We then convert these coordinates to the epoch 2000.0 using the DR9 proper motion of each of their cross matched  {{DR9}} counterparts.  We recalculate the angular separation between each matched USNO-B and  {{DR9}} pair and retain the closest match.  
We discard objects with a minimum angular separation $>$ {{1}}$\farcs0\ (\sim ${{3.5}}$\sigma$). 
The final (one-to-one) cross matched catalog contains  {{43,873,069}} objects, as shown in Figure \ref{fig: flowchart}. 

Unless specified otherwise, references to {{DR9}} and USNO-B objects in the rest of the paper refer to this sample of $\approx$ {{44}} million objects and not to other data those catalogs.

\subsection{Spurious objects in USNO-B}
\label{sec: spurious}

The USNO-B catalog is known to contain many objects that are not astrophysical \citep{Storkey+04, Barron+08}.
Defects or emulsion flaws in the original plates, satellite or meteor trails, and optical artifacts near bright stars are reported as objects in the original catalog. We chose our selection criteria for  {{DR9}} objects in (\S\ref{sec: dr8}) to strongly suppress the inclusion of these artifacts. For example, it is unlikely for plate artifacts to have reliable proper motions and to be matched to point sources in  {{DR9}}.  

In order to quantify the potential contribution of spurious sources, we compared our cross matched catalog to the results of \citet{Barron+08}. They used computer vision techniques to identify diffraction spikes and annular reflection halos among the entries in USNO-B.  They took advantage of the common spatial pattern among these optical artifacts to detect them automatically. They found that $\approx$ 2\% of entries are identified as spikes or halos.  By comparison, 0.06\% of objects in our cross matched catalog are classified as spikes or halos by \citet{Barron+08}. 
This implies that our selection criteria have reduced the rate of spurious sources by a factor of 2/0.06 = 30.  However, this also shows that  $\approx$ 24,000 objects in our catalog are classified as spurious. The automatic classification scheme of \citet{Barron+08} is statistical and some objects classified as spurious may be real. We therefore retain these potentially spurious objects in our catalog and note that they may not be real (see Table \ref{tab: columns} below).

\subsection{Photometric system}
\label{sec: photo systems}

In order to compare accurately the photometry of objects in USNO-B and  {{DR9}}, the photometric measurements must be placed onto a common system.  This can be done by carefully selecting well-calibrated, cross-matched objects in both catalogs and constructing a transformation function through a least-square minimization scheme, as done by \citet{usnob}, \citet{Munn+04} and \citet{Sesar+06}  from a  sample of objects in SDSS Early Data Release, Data Release 1, and Data Release 2, respectively.  

However, the USNO-B photometry suffers from large systematic errors and requires recalibration \citep{usnob, Munn+04, Sesar+06}.   In addition, any relative changes in magnitude between the two catalogs, such as those caused by variability, are not very sensitive to the particular transformation functions.  Therefore in order to avoid the introduction of conflicting photometric conversion systems, we adopt the transformations from the Sloan system to the USNO-B system as originally described by \citet{usnob}.  For each USNO-B band, the transformations are:
\begin{eqnarray}
 O_{SDSS} = g + 0.452 (g-r) + 0.08  \\ 
 E_{SDSS} = r - 0.086 (g-r) - 0.20  \\
 J_{SDSS} = g + 0.079 (g-r) + 0.06 \\
 F_{SDSS} = r - 0.109 (g-r) - 0.09 \\
 N_{SDSS} = i - 0.164  (r-i) - 0.44
\end{eqnarray}
where $g,r,$ and $i$ are the point spread function (PSF) magnitudes reported by  {{DR9}} in the relevant band ($u$ and $z$ band data are not used here).   {{DR9}} reports multiple magnitudes for each object using several different algorithms;  the PSF magnitudes are more accurate than the model magnitudes or composite model magnitudes also listed for each source (see \S\ref{sec: limitations} below).   $(O,E,J,F,N)_{SDSS}$ are the magnitudes of a  {{DR9}} object in the USNO-B system; we use $m_{SDSS}$ as the generic term for these magnitudes.  Similarly, we use $(O,E,J,F,N)_{USNO}$ to refer to the original magnitudes of a USNO-B object in the USNO-B system; we use $m_{USNO}$ as the generic term. 
We calculate the uncertainty in $m_{SDSS}$ using the 1$\sigma$ errors reported by  {{DR9}} in each filter and propagating them according the equations above.  These errors are generically referred to as $\sigma_{SDSS}$.

For several of the historical surveys on which USNO-B is based, different filters were used with plates of the same emulsion.  In principle, a distinct photometric system should be created for each unique combination of emulsion and filter.  However, the USNO-B data require a full recalibration (described below) and the recalibration process implicitly accounts for the effect of different filters.  
We therefore avoid introducing the complication of multiple photometric systems for each band.

\section{Identification of blended USNO-B objects}
\label{sec: pre-flag}

There are several astrophysical objects in USNO-B with photometry that may be adversely affected by their proximity to other objects, either astrophysical or artificial in origin.  It is important to identify such objects with inaccurate photometry in order to minimize their influence on the photometric recalibration process described in \S\ref{sec: recal} below. There are two cases for which the photometry may be inaccurate because of this effect. One case is where the photometry is influenced by proximity to artifacts from very bright stars.  The other case applies to objects that are very close to one another and suffer from adverse effects of blending.  The two cases are discussed below in \S\ref{sec: blend bright} and \S\ref{sec: blended}, respectively, and are summarised in Figure \ref{fig: flowchart}.

\subsection{Proximity to bright star artifacts}
\label{sec: blend bright}

Our selection criteria for  {{DR9}} objects excludes bright stars but does not exclude objects that are near bright stars (step \#4 in \S\ref{sec: dr8}). 
The  {{DR9}} pipeline measures accurately the brightness of objects near bright stars (or reports that  {{it}} cannot be measured reliably). However, USNO-B does not measure these objects accurately and it does not indicate if the measurements are reliable. A visual inspection of photographic plate and  {{DR9}} images (along with  {{DR9}} and original USNO-B magnitudes) shows that there are many objects within a few arcminutes of bright stars that have highly inaccurate USNO magnitudes, even those that are not on diffraction spikes or near the edge of reflection halos. An example of this is shown in Figure \ref{fig: thumbnail bright star}. 

We identify all USNO-B objects that are within an exclusion radius of a bright star.  Because the asymmetric artifacts near bright stars are not well fit to a model, we calculate the exclusion radius empirically from a visual inspection of plate images. 
For each USNO band, we inspect a large sample of images in the Digitized Sky Survey (DSS) available through the SkyView Virtual Observatory\footnote{\href{http://skyview.gsfc.nasa.gov/}{http://skyview.gsfc.nasa.gov/}}. The images are of the same plates (or glass copies) of ones used to create USNO-B.  The images are centered on bright stars in the Tycho-2 catalog \citep{Hog+00}. We chose up to 5 randomly selected stars in 20 narrow magnitude bins for stars brighter than 12th magnitude; we found that stars fainter than this do not have spikes or halos and are well fit by Gaussian point spread functions.  For each image, we estimate the maximum angular radius of the spikes or halo centred on the bright star.  This subjective estimate of the radius is conservative; we chose a radius that was well outside the visible optical artifacts. This method is also insensitive to the lossy compression of the DSS images.

The results of the visual inspection are shown in Figure \ref{fig: bright star width}.  The filled circles show the maximum radius of artifacts for the sample of stars in each magnitude bin. The vertical dotted lines show the full range of estimated radii of artifacts in each bin.  A least-squares model fit to the filled circles is shown as a solid line. We use the parameters of this model to estimate the exclusion radius for every bright star in the USNO-B catalog.   For this calculation, we use the magnitude from Tycho-2 rather than the magnitude reported by USNO-B; USNO-B magnitudes are not reliable for these bright stars.  For USNO bands $O$ and $J$, we used the Tycho $B_T$ magnitude; the $V_T$ magnitude was used for bands $E$, $F$, and $N$.  

Any object that falls on or within the exclusion radius of a bright star is classified accordingly. 
For each USNO-B band, there are $\sim$ 200,000 -  {{700,000}} of these tagged objects in the cross matched catalog ($\approx$ 1\% of entries ).  Note that the practical limitation 
on visually inspecting every bright star implies that there could be objects with inaccurate photometry near bright stars that are missed by this process.

We note that our model fits in Figure \ref{fig: bright star width} compare favourably to the results of \citet{Barron+08} who measured similar parameters with computer vision techniques (cf. their Figures 5 and 6).  However, we cannot use the cleaned USNO-B catalog of \citet{Barron+08} because it only removed objects in narrow annular rings from halos and objects that are part of diffraction spikes; their catalog did not remove real objects that are very close to bright stars. The number of objects that we identify as being near bright stars is more than an order of magnitude greater than those classified as spikes or halos by \citet{Barron+08}.  Through the identification of these objects, we improve significantly the reliability of the photometry of the objects in our catalog.

\subsection{Blends with nearby objects}
\label{sec: blended}

We used a different process to identify objects that are blended with very nearby objects (but are not bright enough to cause optical artifacts). The typical full-width at half-maximum (FWHM) of point sources on the USNO-B photographic plates ($\approx$ 3\arcsec) is much greater than the median $r$-band FWHM of point sources in  {{DR9}} (1\farcs3). This creates two undesirable possibilities when comparing objects in USNO-B with those in DR{{9}}:

\begin{enumerate}
\item{two or more adjacent objects within USNO-B can have overlapping full-width half-maxima.  In this case, the magnitude of one or more of these objects may be severely over- or under estimated in USNO-B.}
\item{one object in USNO-B may be cross matched to a {{DR9}} object that has one or more  other  {{DR9}} objects very nearby. In this case, the single USNO-B object may actually be a blend of two more objects that are resolved in  {{DR9}}. An example of this situation is shown in Figure \ref{fig: thumbnail blended sdss}.} 
\end{enumerate}

We determine the apparent angular sizes of USNO-B objects empirically, using DSS images of randomly selected, isolated point sources spanning a wide range of magnitudes.  We calculate the half-width at half-maximum (HWHM) of more than 175 objects per band from a two-dimensional Gaussian fit to each star. The HWHM was taken to be the larger of the widths in $x$ and $y$, and only circular fits (axis ratio between 0.8 and 1.2) are recorded.  Objects with non-circular widths are not used in this calculation.

The results are shown in Figure \ref{fig: faint star width}.  We find a non-linear negative correlation between HWHM and the USNO-B magnitude, $m_{USNO}$, with a typical range of half-widths of $1\farcs5 - 4\farcs0$.  We fit models consisting of second-order polynomials (with a constant minimum value at faint magnitudes) to the data; the least-squares best fits are overlaid in Figure \ref{fig: faint star width}.  We calculate the HWHM of every object in the cross matched catalog using the best fit model parameters. 

In order to assess whether a USNO-B object is blended in scenario (1) above, we calculate the angular separation between all pairs of USNO-B objects.  An object and its neighbors are tagged as blended if the angular distance to any neighbor is less than the linear sum of the HWHM of the object and its neighbor. In each band, $\approx$ 1.5 million objects ($\approx 5\%$ of entries in our cross matched catalog) are tagged this way.

In order to assess if a USNO-B/{{DR9}} cross matched object is blended under scenario (2), we calculated the angular separation between all pairs of objects within the original {{DR9}} (of which the {{DR9}} objects described above in \S\ref{sec: dr8} are a subset).  
If we had only considered objects in our cross matched catalog, several USNO-B objects would be classified incorrectly as unblended. 
For this superset of objects in {{DR9}}, we select all objects that could have been detected on the original USNO-B plates.  We selected point sources and extended sources, and relaxed the selection criteria for some of the photometric pipeline flags; this superset contains $\approx$ {{200}} million objects (see Appendix \ref{sec: app b} for the full list of selection criteria).

We calculate the HWHM of the superset of {{DR9}} objects using the composite model magnitudes from {{DR9}} converted to the USNO system.
The composite model magnitude is the linear combination of an exponential model fit and a de {{Vaucouleurs}} model fit that best fits the image; the PSF magnitudes are not appropriate because we have included non-stellar sources in the sample.  
We tag a USNO-B object as blended if its angular distance to a  {{DR9}} object (that has an overlapping HWHM with another  {{DR9}} object) is less than the sum of the HWHM of the {{DR9}} objects. 
For each band, {{there are}} $\sim$ 500,000 {{-- 1 million objects}} ($\approx$ 1\%{{-2\%}} of entries) {{that}} suffer from this kind of blending; more than 80\% of these blended objects are tagged as blended in more than one band.

A simpler, alternative approach to identifying blended objects is to require that an object in one catalog is matched to exactly one object in another catalog within a large radius, as done by \citet{Sesar+06}. However, our empirical approach accounts for the variable effective size of objects and retains genuine objects that might have been otherwise rejected. Our method also takes into account the higher angular resolution of the  {{DR9}} images compared to those from which USNO-B were derived.

Another approach is to consider the classification flags provided by USNO-B that distinguish stellar from non-stellar objects.  However, the relatively low accuracy of the classification \citep[$\approx$ 85\%;][]{usnob} and the large number of objects in our cross matched catalog implies that a large number of false objects would be retained and a large number of real objects would likely be removed if this was adopted.

\section{Photometric Recalibration of USNO-B}
\label{sec: recal}

The photometry in USNO-B suffers from systematic inaccuracies that exceed the reported errors.
The authors of USNO-B point out that the photometry is the weakest aspect of the catalog \citep{usnob}.
An illustration of these inaccuracies is shown in Figure \ref{fig: weird histos}.  The Figure shows the distribution of USNO-B and SDSS magnitudes of objects on four plates.  The solid dark lines show the frequency distributions of magnitudes as reported in the original USNO-B catalog. For comparison, the frequency distribution of magnitudes of {{DR9}} objects, cross-matched to USNO-B and placed onto the USNO-B photometric system as described in \S\ref{sec: xmatch} and \S\ref{sec: photo systems}, are shown as red dashed lines.  The magnitudes for  {{DR9}} objects are in continuous distributions with monotonically increasing numbers of fainter objects with a decline near the limiting magnitude of USNO-B objects on the plate, as expected. 

The USNO-B data are distributed quite differently. The panel on the lower right shows a relative deficit of objects with $m_{USNO}$ near 19; this is a common feature for most plates. All four panels show large discontinuous jumps in the number of USNO-B objects in adjacent 0.05-mag wide bins. The lower left panel shows a very large discontinuity with most of the objects on the plate having identical USNO-B magnitudes ($m_{USNO} = 20.70$).  The panel on the upper left shows a wide gap in the distribution with a collection of objects with an unphysical $m_{USNO} \approx 24$ mag.
These unusual properties are likely a consequence of the photometric calibration of the USNO-B catalog.  Due to a lack of available standard stars, \citet{usnob} used calibration curves that were calculated separately for bright and faint stars, and both required considerable extrapolation.  {{The original USNO-B photometric calibration of more than 55\% of our cross-matched objects used faint standards that were not on the same plate as the objects themselves (e.g., using overlapping plates).}}
The deficit of objects in {{the lower right panel of}} Figure \ref{fig: weird histos} near $m_{USNO} \approx 19$ mag is likely due to a mismatch of the calibration curves; the `pileup' of objects at one value of magnitude {{in the other panels}} is likely due to an anomalous calibration curve. 

By taking advantage of the high photometric accuracy of  {{DR9}},  the USNO-B photometry can be recalibrated and the systematic errors can be substantially reduced.  {{However, a robust recalibration that yields accurate photometry can only be carried out for the $\approx$ 44 million USNO-B objects in our cross matched catalog, and not for all $\approx$ 260 million USNO-B objects in the overlap between USNO-B and DR9. The reason is that a number of steps in the recalibration process are dependent explicitly or implicitly on the photometry of the DR9 counterparts (discussed in detail below).}}

We recalibrate all objects in our cross-matched catalog using a recalibration process that shares some similarities to the techniques described by \citet{Munn+04} and \citet{Sesar+06}. 
However, there are a number of key differences between our recalibration process and those of \citet{Munn+04} and \citet{Sesar+06}.  For clarity, we describe our process here and discuss the differences with earlier work in \S\ref{sec: comparison}. 

Our approach to recalibration  is fundamentally based on four assumptions:  that no more than a few percent of the objects vary in magnitude (flux) by a few percent between USNO-B and {{DR9}} epochs; that the photometric uncertainties in  {{DR9}} data are negligible compared to those in USNO-B; that systematic errors are the dominant contribution to the photometric errors in USNO-B; and that those systematic errors are spatially correlated on the sky.  The justification for the last assumption is empirical and is confirmed by \citet{Munn+04} and \citet{Sesar+06}. 
The recalibration is done in two stages; first a correction over small angular scales ($\approx 10\arcmin$) as described in \S\ref{sec: recal 1}, then a correction over larger angular scales ($\approx 6\arcdeg$) as described in \S\ref{sec: recal 2}.  A summary of these steps is shown in Figure \ref{fig: flowchart}.

\subsection{Small-scale correction}
\label{sec: recal 1}

In the first stage of the photometric recalibration, we make corrections to USNO-B objects grouped over small areas of the sky.  Each field imaged as part of each phase of SDSS is uniquely characterized by a run number, field number, and camera column number.  Each of these smallest imaging units subtends 0.034 deg$^2$ and forms the basis for the  photometric calibration.  In our cross matched catalog, there are $\approx$ 500,000 of these imaging units that enclose at least one object.  
If a single SDSS imaging unit happens to enclose objects from more than one USNO-B plate, the SDSS imaging unit is split into separate sub-units to ensure one sub-unit falls on a single USNO-B plate.
For each sub-unit and for each USNO-B band, we perform the following steps:

\begin{enumerate}
\item{identify a set of objects that we define as calibration objects, which each satisfy the following conditions:}
\begin{itemize}

\item{$m_{SDSS} < (m_{lim}$~-~0.5 mag) , where $m_{lim}$ is the limiting magnitude of the whole USNO-B plate.  The limiting magnitude is defined empirically such that 95\% of objects on the plate satisfy $m_{USNO} < m_{lim}$.  The mean value of $m_{lim}$ is 20.6, {{19.6}},  {{21.0}}, {{19.7}} and 18.7 mag for bands $O,E,J,F,$ and $N$, respectively. 
To determine $m_{lim}$, we only consider objects that are identified as not being near bright stars (\S\ref{sec: blend bright}), are not blended (\S\ref{sec: blended}), {{are not spike or halo artefacts (\S\ref{sec: spurious})}}, and are brighter than  {{magnitude 22.0}}.  The  {{constraint}} $m_{USNO} < 22$ avoids several objects on some plates that have unreasonably faint magnitudes (see Figure \ref{fig: weird histos}).  
For the  {{54}} plates  {{with}} less than 20 objects  available to determine $m_{lim}$ ($<$ 2\% of the total number of plates), we use the average value of $m_{lim}$ for all plates in that band.  We select objects 0.5 mag brighter than $m_{lim}$ to  avoid objects near the faint end of the magnitude distribution where the USNO-B magnitudes are less reliable.
}

\item{the object is identified as not being near bright stars (\S\ref{sec: blend bright})  {{nor}} blended (\S\ref{sec: blended}), {{nor a spike or halo artefact (\S\ref{sec: spurious})}} and}
\item{$(u-g) >0.7$. This criterion avoids blue objects that are potentially variable quasars \citep{Richards+02}}.
\\

On average, there are 45 calibration objects for each sub-unit (per band); within each USNO-B band, 75\% - 90\% of sub-units have four or more calibration objects.

\end{itemize}
\item{if there are four or more calibration objects, determine the coefficients $A,B,C$ that minimize the absolute difference $|m^{\prime}_{USNO} - m_{SDSS}|$ among the calibration objects,  
where
\begin{equation}
\label{eq: stage 1}
m^{\prime}_{USNO} = A~m_{USNO} + B~(color) + C
\end{equation}
is the small-scale recalibrated magnitude of the calibration objects, $m_{USNO}$ is the original USNO-B magnitude, $color = (g-r)$ for bands $O,E,J,F$ and $color = (r-i)$ for band $N$. This model enables corrections for low order non-linearities and zero point errors in the original USNO-B calibration. 
The minimization was performed with the Levenberg-Marquardt algorithm \citep{Markwardt09}. 
We found that the introduction of more than three parameters (e.g., second-order magnitude or color terms) did not significantly improve the model fits.  An example of a best fit model with 55 calibration objects is shown in Figure \ref{fig: stage 1 recal}.  Note that these parameters are not all independent; $A$ and $C$ are generally anti-correlated.} 

\item{if a convergent solution is found, use equation (\ref{eq: stage 1}) to calculate $m^{\prime}_{USNO}$ for {\it{all}} objects in a given band belonging to this imaging sub-unit (using the parameters from the best fit model). 
In order to avoid adverse effects of atypical model fits, this adjustment is made only if the model parameters $A, B $ and $C$ and the standard deviation of ($m^\prime_{USNO} - m_{SDSS}$), $\sigma_m$, are not significantly different than their mean values for all sub-units in that band.  Frequency distributions of $A,B,C, $ and $\sigma_m$ are shown in Figure \ref{fig: step 1 coeffs}.  The values near the mean are well fit by a Gaussian distribution; the shaded areas in the figure show the values that are within 5 standard deviations from the mean.
A revised value of $m^{\prime}_{USNO}$ is calculated only if $A, B, C$, and $\sigma_m$ are all within these shaded areas, otherwise no recalibration is possible and all objects on the sub-unit are discarded in that band (see Figure \ref{fig: flowchart}).   
This criterion is met by $\approx$ 95\% of sub-units for each band and it ensures that coefficients that are derived from poor model fits, or from sub-units with a small number of calibration objects or with an unusual magnitude distribution are discarded.  We found that using a bootstrapping technique or a $\chi^2$ goodness-of-fit metric was not sufficient to exclude anomalous  fits. Discarded objects are removed from the catalog (only in the problematic USNO-B band) and not retained for the second stage of recalibration to be described in \S\ref{sec: recal 2}.}

\end{enumerate}

An illustration of the improvements to USNO-B magnitudes resulting from the small angular scale recalibration is shown in Figure \ref{fig: thumb demo}. {{The Figure shows corrections of several magnitudes and it highlights the importance of identifying the appropriate imaging sub-units. In Table \ref{tab: coeffs}, we provide the full list of coefficients $A, B, C$ and $\sigma_m$ for objects that are processed by the steps described above (i.e., discarding the 5$\sigma$ outliers). The Table contains 2,549,344 rows, one for each unique imaging sub-unit.}} 

\subsection{Large-scale correction}
\label{sec: recal 2}

After removing systematic errors on small angular scales in the first stage of recalibration, there is a common pattern in the residual magnitudes $m^{\prime}_{USNO} - m_{SDSS}$ as a function of $m_{SDSS}$ for objects in a given band on the same USNO-B plate.  Several examples of this pattern are shown in Figures \ref{fig: stage 2 sample 1} \& \ref{fig: stage 2 sample 2}.   The underlying cause for this pattern is not clear, but it is likely an artifact of the original photometric calibration of the USNO-B catalog (see \S\ref{sec: usno}).  

In order to remove this systematic error, we calculate the median deviation of $m^{\prime}_{USNO} - m_{SDSS}$ as a function of $m_{SDSS}$ for each plate. Examples of this median deviation are shown as solid green lines in Figures \ref{fig: stage 2 sample 1} \& \ref{fig: stage 2 sample 2}.  To evaluate the median deviation in a robust way, we use an initial minimum bin width of 0.5 mag and a bin separation of 0.1 mag. The median deviation and the interquartile range (IQR) of the data are recorded for each bin. If there are fewer than 10 objects in a bin, the IQR is flagged as being derived from a `sparse' bin and the bin width is increased incrementally in units of 0.05 mag until 10 objects are in the bin. The median deviation is recalculated each time the bin width is increased. This procedure minimizes the influence of blended or large amplitude variables, and it ensures that a continuous function is identified for each plate, even in cases where the data are sparse.  The median deviation is removed from all objects on a plate with the deviations interpolated between adjacent bin centers.  For objects with magnitudes below or above the minimum and maximum bin center, respectively, the deviation is extrapolated from the four closest bin centers.  The resultant magnitude, $m^{\prime\prime}_{USNO}$, is the final recalibrated magnitude in a given band. 
 
We estimate the photometric uncertainty of each recalibrated object from the IQR in each (original) bin. 
The IQR is a robust statistic that is insensitive to large outliers, some of which may correspond to real variable stars and quasars. 
For normally distributed data, the Gaussian width $\sigma$ is equivalent to $\approx$ 0.72~IQR.  
In each bin, we assign 0.72$~$IQR  to $\sigma_{USNO}$, the uncertainty for the recalibrated objects in the original bin.  The distribution of uncertainties may be discontinuous, depending on the population of objects in the narrow bins.  These {{uncertainties}} are shown in Figures \ref{fig: stage 2 sample 1} \& \ref{fig: stage 2 sample 2} as solid red lines; an asterisk is overlayed on the red line for bins flagged as `sparse'.  
There are $\approx$ 10,000 objects in each band that are flagged as `sparse' ($<$ 0.03\% of the total). We caution that uncertainties flagged as `sparse' may be over- or underestimates. 

\begin{figure}
\includegraphics[scale=0.45]{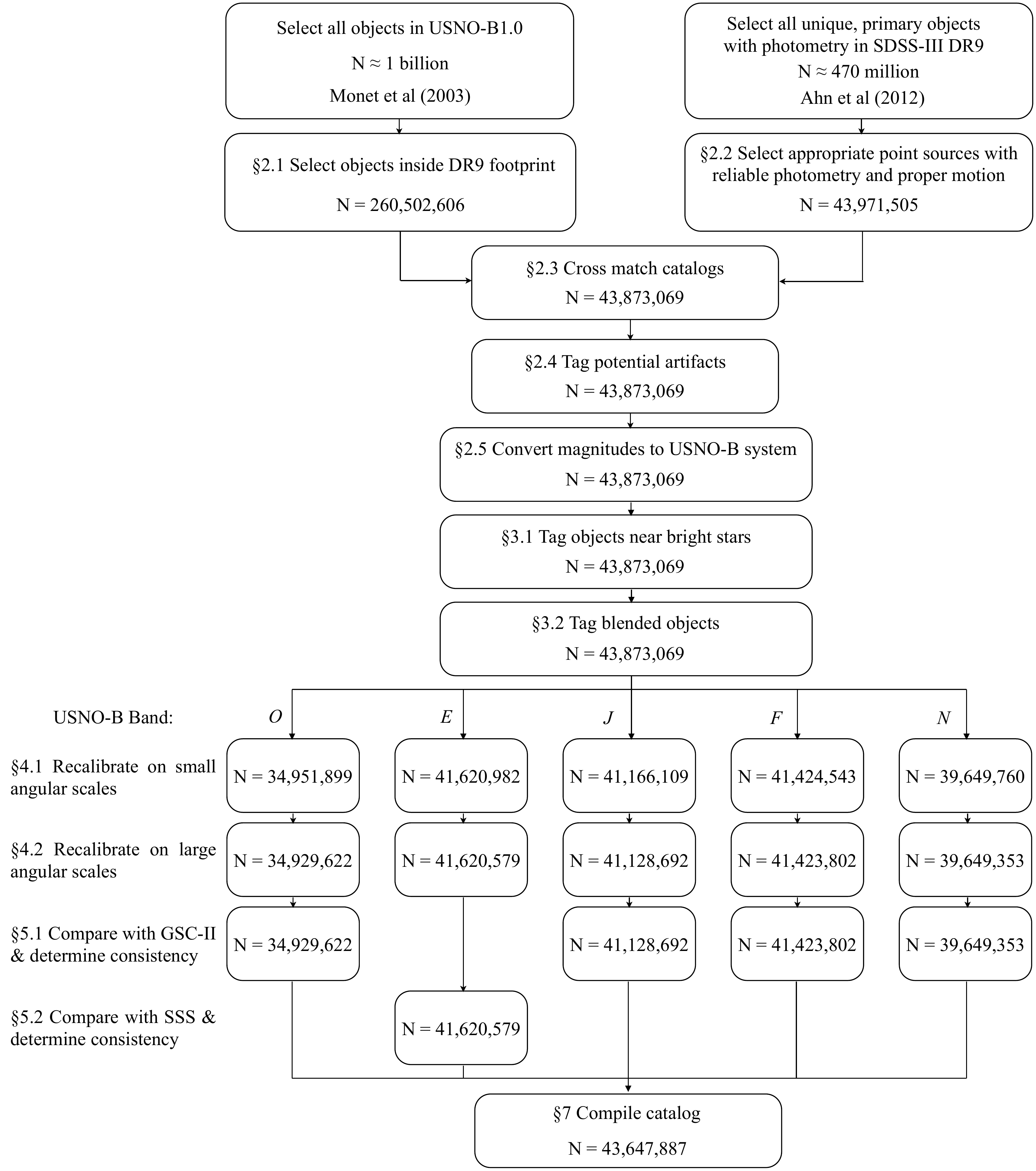}
\caption{Flowchart outlining the process by which the catalog was created.  For each step, the number of objects that remain after the step was applied is given as $N$. For some steps, no objects are discarded and $N$ does not change between consecutive steps. A reference to the original catalogs or relevant section in the paper is also provided. 
\label{fig: flowchart}
}
\end{figure}

\begin{figure}
\includegraphics[scale=0.55,angle=90]{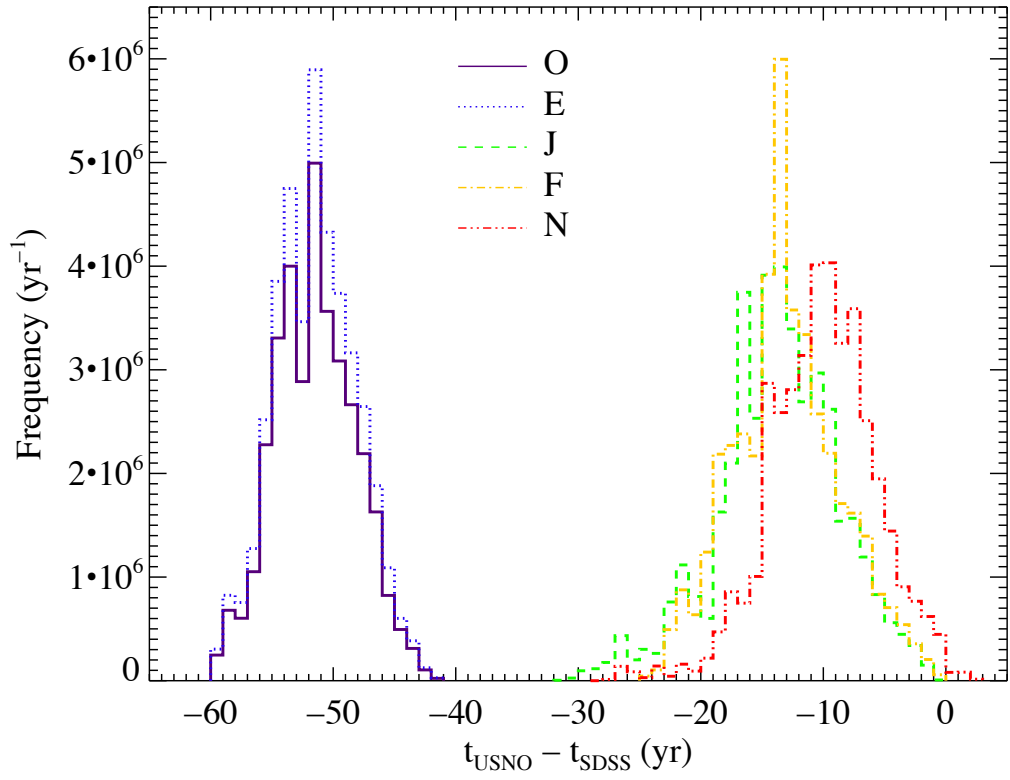}
\caption{Histogram of the difference between the epochs of cross-matched point sources between USNO-B and DR{{9}}. Objects are grouped together by the USNO-B band shown in the legend.
\label{fig: histo epochs}
}
\end{figure}

\begin{figure}
\includegraphics[scale=0.6]{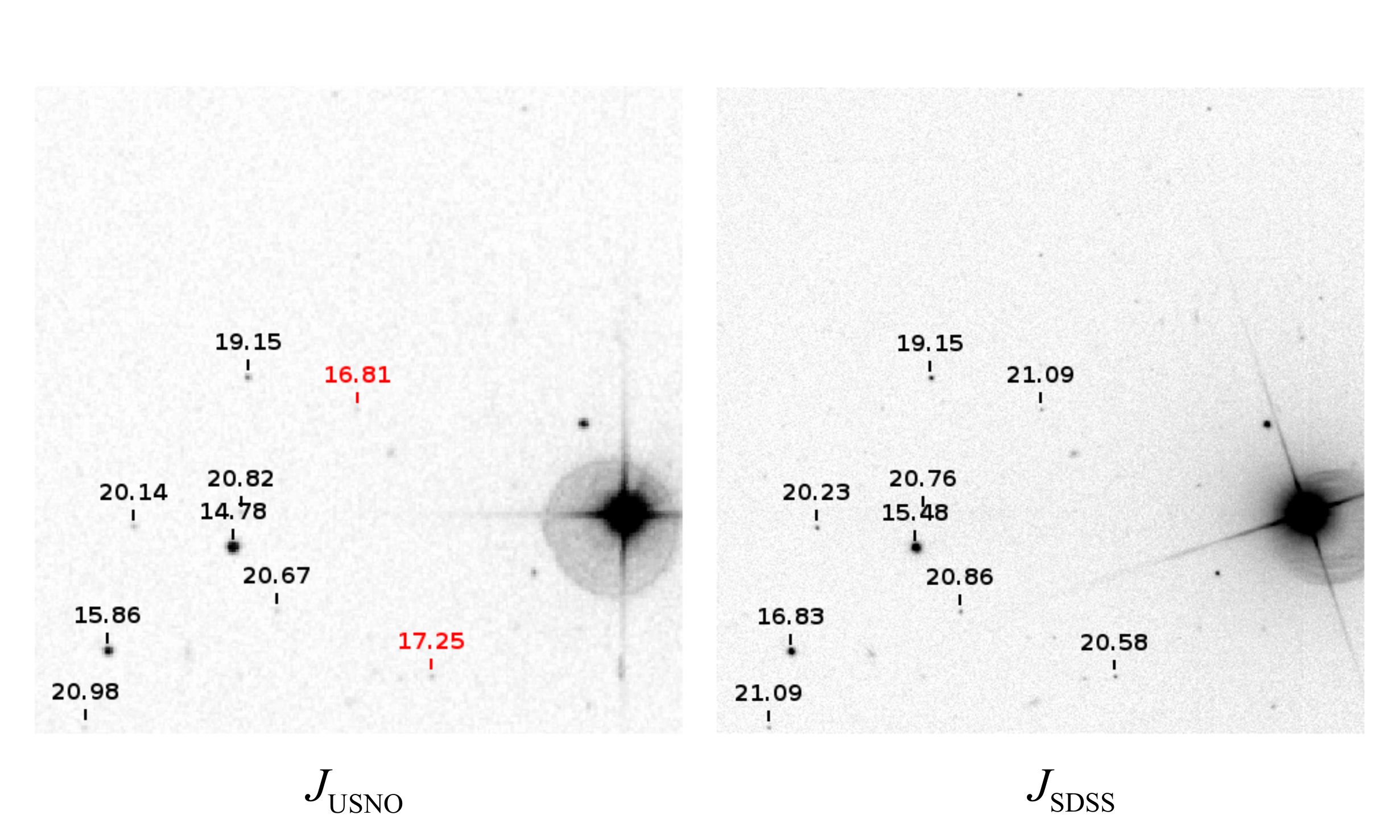}
\caption{Two images, each 5\arcmin\ on a side, from the DSS POSS-II $J$ band (left) and SDSS DR{{9}} $g$ band (right) toward ($\alpha, \delta) = (358.3130\arcdeg, +2.3446\arcdeg)$, illustrating the influence of proximity to bright stars on USNO-B magnitudes. North is up and east is to the left. The tick marks and labels show the location and magnitude of objects in our cross-matched catalog described in (\S\ref{sec: xmatch}). The magnitudes from USNO-B ($m_{USNO}$) and magnitudes from DR{{9}} converted to the $J$ band ($m_{SDSS}$) are shown on the left and right panels, respectively. 
The magnitudes labeled in red are identified automatically and they correspond to objects that are within the exclusion radius of a bright star. 
None of the objects appear to vary in brightness between the images. However, two of the objects labelled in red have USNO-B magnitudes that exceed their DR{{9}} counterpart by more than 3 magnitudes (see \S\ref{sec: blend bright}).
\label{fig: thumbnail bright star}
}
\end{figure}

\begin{figure}
\includegraphics[scale=0.7,angle=90]{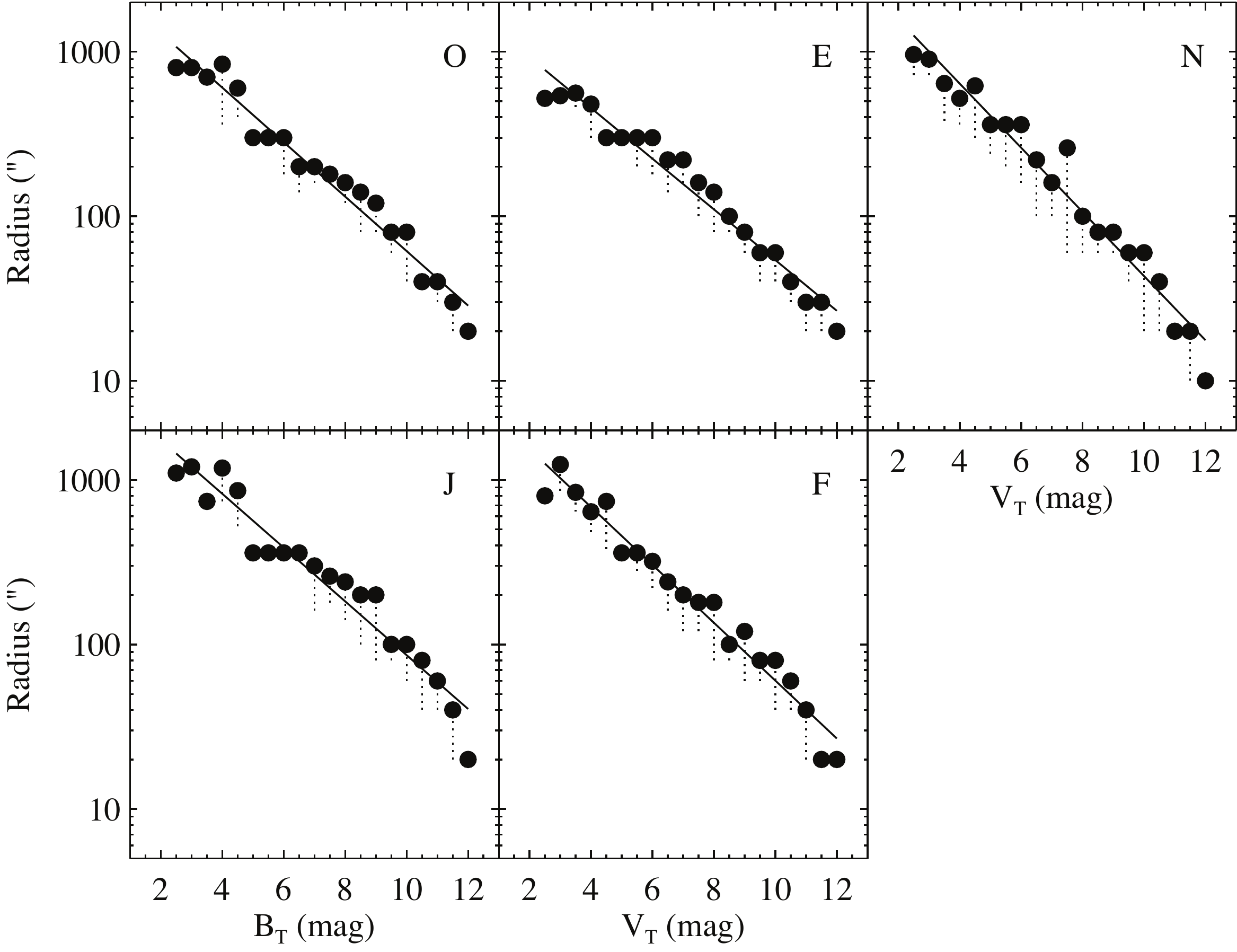}
\caption{Exclusion radius of bright stars on USNO-B photometric plates as a function of their Tycho-2 magnitude. The USNO-B band is labeled in the upper right corner of each panel. The solid circles represent the maximum radius among the stars in each narrow magnitude bin; the vertical dashed lines show the full range of radii. The solid line shows the least-squares best fit of an exponential function to the data.} 
\label{fig: bright star width}
\end{figure}

\begin{figure}
\includegraphics[scale=0.7,angle=90]{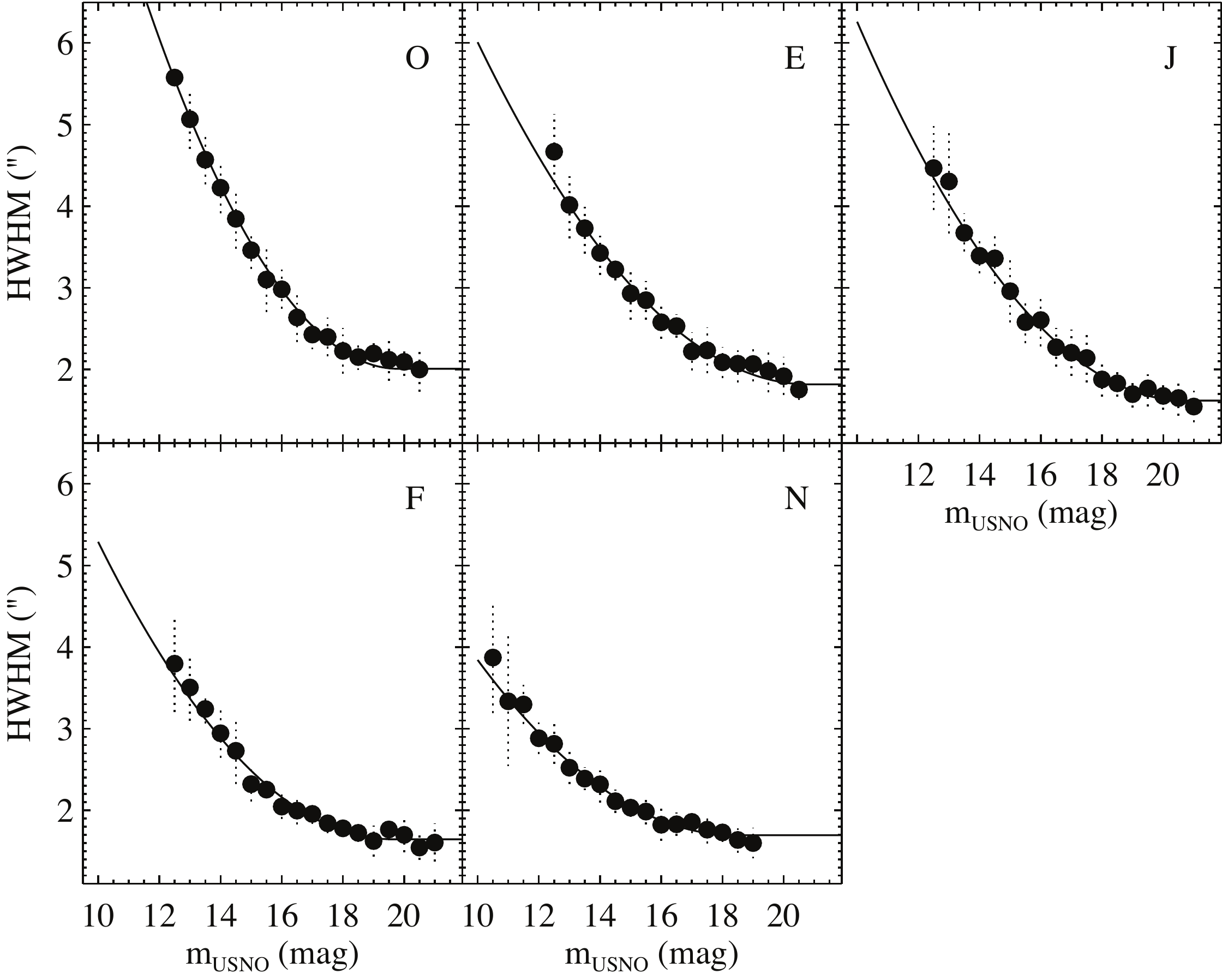}
\caption{Half-width at half-maximum of images of stars on USNO-B photometric plates as a function of stellar magnitude. The solid circles represent the average measurement among a set of randomly selected, isolated stars in each magnitude bin; the vertical dashed lines are the root-mean-square deviation from the mean. The solid line shows the best fit to data with a second-order polynomial (fixed at a constant value at the faintest magnitudes; see \S\ref{sec: blended}). 
\label{fig: faint star width}
}
\end{figure}

\begin{figure}
\includegraphics[scale=0.45]{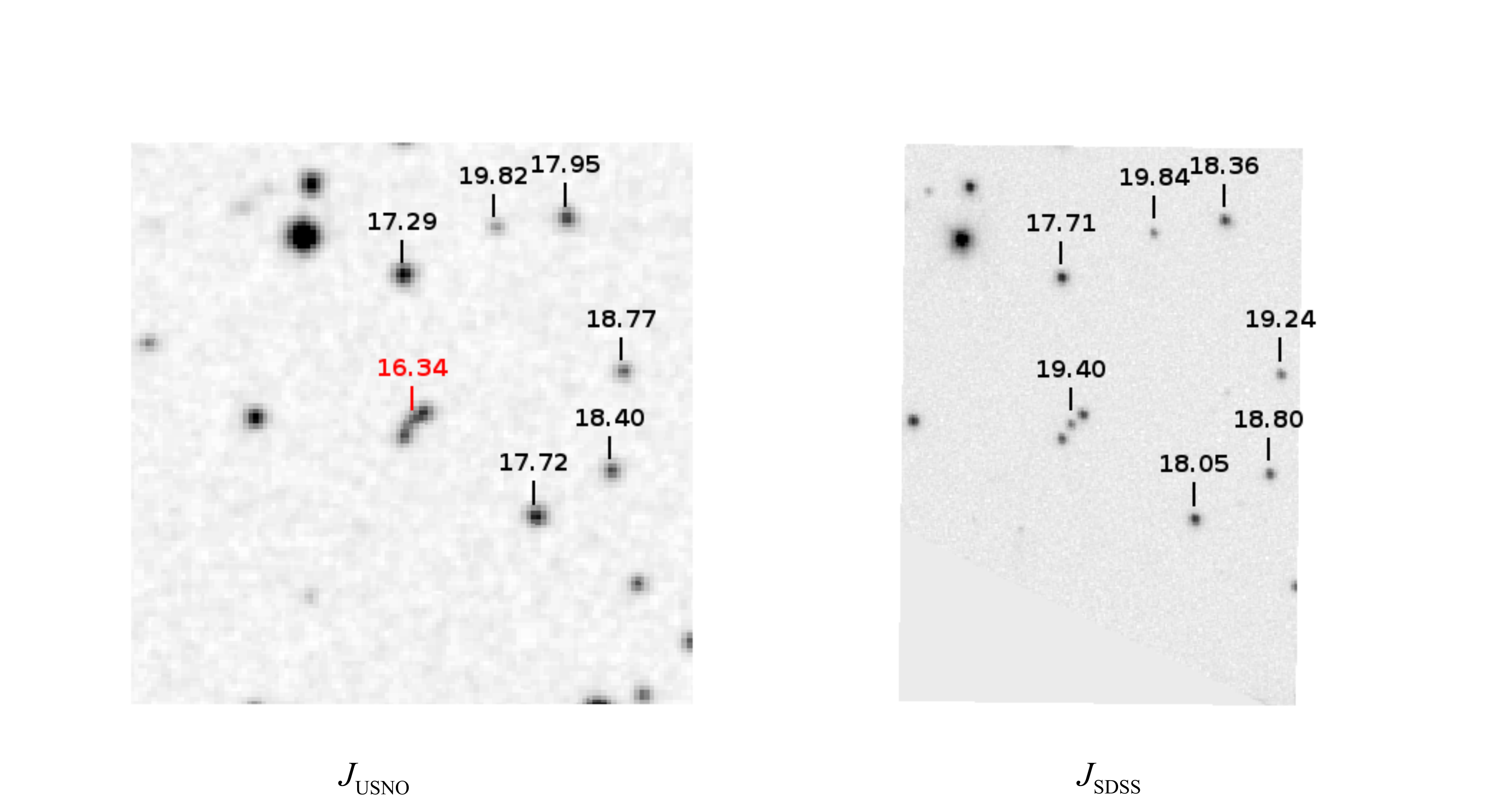}
\caption{Two images, each {{up to}} 2\arcmin\ on a side, from the DSS POSS-II $J$ band (left) and SDSS DR{{9}} $g$ band (right) toward ($\alpha, \delta) = (96.2109\arcdeg, +36.6990\arcdeg)$; north is up and east is to the left. The tick marks and labels are as for Figure \ref{fig: thumbnail bright star}, except that the object labeled in red at the center is identified automatically as being blended with objects that are resolved in SDSS (see \S\ref{sec: blended}).
\label{fig: thumbnail blended sdss}
}
\end{figure}

\begin{figure}
\includegraphics[scale=0.6,angle=90]{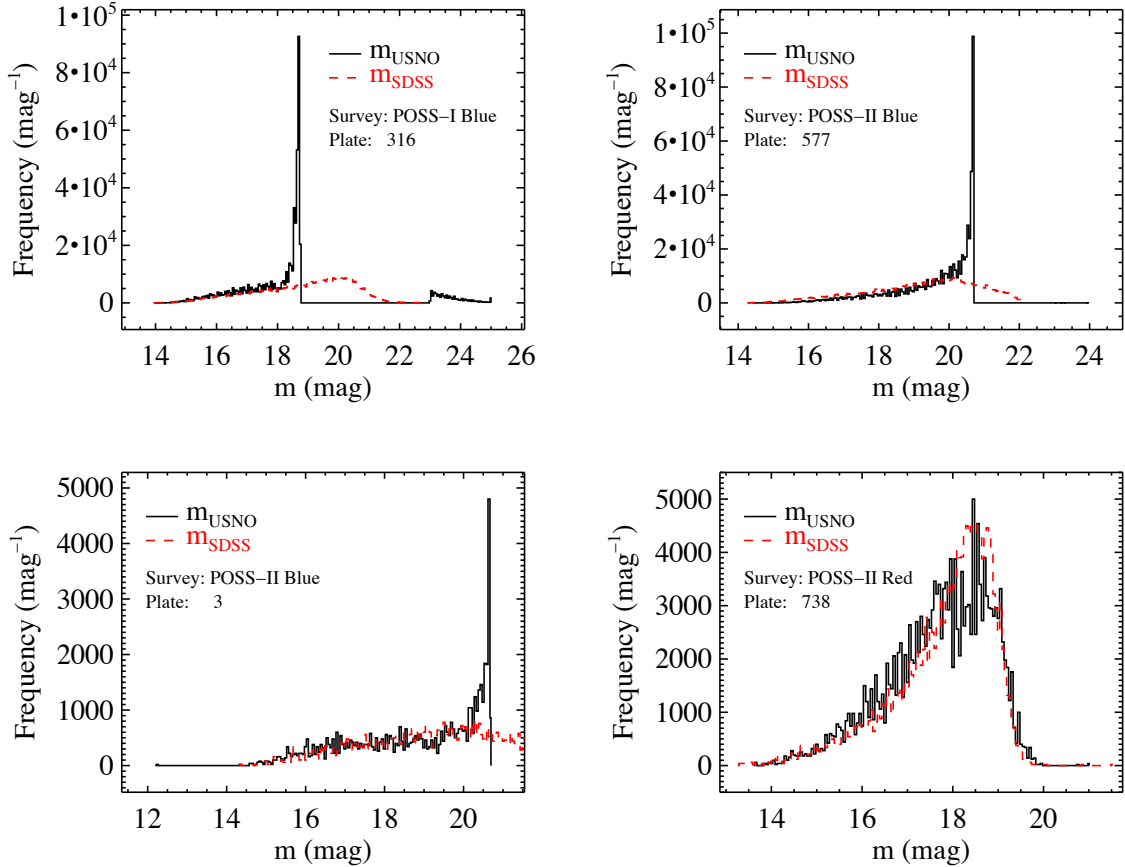}
\caption{Histograms of the USNO-B and SDSS magnitudes of objects on a sample of USNO-B plates. 
Original USNO-B magnitudes are shown as solid dark lines ($m_{USNO}$); magnitudes of cross matched objects from SDSS DR{{9}} (converted to the USNO system) are shown as dashed red lines ($m_{SDSS}$).  In each individual panel, the area under each curve is the same. The corresponding USNO-B survey name and plate number is labelled in each panel. 
{{All of these plates were calibrated using faint photometric standards that were one or more plates away (on overlapping plates)}}. The large discontinuities in adjacent bins among USNO-B magnitudes illustrate the systematic errors in USNO-B photometry. 
\label{fig: weird histos}
}
\end{figure}

\begin{figure}
\includegraphics[scale=0.7,angle=90]{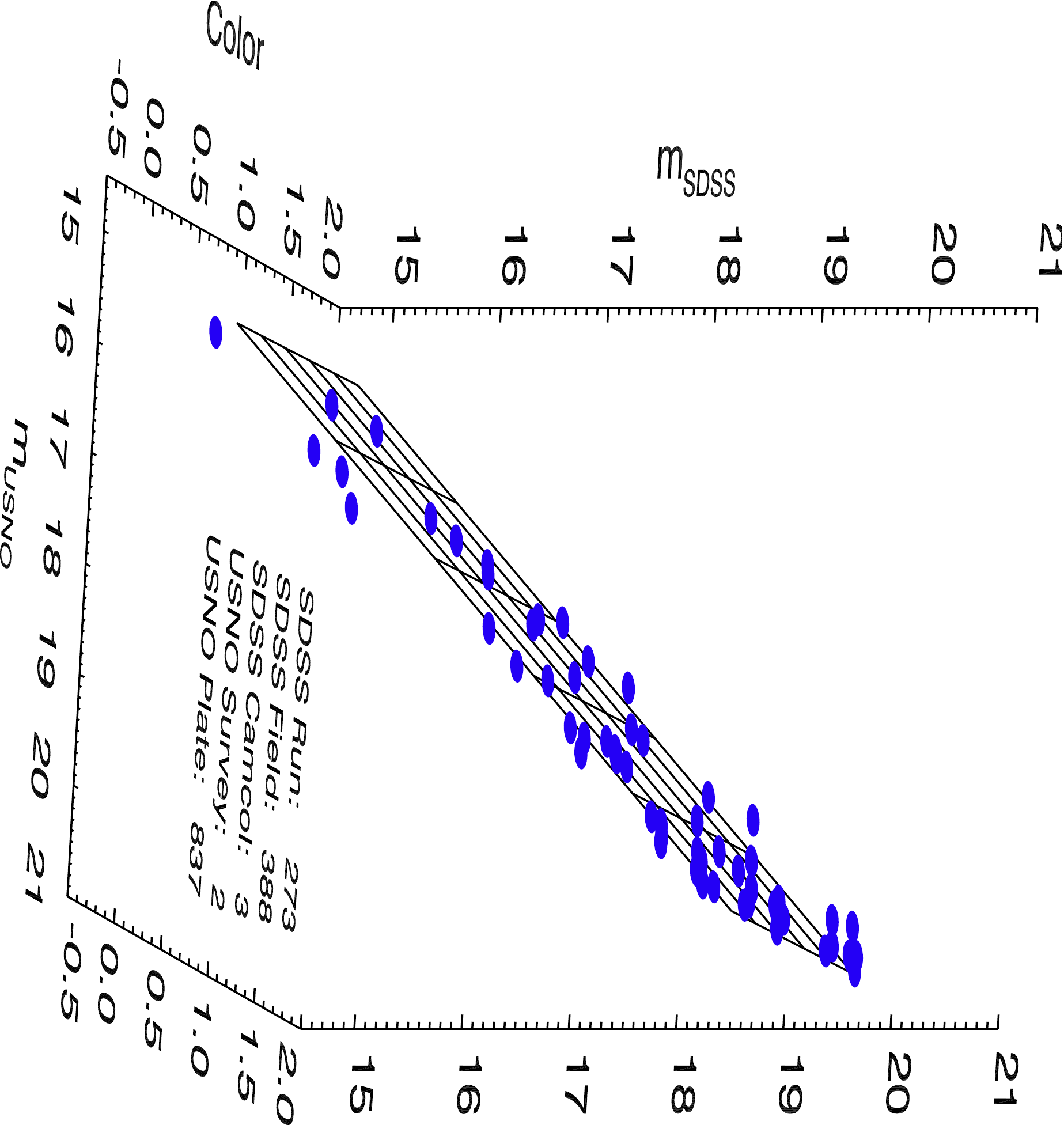}
\caption{Demonstration of the first stage of photometric recalibration of USNO-B magnitudes.  The blue dots show the original USNO-B magnitudes ($m_{USNO}$), the magnitudes of the cross matched objects in SDSS DR{{9}} converted to the USNO system ($m_{SDSS}$), and the colors (from SDSS DR{{9}}), for a set of 55 calibration objects in a small area of sky in the $J$ band.   The three-parameter best fit to the data ($A = 0.93, B = 0.09, C = 0.75$) is shown as a hatched plane (see \S\ref{sec: recal 1} for details).   
\label{fig: stage 1 recal}
}
\end{figure}

\begin{figure}
\includegraphics[scale=0.6,angle=90]{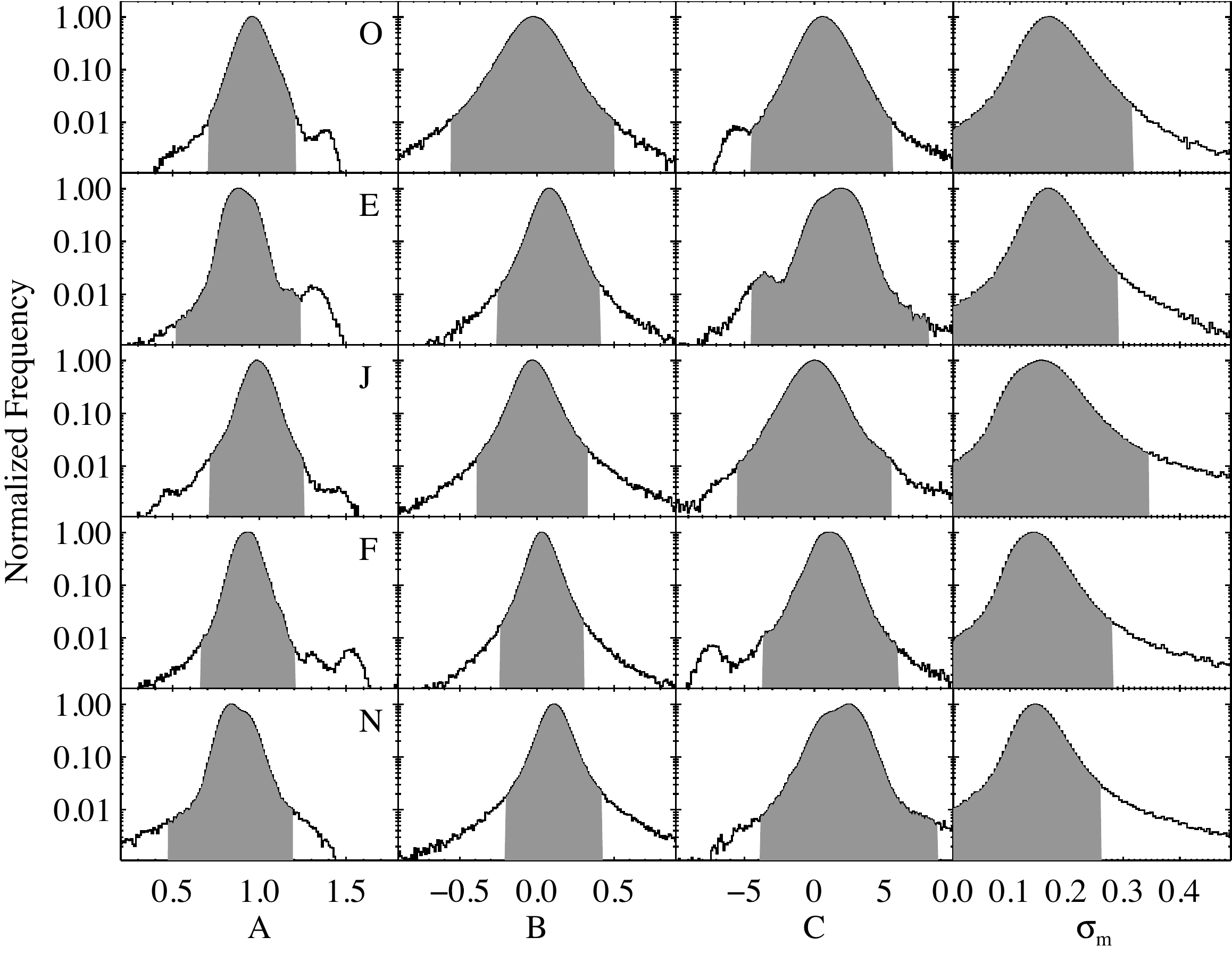}
\caption{Histograms of the best fit model parameters $A, B, C$ and $\sigma_{m}$ for all imaging sub-units. The USNO-B band is labelled in the leftmost panel of each row.  The shaded areas show the values of the parameters that are within 5 standard deviations from the mean.  For each band, objects from sub-units that have one or more model parameters outside the shaded regions are discarded (see \S\ref{sec: recal 1} for details). 
\label{fig: step 1 coeffs}
}
\end{figure}

\begin{figure}
\includegraphics[scale=0.35]{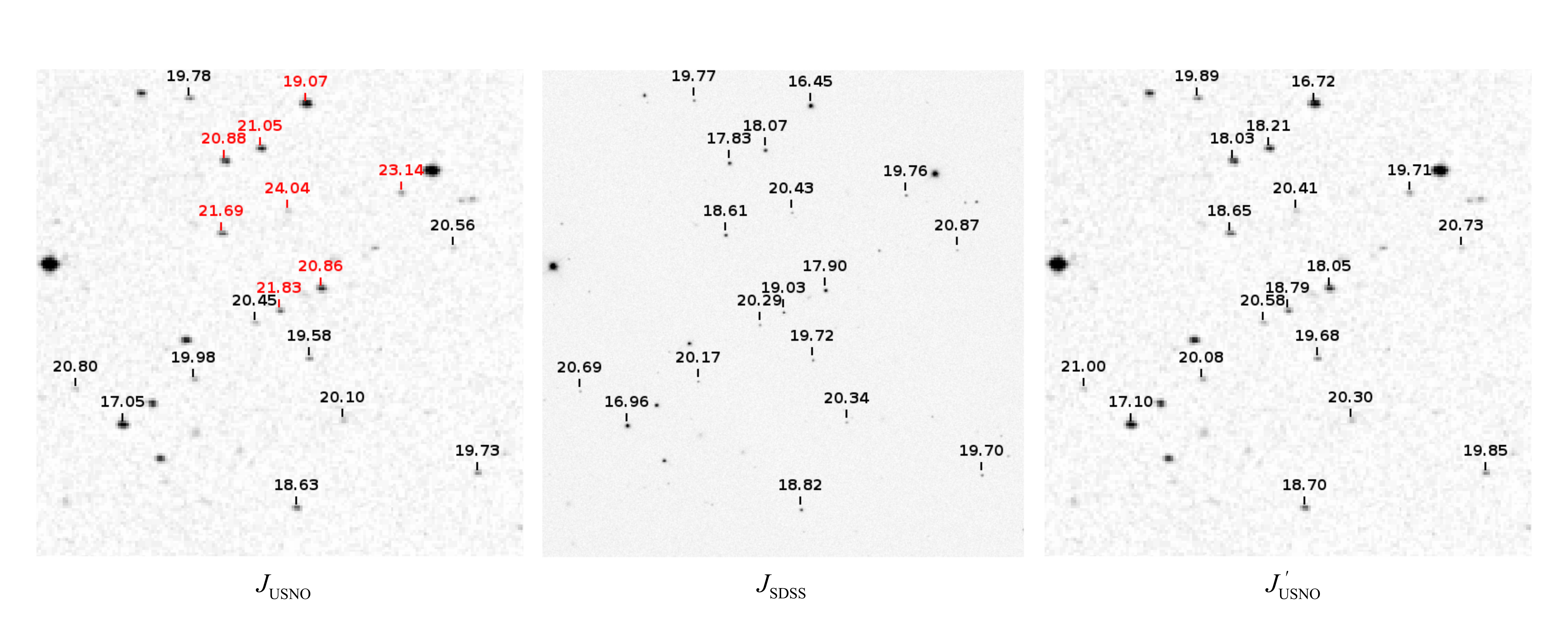}
\caption{Three images, each 5\arcmin\ on a side, toward $(\alpha, \delta) = (223.5665\arcdeg, -7.6213\arcdeg)$ from the DSS POSS-II $J$ survey (left, right), and SDSS $g$ band (center), illustrating the small-scale recalibration. The orientation, tick marks, and labels are as for Figure \ref{fig: thumbnail bright star}, except that the labels on the rightmost panel shows the USNO-B magnitudes after the small-scale recalibration (the images on the leftmost panel and rightmost panel are identical).
In the leftmost panel, the objects labelled in red have $|m_{USNO} - m_{SDSS}| > 2$~mag and they all belong to the same imaging sub-unit; the objects labelled in black all belong to a different imaging sub-unit. 
A visual comparison of the POSS-II and DR{{9}} images show no large amplitude variables; after recalibration all objects shown here have $|m^{\prime}_{USNO} - m_{SDSS}| < 0.3$~mag. 
The figure shows the importance of identifying unique combinations of plates and imaging units for recalibration. 
\label{fig: thumb demo}
}
\end{figure}

\begin{figure}
\includegraphics[scale=0.6,angle=-90]{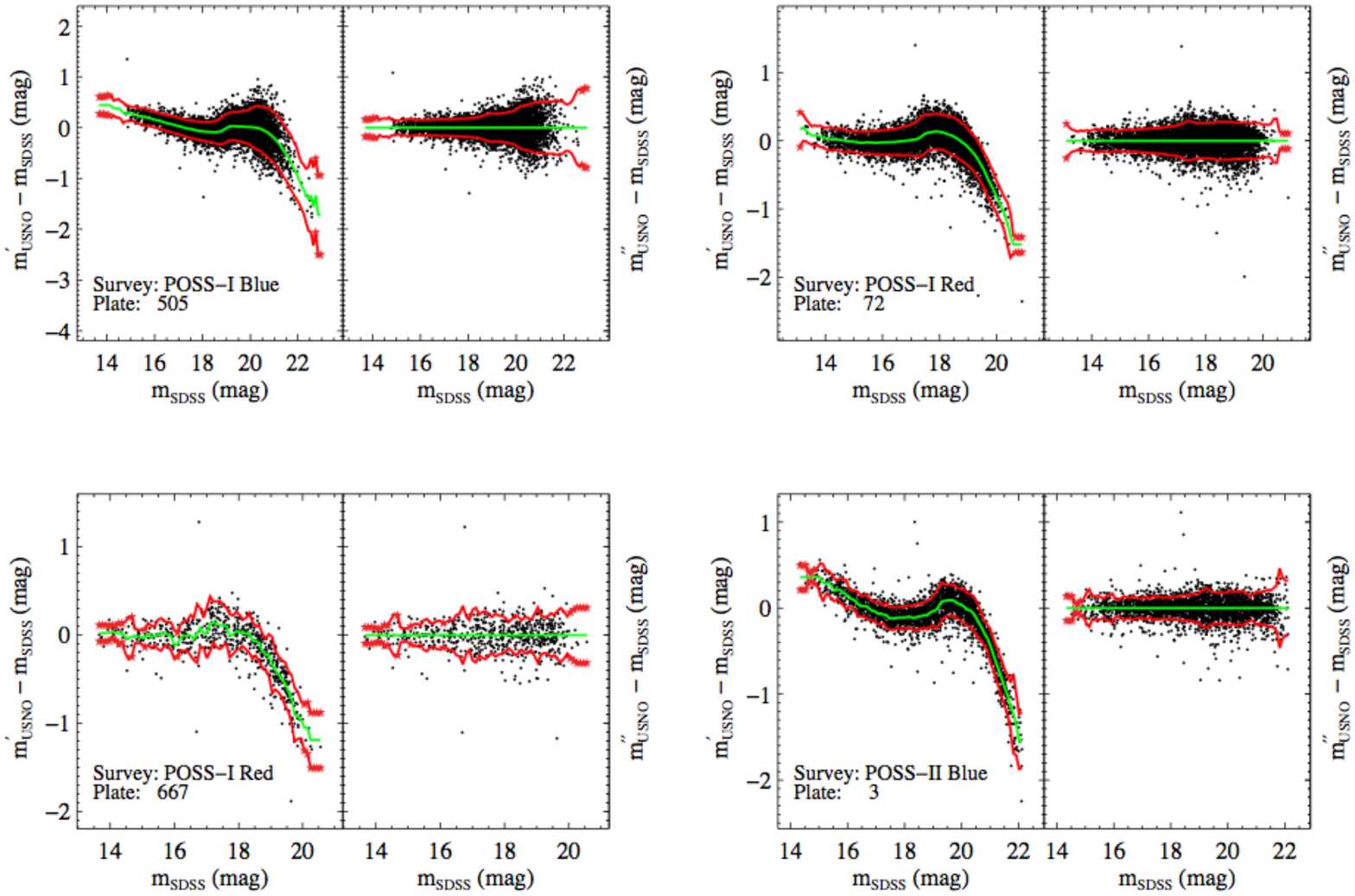}
\caption{Illustration of the second stage of the recalibration process for four typical USNO-B plates. 
The left hand side of each panel shows $m^{\prime}_{USNO}$ - $m_{SDSS}$ as a function of $m_{SDSS}$, where $m^{\prime}_{USNO}$ is the USNO-B magnitude after the first stage of the recalibration.
The right hand side of each panel shows $m^{\prime\prime}_{USNO}$ - $m_{SDSS}$ as a function of $m_{SDSS}$ on the same vertical scale as on the left side, where $m^{\prime\prime}_{USNO}$ is the USNO-B magnitude after the second stage of the recalibration.
The solid green and red lines show the median deviation and 3$\sigma_{USNO}$ dispersion around this median, respectively.
The red asterisks show places where $\sigma_{USNO}$ has been derived from a sparse bin.
Note the common pattern among all plates, with negative slopes and inflection points for the distribution near $m_{SDSS} \sim 19$.
The median deviation has been subtracted from $m^{\prime}_{USNO}$ to calculate $m^{\prime\prime}_{USNO}$; this is shown in the right hand side of the panels.  For clarity, a maximum of 5,000 randomly selected points are shown for each plate (see \S\ref{sec: recal 2} for details).
}
\label{fig: stage 2 sample 1}
\end{figure}

\begin{figure}
\includegraphics[scale=0.6,angle=-90]{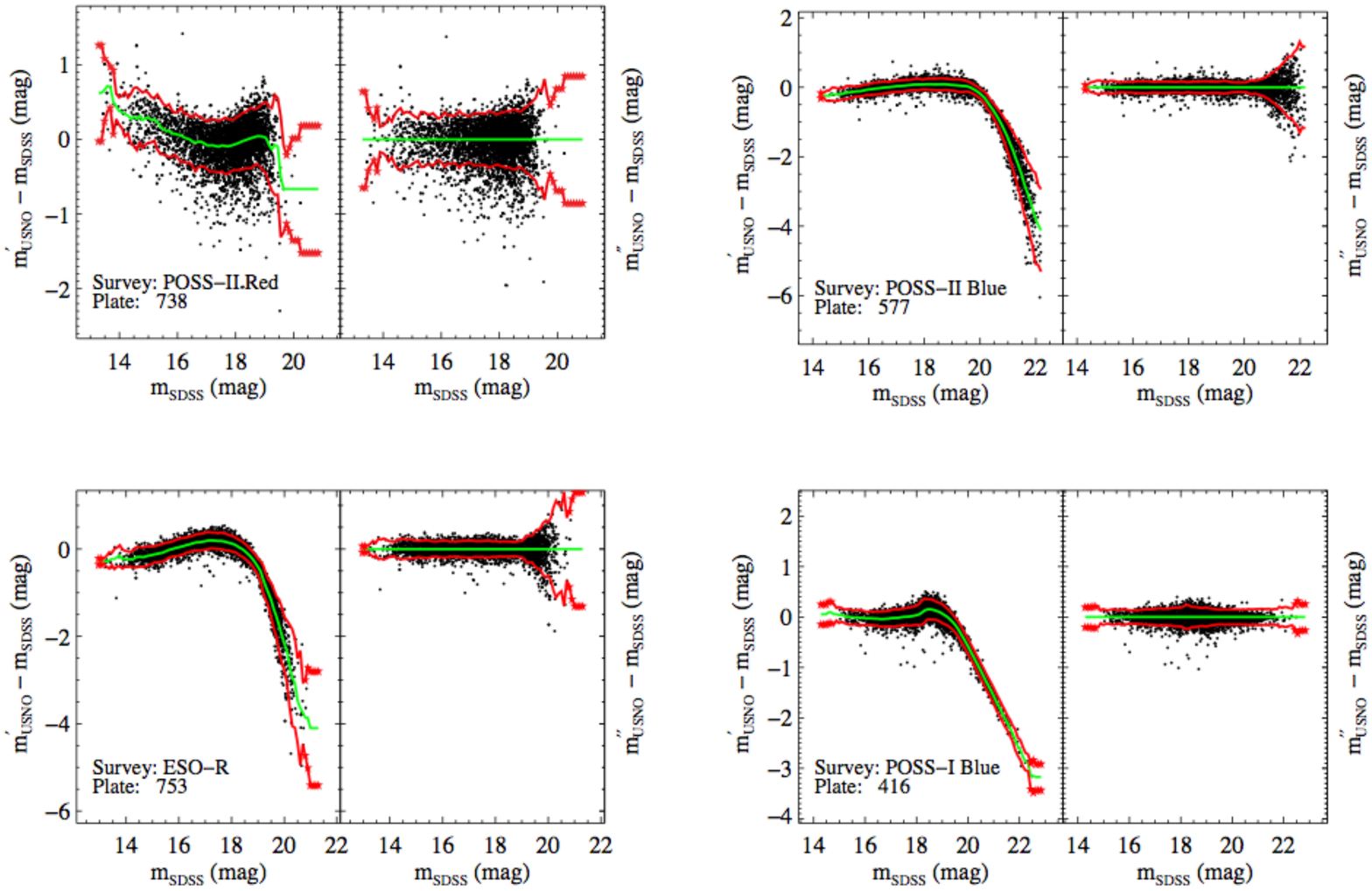}
\caption{Same as Figure \ref{fig: stage 2 sample 1}, but for four plates that exhibited some of the largest median deviations of $m^{\prime}_{USNO}$ - $m_{SDSS}$.
\label{fig: stage 2 sample 2}
}
\end{figure}

In order to ensure the robustness of the values for the median deviation and $\sigma_{USNO}$, we only perform this second stage of recalibration for plates that have at least 10 objects that passed the first stage of recalibration as described in \S\ref{sec: recal 1}.  We also impose the constraint that the first stage recalibrated magnitude, $m^{\prime}_{USNO}$, is less than 22.0; this avoids both objects near the completeness limit and also the few plates that have clusters of objects with unphysical $m_{USNO} \approx 24$. The very few objects that fail to meet these criteria ($<$ 0.05\% per band) are removed from the catalog, as shown in Figure \ref{fig: flowchart}.

The right hand side of each panel in Figure \ref{fig: stage 2 sample 1} shows the results of this second stage of recalibration. Most of the adjustments are of the order of a few tenths of a magnitude. However, as shown in Figure \ref{fig: stage 2 sample 2}, some of the corrections can be very large. For plates where there is a `pileup' of USNO magnitudes at one specific value, the adjustments can be several magnitudes; the maximum correction to any one object in the catalog is  {{4.5}} mag. The total number of fully recalibrated objects is $\approx$ {{35-42}} million per band; see Figure \ref{fig: flowchart} for the exact numbers. 

\section{Identification of anomalous USNO-B magnitudes}
\label{sec: consistent}

\begin{figure}
\includegraphics[scale=0.55]{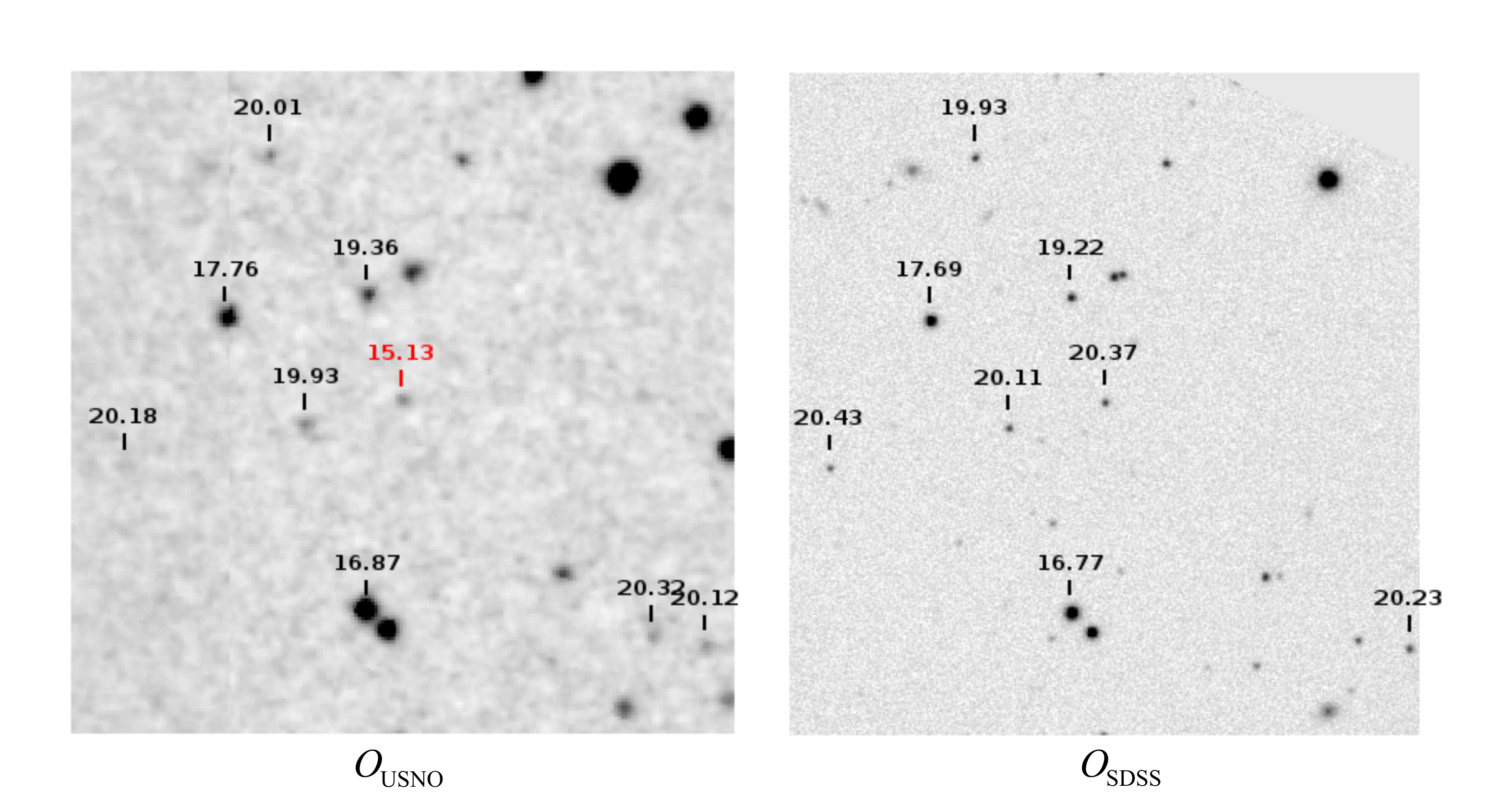}
\caption{Two images,  {{3\farcm5}}\ on a side, toward {{$(\alpha, \delta) = (214.9278\arcdeg, -18.4581\arcdeg)$}} from the DSS POSS-I $O$ survey (left), and SDSS $g$ band (right), illustrating an anomalous USNO-B magnitude.
The orientation, tick marks, and labels are as for Figure \ref{fig: thumbnail bright star}, except that
an isolated, unblended object at the center of the image is labelled in red.
This object has a USNO-B magnitude that differs from its DR{{9}} counterpart by {{more than 5}} magnitudes. 
However, the images show that the magnitude has not changed significantly. 
All of the other cross matched objects have magnitudes that differ between USNO-B and DR{{9}} by $\lesssim 0.5$ mag, and therefore the object in the center is not accurately recalibrated (the recalibration give $m^{\prime\prime}_{USNO}$ = {{15.75}} for this source).  Unlike USNO-B, the magnitude of this object in GSC-II is  consistent with the magnitude in DR{{9}} (coincidentally, $m_{SDSS}$ = $m_{GSC}$ = {{20.37}}). 
\label{fig: thumbnail inconsistent}
}
\end{figure}

A visual inspection of some photographic plates reveals a number of isolated, unblended point sources with USNO-B magnitudes that are inconsistent with the magnitudes of nearby objects of similar appearance on the same plate.  For example, there are a number of cases where two sources, separated by a few arcminutes, appear to be the same brightness on the plate. However, USNO-B reports that one of the objects is several magnitudes brighter or fainter than its neighbor.  An example of this is shown in Figure \ref{fig: thumbnail inconsistent}, where two objects of apparently the same brightness have $|m_{USNO} - m_{SDSS}| >$ {{5}} mag. 
Because the other objects on the same imaging sub-unit have relatively low values of $|m_{USNO} - m_{SDSS}|$, the recalibration process does not adjust $m_{USNO}$ accurately for this anomalous source.

The origin of this inconsistency is not clear; the effect does not appear to be correlated with other properties (plate position, proximity to bright stars, etc.). 
There is a slight tendency for the original USNO-B magnitude of these objects to be overestimated rather than underestimated, i.e.\ they are too faint ($\approx$ 60\% of inconsistent objects have $m^{\prime\prime}_{USNO} - m_{SDSS} < 0.0$).
We found that objects suffering from this inconsistency can be identified automatically by checking if $m^{\prime\prime}_{USNO}$ is consistent with two other catalogs derived from the same photographic plates: the Guide Star Catalog 2.3.2 \citep[GSC-II;][]{Lasker+08} and the SuperCOSMOS Sky Survey \citep[SSS;][]{SSS-Overview}.  
These catalogs report the survey name for each object, but not the plate identification number.
Because many of the plates overlap on the sky, the coordinates of an object are not sufficient to infer the plate identification. Therefore we could not use these catalogs in the recalibration process described in \S\ref{sec: recal 2} above.  However, these catalogs can be used to identify most of the objects that suffer from the unusual inconsistency illustrated in Figure \ref{fig: thumbnail inconsistent}.  Our approaches to check our results against GSC-II and SSS are described in \S\ref{sec: comp gsc} and \S\ref{sec: comp sss}, respectively, and are summarised in Figure \ref{fig: flowchart}.

\subsection{Comparison to GSC-II}
\label{sec: comp gsc}

GSC-II is an all-sky astrometric and photometric catalog derived from digital scans of photographic plates from the Palomar and UK Schmidt telescopes \citep{Lasker+08}.  Although it is based on many of the same plates as USNO-B, the scanning, processing, calibration, etc.\ of GSC-II were performed independently.  Among other data, GSC-II provides photometry in four optical bands corresponding to the USNO $O, J, F, $ and $N$ bands and epochs; it does not include data for band $E$. 

A total of $\approx$ 240 million objects in GSC-II fall within a $1^\circ$ wide boundary surrounding the areal footprint of  {{DR9}}.  In order to assess whether the USNO-B magnitudes are consistent with GSC-II, objects in the two catalogs must be cross matched.  However, the coordinates in GSC-II are not all reported at a single epoch and no proper motions are provided. We first match GSC objects to  {{DR9}} objects, using the coordinates of  {{DR9}} objects at epoch 2000.0 and with a large matching radius of 30\arcsec\ (many-to-many).  We then convert  GSC-II coordinates to  epoch 2000.0 using the absolute proper motion of each object's  {{DR9}} counterpart. The radial distances between the GSC-II and {{DR9}} objects are recalculated, and only the closest object with an angular separation  {{$\le$ 1}}\farcs0 is recorded in the final (one-to-one) cross matched catalog between GSC-II and  {{DR9}} of $\approx$  {{44}} million objects. {{We note that GSC-II has a very small number of objects that share identical celestial coordinates but have distinct identifications and photometry. In these rare cases, we cross-matched the DR9 object with the corresponding GSC-II object that had the lowest alphanumeric GSC-II identification.}}

The GSC-II objects are then matched to USNO-B objects by identifying their common match in  {{DR9}} (c.f.\ \S\ref{sec: xmatch}).  We imposed additional requirements that the photometric error reported by GSC-II, $\sigma_{GSC}$, is non-null (i.e., $0.0 < \sigma_{GSC} < 9.9$), and that the GSC-II photometry for a given object comes from the same survey as the corresponding USNO-B photometry (e.g. Palomar, UK Schmidt, or ESO/SERC).  The fraction of objects in our USNO-DR{{9}} cross-matched catalog (derived in \S\ref{sec: xmatch}) that were {{recalibrated and}} not matched to a GSC-II object is 5\% - 10\%, depending on the USNO-B band. 

Because of differences in processing and calibration, the photometric systems of USNO-B and GSC-II are not identical. However, \citet{Sesar+06} showed that the differences between the two systems in the USNO-B $J$ and $F$ bands are very small; the typical discrepancy between the two is $\lesssim$ 0.1 mag.  
This difference is small compared to the inconsistencies that we are checking for, so we did not convert the GSC-II photometry to the USNO-B system. 

For each USNO-B band, 
we classify a USNO-B object as ``consistent" with its matched GSC-II counterpart if either of the following conditions are met:

\begin{itemize}

\item{ $|m^{\prime\prime}_{USNO} - m_{GSC}| \le (\sigma_{USNO} + \sigma_{GSC})$, where $m_{GSC}$ and $\sigma_{GSC}$ are the reported magnitude and error of an object in GSC-II in the {{relevant}} band, respectively. This condition accounts for objects with large uncertainties in their magnitudes. We did not add the errors in quadrature for two reasons: the source of the errors are largely systematic (not random), and they are derived from the same plates, so they may not be independent.}

\item{ $|m^{\prime\prime}_{USNO} - m_{GSC}| \le 0.5$ mag.  This condition implicitly imposes a lower limit to the total uncertainty on the two individual magnitude measurements. This prevent{{s}} objects with magnitudes that are within the systematic uncertainty of both catalogs from being classified as inconsistent.
The GSC-II catalog reports stellar photometry accurate to 0.13 - 0.22 mag; for individual objects, magnitude errors are approximate and conservative estimates. However, the mean value of $\sigma_{GSC}$ in each band is strongly peaked at $\approx$ 0.4 mag, with a minimum value of 0.05 mag. By comparison, the mean value of $\sigma_{USNO}$ is 0.15 mag but is as low as $\approx$ 0.02 mag for some objects. 
}

\end{itemize}

Approximately 3\% - 7\% of objects fail the above tests (depending on the USNO-B band) and are hence are identified as having inconsistent magnitudes between GSC-II and USNO-B. A larger fraction of these inconsistent objects have USNO magnitudes that also differ significantly from their  {{DR9}} counterpart: among objects with $|m^{\prime\prime}_{USNO} - m_{SDSS}| > 0.5$, the fraction of objects classified as inconsistent is 15\% - 30\%. 
Within this subset of inconsistent objects, more than 85\% have GSC-II magnitude{{s}} that are consistent with SDSS. 
These results show that this inconsistency, i.e.~bad USNO-B magnitudes of unknown origin, is a major contributor to the potential false identification of variable objects between USNO-B and SDSS.

Inconsistency between USNO-B and GSC-II objects is indicated in our catalog; this tagging process did not remove any objects.  A null tag is used to indicate USNO-B objects that were not matched to a GSC-II object  in a given band (see \S\ref{sec: results} for details).

\subsection{Comparison to SSS}
\label{sec: comp sss}

The comparison with GSC-II in \S\ref{sec: comp gsc} above allowed us to identify anomalous USNO-B magnitudes in bands $O,J,F$ and $N$. For band $E$, we use the SuperCOSMOS Sky Survey \citep[SSS;][]{SSS-Astrometry,SSS-Images,SSS-Overview} for this purpose.
The SSS is also based on digitized scans of photographic plates from the Palomar, ESO, and UK Schmidt telescopes \citep{SSS-Astrometry,SSS-Images,SSS-Overview}.  
{{Access to SSS data is available through the SuperCosmos Science archive\footnote{\href{http://surveys.roe.ac.uk/ssa/}{http://surveys.roe.ac.uk/ssa/}}.
}}

We used the archive to cross match USNO-B  {{objects}} with {{those in}} the SSS catalog.  The cross matching is restricted to SSS objects of reliable quality in the $E$ band (quality flag = 0).  The SSS reports coordinates at the epoch of the plate from which positions were derived; USNO-B coordinates are all at epoch 2000.0. In order to account for proper motions, the cross matching was done in multiple stages. First, objects in USNO-B that are less than 30\arcsec\ away from objects in the SSS are matched to one another (many-to-many); for the USNO-B objects, the coordinates reported by USNO-B are used.  Second, the coordinates of the matched SSS objects are converted to epoch 2000.0 using the proper motions reported by the SSS.  This conversion was only applied to SSS objects that have high signal-to-noise proper motions ($> 3 \sigma$ in each coordinate) and have a valid $\chi^2 < 2$ ($\chi^2$ is a measure of the goodness-of-fit of the proper motion based on merged sources from multiple plates).  No coordinate conversion was applied to SSS objects with unreliable proper motions.  Finally, the closest matches between SSS and USNO-B objects within a {{1}}\farcs0 ($\approx$ {{5}}$\sigma$) radius were recorded. 

The SSS catalog does not report magnitude uncertainties for individual objects. Across the entire catalog, the absolute accuracy in the photometry is reported as $\sim 0.3$ mag for objects with $m > 15$ mag.  
Following our discussion in \S\ref{sec: comp gsc} above, a USNO-B object is classified as consistent with the magnitude of its SSS counterpart if $|m^{\prime\prime}_{USNO} - m_{SSS}| \le (\sigma_{USNO} + 0.3)$ or if $|m^{\prime\prime}_{USNO} - m_{SSS}| \le 0.5$ mag.   The fraction of $E$-band object with $|m^{\prime\prime}_{USNO}-m_{SDSS}| > 0.5$ mag that fail this test and are thus tagged as inconsistent is  {{43\%}}, considerably higher than the fraction of inconsistent objects in GSC-II (15\% - 30\%).  Because the origin of the inconsistency is not known, it is not known why this fraction is higher in the $E$ band relative to the other bands.  
{{One possibility is that strong spectral features in some variables stars (e.g. H$\alpha$ or TiO from Mira variables) are not accounted for in the conversion from the SDSS to USNO-B photometric system defined in \S\ref{sec: photo systems}.
}}

\section{Results}
\label{sec: results}

The recalibrated USNO-B magnitudes, SDSS DR{{9}} counterparts, tags for data quality and other relevant data are compiled into one catalog. The catalog contains {{43,647,887}} unique objects, each occupying one row in the table.  Each row has information for an object in at least one USNO-B band; the fraction of rows with information for more than one, two, three, or four bands is {{99.8\%, 98.4\%, 92.0\% and 65.3\%}}, respectively.
A complete description of all  {{51}} columns in the catalog is given in Table \ref{tab: columns}. Table \ref{tab: sample} shows a sample of the catalog.

{{The catalog, when combined with data in Table \ref{tab: coeffs}, contains enough information to enable users to reproduce our results from the originating catalogs or to change our recalibration scheme to create a modified catalog. 
For example, we include values of $m^\prime_{USNO}$ (from \S\ref{sec: recal 1}); these can be used to calculate our large-scale corrections (from \S\ref{sec: recal 2}) from $m^{\prime\prime}_{USNO}$ - $m^\prime_{USNO}$. Our catalog may also be used to calculate revised proper motions by using the identities and coordinates of cross matched objects in USNO-B and DR9.
}}

For convenience, we have combined several of the tags indicating proximity to bright stars, blending, etc., into one bit mask called a `quality flag' which is described in Table \ref{tab: bitmask quality}. The flag is composed of four bits, is represented as an integer in base 10, and can be filtered using bitwise operators. For example, if an object is within the exclusion radius of a bright star and is blended with one or more DR{{9}} objects and, but is {\it{not}} blended with USNO-B objects and has a $\sigma_{USNO}$ that is {\it{not}} derived from sparse data, its quality flag is $2^0 + 2^2 = 5$.

\subsection{Quantitative Improvement to USNO-B Photometry}

With the assumption that only a few percent of objects are genuinely time-variable between the epochs of the USNO-B and DR{{9}}, a good quantitative measure of the accuracy of USNO-B photometry is the frequency distribution of $m_{USNO} - m_{SDSS}$.  The width of this distribution is a joint measure of the photometric accuracy and the mean level of variability among variable stars and quasars.  The symmetry of the distribution is a measure of selection bias in the two catalogs,  especially arising from the differences in sensitivity.  Because SDSS is more than one magnitude deeper than the photographic plate surveys, it may be more likely that a brighter USNO-B objects is matched to a fainter DR{{9}} objects compared to the reverse scenario.  (We minimized this bias by incorporating the limiting magnitude of each plate into the first stage of our recalibration in \S\ref{sec: recal 1}.)

\begin{figure}
\includegraphics[scale=0.6,angle=90]{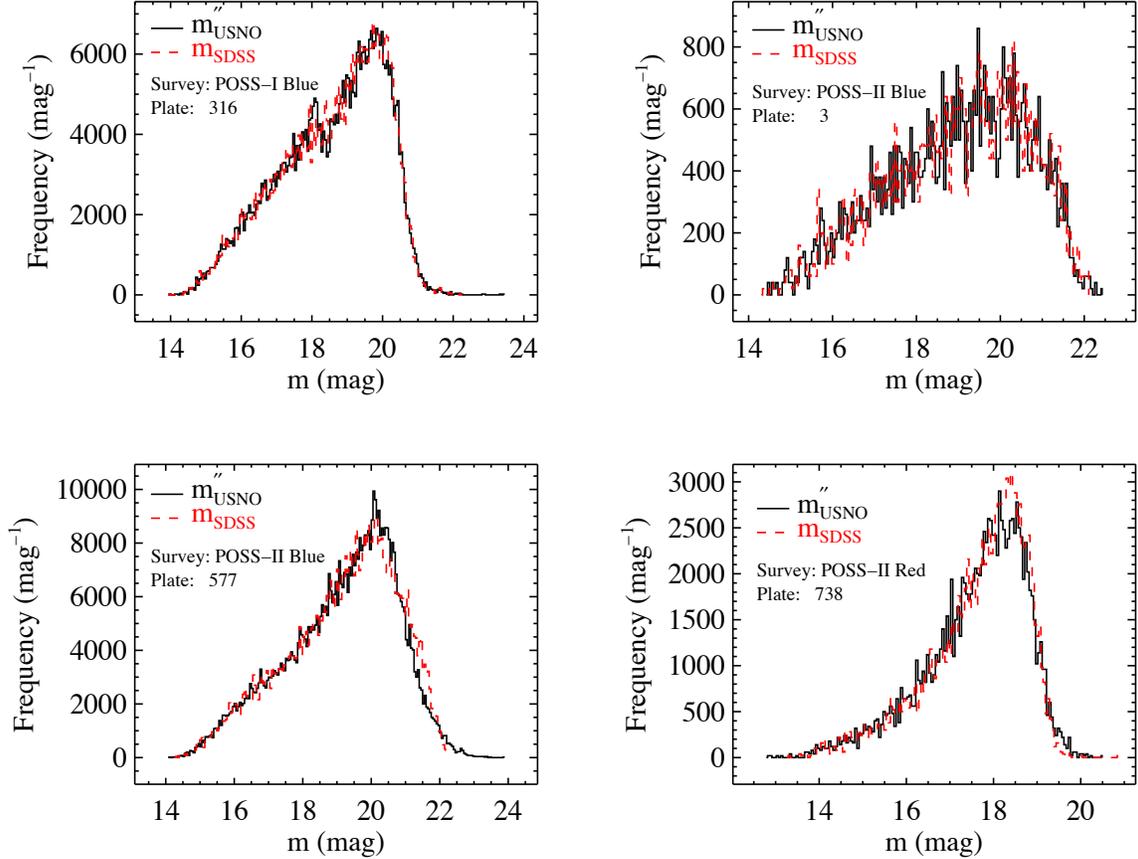}
\caption{Histogram of the recalibrated USNO-B magnitudes and SDSS magnitudes from the {\it{same}} sample of USNO-B plates as shown in Figure \ref{fig: weird histos}. 
Recalibrated USNO-B magnitudes are shown as solid dark lines ($m_{USNO}$); magnitudes of cross matched objects from SDSS DR{{9}} (converted to the USNO system) are shown as dashed red lines ($m_{SDSS}$).
In each individual panel, the area under each curve is the same.  Note that the area under each curve is smaller than area under the curves shown in Figure \ref{fig: weird histos} for the same plates. 
The reason is that some of the objects used in Figure \ref{fig: weird histos} have been discarded as part of the recalibration process described in \S\ref{sec: recal}.
\label{fig: weird histos recal}
}
\end{figure}

An illustration of the quantitative improvement that we have provided for USNO-B photometry is shown in Figure \ref{fig: weird histos recal}. The Figure shows the distribution of recalibrated magnitudes (solid black) and of the SDSS DR{{9}} magnitudes (dashed red) of the same objects on four photographic plates. The plates are identical to the ones represented in Figure \ref{fig: weird histos}.  After recalibration, the distributions are very similar and the large discontinuities seen in Figure \ref{fig: weird histos} are absent. 

\begin{figure}
\includegraphics[scale=0.65,angle=90]{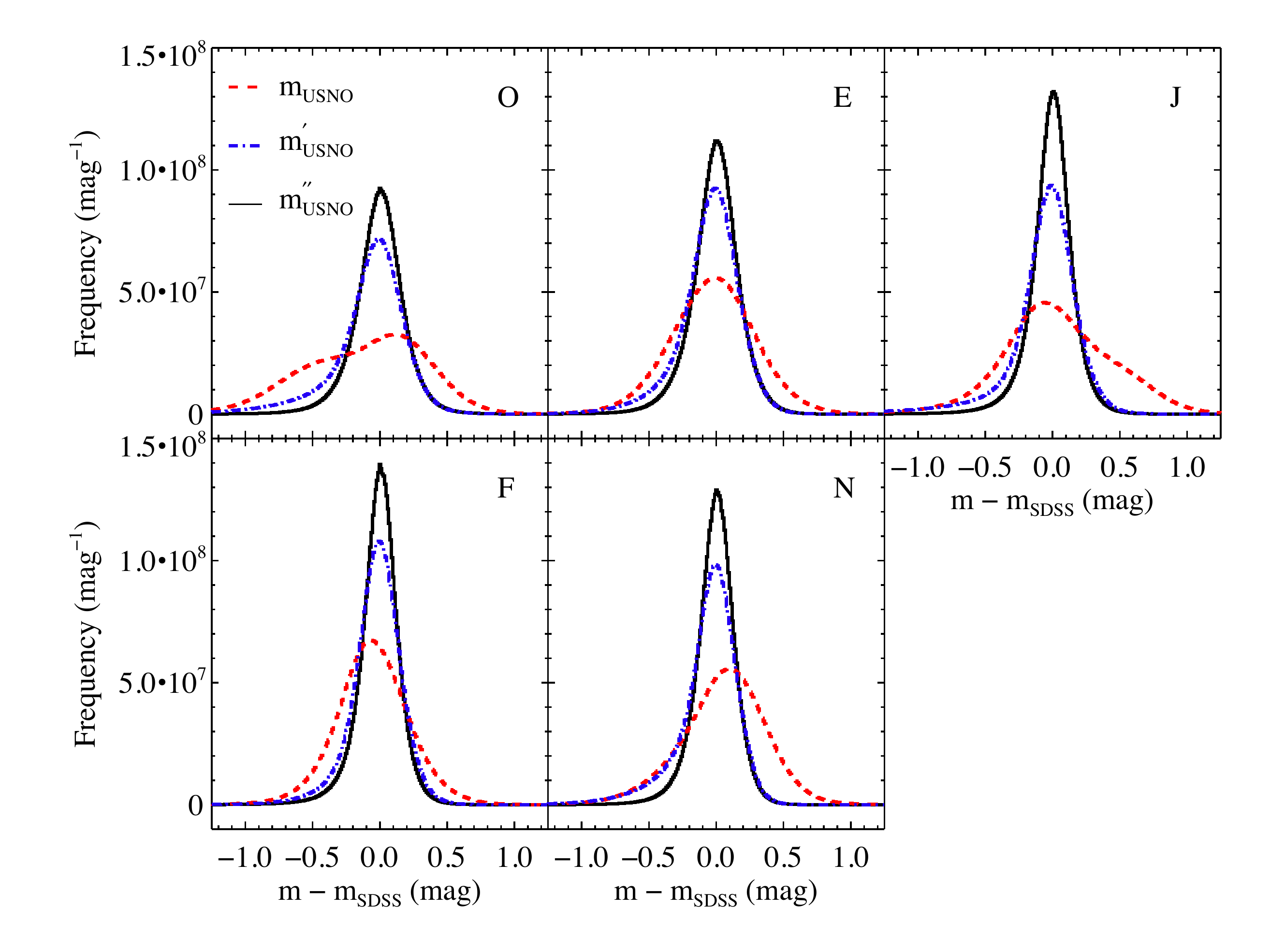}
\caption{Histograms of the difference between USNO-B and DR{{9}} magnitudes in our catalog. 
The original USNO-B magnitudes ($m_{USNO}$), first stage recalibrated magnitudes ($m^{\prime}_{USNO}$), and final recalibrated magnitudes ($m^{\prime\prime}_{USNO}$) are shown as dashed red, dot-dashed blue, and solid black lines, respectively. The relevant USNO-B band is indicated by the label in the upper right of each panel. 
Only objects that passed all stages of recalibration are shown, i.e.~the areas under the curves in a given panel are the same.
\label{fig: histo summary}
}
\end{figure}

\begin{figure}
\includegraphics[scale=0.6,angle=90]{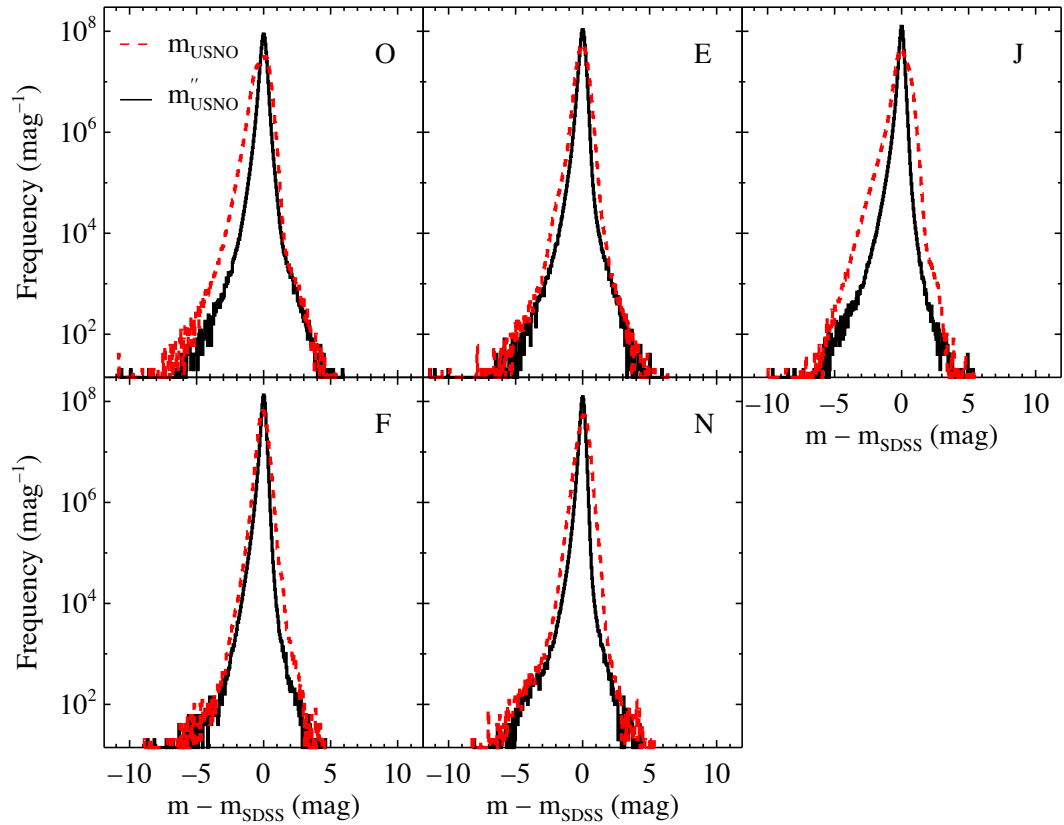}
\caption{Same as Figure \ref{fig: histo summary}, but with different axis ranges and scales. For clarity, the dot-dashed blue line has been omitted. The x-axis range has been expanded to show the full range of the data.  The y-axis is on a logarithmic scale with the lower end of scale chosen to show bins that have as few as one object in them.  
\label{fig: histo alt}
}
\end{figure}

Other illustrations of the quality of our USNO-B photometric recalibration are shown in Figures \ref{fig: histo summary} and \ref{fig: histo alt}. Figure \ref{fig: histo summary} show the frequency distribution of the difference between USNO-B and SDSS magnitudes. 
Only objects that passed all stages of recalibration are shown, i.e. the area under the curves are identical for each band. 
Compared to the original USNO-B data (dashed red), the fully recalibrated data (solid black) are more symmetric, are much narrower and have a mode much closer to zero. 
Figure \ref{fig: histo alt} shows the same data as Figure \ref{fig: histo summary}, but with expanded axes that show the full distribution of all data values. 

None of the frequency distributions in Figures \ref{fig: histo summary} and \ref{fig: histo alt} are well described by a Gaussian distribution; this makes it difficult to quantify the improvement in the USNO-B photometry with a single number.  Even the recalibrated data contains many outliers at very large values of $|m^{\prime\prime}_{USNO} - m_{SDSS}|$, as shown in Figure \ref{fig: histo alt}. The vertical scale in this Figure shows bins with as few as one object in them; the horizontal scale shows the full range of $m^{\prime\prime}_{USNO} - m_{SDSS}$.  In each band, there are  {{27,000 -- 64,000}} recalibrated objects with $|m^{\prime\prime}_{USNO} - m_{SDSS}| \ge 1.0$ mag (compared to 150,000 -- {{1.3}} million for the original USNO-B magnitudes). 
The interquartile range of the recalibrated data are a factor of {{3.0, 1.9, 3.0, 2.0, and 2.3}} smaller than for the unrecalibrated data in bands $O,E,J,F,$ and $N$, respectively.  The standard deviation of $|m^{\prime\prime}_{USNO} - m_{SDSS}|$  is $\approx$ 0.1 mag, which is about a factor of two smaller than for the unrecalibrated data.

Many of the outliers shown in Figure \ref{fig: histo alt} are not likely to be real, large amplitude variable stars and quasars, but some of them are (see \S\ref{sec: vars} below). Figure \ref{fig: histo culled} shows a comparison of the distribution of $m_{USNO} - m_{SDSS}$ 
when additional constraints on the quality flags are imposed. 
All of the original magnitudes, all of the recalibrated magnitudes, and only recalibrated magnitudes of objects that are unblended and consistent with other catalogs are shown as dashed red, dotted blue, and solid black lines, respectively.   
The areas under the curves in each panel are not the same; the solid black line represents a subset of the other data sets.  The Figure shows that most of the outliers are blended objects or are inconsistent with other catalogs.  Compared to the full set of recalibrated data, the number of objects with $|m^{\prime\prime}_{USNO} - m_{SDSS}| \ge 1.0$ mag is reduced by almost an order of magnitude, down to {{2,000 -- 13,000}} in each band ($\approx$ 0.02\% of the total).

\subsection{Comparison to other work} 
\label{sec: comparison}

Other groups have undertaken efforts to improve the photometry of a large fraction of USNO-B using data from SDSS. The proper motion work of \citet{Munn+04} describes a photometric recalibration algorithm and discusses preliminary results based on data from SDSS DR1. They refer to the work of \citet{Sesar+06}, which describes a similar algorithm in more detail using data from SDSS DR2.  The Catalog Archive Server for all SDSS data releases since DR4 \citep{DR4} includes a table that provides recalibrated magnitudes of USNO-B objects.  According to the database schema, this table is based on the recalibration work of \citet{Munn+04}.

There are several key differences between the algorithms of \citet{Munn+04} and \citet{Sesar+06} compared to our work.  Our selection of data from the SDSS data is more restrictive and is designed to suppress the inclusion of spurious USNO-B objects.  We identify objects that may suffer from blending effects in a robust way, and exclude them as calibration objects.  When determining the blended status of objects, we  account for the variation of object size with magnitude.  We also account for the variable magnitude limits on each plate rather than adopting a fixed cutoff for the entire catalog.  During the recalibration, we reject bad model fits based on the distribution of parameters from all model fits.  We have found a number of causes for the spurious nature of large amplitude variable candidates and have implemented automated techniques for identifying them.  We compare our results to other historical catalogs based on the same photographic plates as an additional check on the quality of the recalibration. 
By using the tags for unblended and consistent objects, most of the large amplitude variable objects in our catalog are real (see \S\ref{sec: limitations} below). Finally, we provide object identifications and epochs in our catalog that enable the analysis of variable objects in a convenient way.  None of these were done by \citet{Munn+04} or \citet{Sesar+06}.

Our catalog is significantly larger compared to the catalog of \citet{Sesar+06}, primarily because of the larger areal coverage of DR{{9}} relative to DR2.
In each band, there are $\approx$  {{30 -- 40}} million objects in our catalog that have photometry that is likely to be reliable (quality flag equal to 0). 
This number is about a factor of  {{15}} larger than the number of objects in \citet{Sesar+06} that are reported to have good photometry.

Quantitative illustrations of the differences between our results and previous work are shown in Figures \ref{fig: histo sesar} and \ref{fig: histo aihara}.  
Figure \ref{fig: histo sesar} compares our results with the publicly available data from \citet{Sesar+06} in the $O$ and $E$ bands; data for the other USNO-B bands is not publicly available.  We selected objects from \citet{Sesar+06} that are identified by them as having good photometry. We cross match those $\approx$  {{2}} million objects with those in our catalog (one-to-one match with a  {{1\farcs0}}  radius).  In order to maintain consistency with our recalibration process, we calculate $m_{SDSS}$ for the \citet{Sesar+06} objects using the photometry from DR2, and we define $m^{S06}_{USNO}$ as the USNO-B magnitude recalibrated by \citet{Sesar+06}. 
In Figure \ref{fig: histo sesar}, the dashed red and dotted blue lines show $m_{USNO} - m_{SDSS}$ from the original USNO-B catalog and $m^{S06}_{USNO} - m_{SDSS}$, respectively, for the same objects (the area under the curves are identical).
The solid black line shows $m^{\prime\prime}_{USNO} - m_{SDSS}$ from our catalog for the same objects, but with the constraint that the objects are tagged as unblended (i.e.\ the area under the black curve is smaller than those under the red and blue curves). 
For the $O$ {{and $E$}} band, {{we applied}} the additional condition that the objects are consistent with GSC-II {{and SSS, respectively.}}  
The Figure shows that for objects with $|m^{\prime\prime}_{USNO} - m_{SDSS}| \gtrsim 0.3 $~mag, our recalibrated data have a distribution that is significantly narrower and more symmetric than those recalibrated by \citet{Sesar+06}. This implies that for objects with large values of $|m^{S06}_{USNO} - m_{SDSS}|$, many of the recalibrated magnitudes reported by \citet{Sesar+06} are likely to be inaccurate. 
The claim by \citet{Sesar+06} that their candidate variables are not dominated by spurious sources is not likely to be true for the most highly variable candidates.

Figure \ref{fig: histo aihara} compares our results with the recalibrated USNO-B magnitudes available through DR{{9}} \citep[][hereafter A1{{2}}]{DR9}. 
The dashed red and dotted blue lines show $m_{USNO} - m_{SDSS}$  and $m^{A12}_{USNO} - m_{SDSS}$ from A1{{2}}, respectively, for the same objects (the area under the curves are identical).
Note that A1{{2}} do not indicate the quality of the recalibrated magnitudes; we show the recalibrated USNO-B magnitude of every object in our catalog (but with $m^{A12}_{USNO}$ from A1{{2}}). 
The solid black line shows $m^{\prime\prime}_{USNO} - m_{SDSS}$ from our catalog for the same objects, but with the constraint that the objects are tagged as unblended and consistent with other catalogs (the area under the black curve is smaller than the red and blue curves).
The figure shows that many of the recalibrated magnitudes reported by A1{{2}} are inaccurate and they offer little or no improvement over the original USNO-B magnitudes, especially for objects with $m^{A12}_{USNO} - m_{SDSS} < -0.5$ mag.

\section{Limitations of the Catalog}
\label{sec: limitations}

Our selection criteria and recalibration process are designed to maximise the size and reliability of the catalog while minimizing the presence of spurious objects or unreliable photometric measurements.  However, there is a very small percentage of objects with inaccurate photometry in our catalog.  In order to quantify the degree of contamination, we visually inspected DSS and SDSS images centered on the objects with the 100 highest values of $|m^{\prime\prime}_{USNO} - m_{SDSS}|$ (i.e., $|m^{\prime\prime}_{USNO} - m_{SDSS}| \gtrsim 2$ mag).
{{We only considered}} objects that are {{not spikes or halos (\S\ref{sec: spurious})}}, {{are}} unblended (\S\ref{sec: blended}), {{are}} away from bright stars (\S\ref{sec: blend bright}),  {{are}} consistent with external catalogs (\S\ref{sec: consistent}), {{and have reliable uncertainties (\S\ref{sec: recal 2})}} .  
We find that {{32\%}} of these objects have abnormal {{properties that strongly indicate that}}  they are not genuine variables. The most common {{abnormal properties}} are, in descending order:

\begin{enumerate}

\item{the object overlaps with an artifact on a USNO-B plate and was not removed by our selection criteria (\S\ref{sec: dr8}) or recalibration (\S\ref{sec: recal}); this affects {{16}}\% of the sample. Two examples of this are shown in Figure \ref{fig: thumbnail artefacts}. These objects could be detected automatically through PSF fitting to images or through other computer vision techniques. Note that these artifacts do not appear to be associated with bright stars and therefore are not classified as artifacts by \citet{Barron+08}.
}

\item{the proper motion of the object or of other objects very nearby is inaccurate; this affects {{5}}\% of the sample. 
This can lead to one of two adverse scenarios. One scenario is where an object is blended at the epoch of the USNO-B measurement (but reported as a single object in the USNO-B catalog). However, it is not blended at the DR{{9}} epoch because the blended object has moved outside the blending `radius'.  Therefore it is not identified as a blended object.  This usually happens for objects with underestimated proper motions; see Figure \ref{fig: thumbnail proper motion} for an example.
Another situation is where a stationary object is reported to have a large proper motion in USNO-B and, as a result, is matched to the wrong DR{{9}} object.  Almost all of the objects that suffer from either of these effects are from the earliest epoch measurements in USNO-B (i.e., $O$ and $E$ bands).
}

\item{the {{brightness}} of the object in USNO-B  is  {{overestimated}} because the object is blended with one or more very faint stars, or with an extended galaxy, or with a bright star. This affects {{5}}\% of the sample.
{{These objects are not tagged as blended because the objects they are blended with were not included in the `superset' of DR9 objects described in \S\ref{sec: blended} and Appendix \ref{sec: app b}. However, relaxing some of the selection criteria would introduce a large number of objects that are either a) not astrophysical,  or b) do not have reliable photometry.}}
}

\item{the USNO-B magnitude is inconsistent with nearby objects of similar brightness on the same plate.  This applies to 4\% of the sample {{and occurs when the GSC-II or SSS magnitude is also incorrect.}
}}

\item{{the radial profile of the light from the object appears to be more consistent with a galaxy rather than a star or quasar. This typically occurs for objects with profiles that have a bright core and symmetric, extended wings; it is found in 1\% of the sample. In these cases, the flux reported by USNO-B is overestimated and it is not appropriate to use the PSF magnitude in SDSS.}}

\end{enumerate}

We compared the number of spurious large amplitude variables in our catalog with the results of \citet{DR9}.
There are a factor of {{240}} more objects with $|m^{A12}_{USNO} - m_{SDSS}| > 2.5$ mag in \citet{DR9} compared to {{the number of unblended and consistent}} objects our catalog. 
If we assume that a) there are the same number of genuine large amplitude variables in the two catalogs and b) {{67}}\% of objects in our catalog with $|m^{\prime\prime}_{USNO} - m_{SDSS}| > 2.5$ mag are genuine, then the  probability that a large amplitude variable candidate is genuine is a factor of  {{240/0.67}} $\approx$ {{360}} higher for our catalog than for \citet{DR9}.
A similar comparison of our data with the \citet{Sesar+06} catalog shows that the probability that a large amplitude variable candidate is  genuine is a factor of 58 times higher for our catalog than for \citet{Sesar+06}. 
However, we caution that a visual inspection of plate images is still recommended in order to rule out a non-astrophysical cause for  apparent variability of any particular object.

Without visually inspecting all the images, it is difficult to estimate precisely the fraction of variable sources across the catalog that are real. However, we manually inspected a random sample of 500 objects with $|m^{\prime\prime}_{USNO} - m_{SDSS}| > 0.5$ mag and found that for 98\% of them, the variability could not be attributed to any of the reasons listed above. This implies that the probability that a variable candidate is genuine is significantly larger for objects with smaller values of $|m^{\prime\prime}_{USNO} - m_{SDSS}|$. 

Another limitation to our catalog is that it can not be used to identify objects that are transient, e.g. objects that are detected in one of USNO-B or DR{{9}}, but not in the other.   Therefore there is an upper limit to the largest change in magnitude that we can identify (about 6 magnitudes). {{Most of}} the {{reliable}} objects in the catalog {{($>$ 99\%)}} are bounded by 14 $\lesssim m^{\prime\prime}_{USNO} \lesssim$ 20; the bright end is {{usually}} set by the condition that objects are not saturated in DR{{9}} and the faint end is set by the sensitivity of the photographic plates from which USNO-B are derived. In principle, transients could be identified by searching for bright objects in DR{{9}} that have no USNO counterparts. The reverse scenario would be overwhelmed with artifacts and other spurious sources. We also note that our avoidance of extended sources means that this work is not sensitive to detecting extragalactic supernovae associated with resolved galaxies.

\section{Suggested Applications}
\label{sec: applications}

\subsection{Historical photometry}

One principal application of the catalog is that of an accurate historical archive. 
The catalog provides an accurate photometric record of stars and quasars over $\approx$ 1/4 of the sky 
over the past $\gtrsim$ 60 years.  With the growing number of wide-field surveys at all wavelengths, the catalog is a valuable resource for retrieving accurate optical magnitudes and colors over long time periods.  A star or quasar that is in the original USNO-B catalog and within the DR{{9}} footprint, but is {\it{not}} in our catalog implies that the USNO-B photometry is not likely to be correct. 
Figure \ref{fig: histo mags} shows the distribution of recalibrated magnitudes ($m^{\prime\prime}_{USNO}$) for all objects in our catalog (solid black) and for those that  {{are candidate variables at the 4$\sigma$ level (dashed blue; see \S\ref{sec: vars} below)}}.  The Figure shows that the sensitivity of USNO-B {{(the mode of the frequency distribution)}} is approximately 20.0, 18.5, 20.0, 19.0, and 17.5 mag for bands $O,E,J,F,$ and $N$, respectively. 
In order to minimise the probability that an object in our catalog is spurious, we recommend applying a filter that only selects objects with an artifact tag equal to 0, a quality flag equal to 0 and  consistency flag that indicates the photometry is in agreement with other catalogs. 

To aid the comparison of our catalog to other surveys, we have derived the transformation from the (recalibrated) USNO-B photometric system to SDSS.
This transformation was calculated by selecting cataloged objects that are unblended, not near artifacts of brights stars, and have reliable uncertainties.
We minimized the absolute difference between the DR{{9}} magnitude and ($m^{\prime\prime}_{USNO}$) and calculated the following transformation functions:
\begin{eqnarray}
 g = O - 0.812 (O-J) - 0.19 \label{eq: usno1} \\ 
 r = E - 0.617 (E-F) + 0.23 \label{eq: usno2} \\ 
 i = N + 0.235 (F-N) + 0.33 \label{eq: usno3}
\end{eqnarray}
where $g,r,i$ are PSF magnitudes from DR{{9}} in the SDSS system, and $O,E,J,F,N$ are the recalibrated USNO-B magnitudes in the USNO-B system.
We caution that the transformation may be inaccurate for objects with USNO-B colors that deviate significantly from 0.0.  After the transformation, the standard deviation of the difference between the SDSS and USNO-B magnitudes is $\approx$ 0.12 mag. These transformation functions are provided as a convenience to users of the catalog; the values of $g,r,$ and $i$ from equations \ref{eq: usno1} - \ref{eq: usno3} are not given in the catalog.

\subsection{Variable stars and quasars}
\label{sec: vars}

Another principal application of the catalog is the identification and analysis of stars and quasars that vary optically over time scales of up to 60 years (see Figure \ref{fig: histo epochs}).  The combination of the large sky coverage, sensitivity, and long time scales makes the catalog suitable for discovering optical outbursts from rare objects such as dwarf novae \citep{Wils+10} and FU Ori stars \citep{HK96}, or the dimming of rare R CrB stars \citep{Clayton12}.  The size of the catalog makes it suitable for studying the ensemble properties of variable quasars or subclasses of common variable stars (e.g. RR Lyrae, Miras, eclipsing binaries, etc.) with a photometric accuracy of $\approx$ 0.1 mag. 

An accurate classification of the variable objects (not provided by our catalog) is required for a robust analysis, and such a classification is best done through multi-wavelength photometry and spectroscopy.  We note that approximately  {{570,000}} objects in our catalog ($\approx$  {{1.3}}\%) have optical spectra available through DR9 that could be used to aid in the classification.
The classification could also be assisted through the repeated observations in SDSS; there are one or more additional photometric measurements in DR{{9}} for $\approx$ 2.8 million objects in  {{our}} catalog ($\approx$  {{6}}\%).

However, single-epoch optical colors are still a useful indicator of an object's classification \citep[e.g.,][]{Richards+02, Ivezic+05}. 
Figure \ref{fig: all stars cmd} shows a color-color diagram of $\approx$  {{37}} million objects in our catalog, using the single-epoch $g-r$ and $u-g$ colors from SDSS DR{{9}}.  The magnitudes have been corrected for interstellar extinction using the dust model of \citet{SFD98} and assuming the dust acts as a foreground screen to each object. To avoid areas of the sky where the dust correction is very large and therefore uncertain (e.g. near the Galactic plane), we only show objects with extinction corrections of 2.0 magnitudes or less in each SDSS band.
The color bar scale shows the density of objects in color-color space. 
The scaling is non-linear ($\propto$ density$^{0.25}$) and was chosen to show the stellar locus in green, yellow, and red. 
The dashed red lines outline the boundaries defined by \citet{Richards+02} that are used to classify point sources.  {{The locus of points near $(u-g = 0.1,~g-r = 0)$ are consistent with low-redshift quasars, and the locus of points near $(u-g = 1.0,~g-r = -0.2)$ are consistent with RR Lyrae stars. }}

Figure \ref{fig: var stars cmd} again shows the density of cataloged objects in color-color space, but only for a subset of the objects in Figure \ref{fig: all stars cmd}. This subset consists of  {{74,062}} objects that have significantly different magnitudes in USNO-B compared to DR{{9}} in the USNO-B $J$ band. Only objects that are variable at the {{4}}$\sigma$ level, i.e. $|J_{USNO} - J_{SDSS}| \ge ${{4}}$ \sqrt{\sigma^2_{USNO}+\sigma^2_{SDSS}}$, are shown.  
Each object is also constrained to have a quality flag equal to 0 and to be consistent with GSC-II.  
(The distributions of variables in color-color space for other USNO-B bands are very similar to Figure \ref{fig: var stars cmd} and hence are not shown.)

The fraction of objects with colors that are consistent with low-redshift quasars is  {{35}} times higher for the variables ({{27.8}}\%)  than for the catalog as a whole (0.8\%) 
Quasars are well known to vary in the optical over a range of time scales \citep[e.g.,][]{Giveon+99} and this is consistent with our results.   
{{In addition, more than 80\% of the 4$\sigma$ candidate variables that have spectra in DR9 are classified spectroscopically as quasars; this fraction is four times higher than the number of quasars among all objects with DR9 spectra.}}
{{A quantitative}}  comparison {{between the two figures}} also shows that the number of objects with colors consistent with RR Lyrae stars is {{five}} times higher for the variables ({{2.6}}\%) than for the catalog as a whole (0.5\%).  These results show that the combination of variability and single-epoch colors could be used to increase the classification accuracy.  We note that our results are consistent with a similar analysis performed by \citet{Sesar+06}, however our sample size of candidate variables ($\approx$ {{74,000}}) is a factor of {{25}} larger than their sample ($\approx$ 3,000).

With such a large catalog of candidate variables, it is interesting to note how many of these objects may be new discoveries.
There are {{247,511}} unique objects that meet our criteria for {{4}}$\sigma$ variables in at least one of the USNO-B bands{{; the distribution of $m^{\prime\prime}_{USNO}$ for these objects is shown in Figure \ref{fig: histo mags} as a dashed blue line. Note that the criteria for 4$\sigma$ variables in our catalog includes two additional requirements: that the quality flag is equal to zero, and the consistency flag is set to true.  
We include a column in our catalog (Table \ref{tab: sample})  that indicates if an object meets these criteria.
The distribution of $m^{\prime\prime}_{USNO} - m_{SDSS}$ as a function of $m_{SDSS}$ for these variable candidates is shown in Figure \ref{fig: delta mag vars}.  }}  

We compare our candidate variable objects with those in the American Association of Variable Star Observers (AAVSO) International Variable Star Index\footnote{\href{http://www.aavso.org/vsx/}{http://www.aavso.org/vsx/}}
\citep[VSX;][]{VSX}. The VSX is a compilation of entries in several catalogs of optically variable stars, e.g., the General Catalog of Variable Stars \citep{GCVS}, the Northern Sky Variability Survey \citep{Wozniak+04}, and the Optical Gravitational Lensing Experiment \citep{Udalski+08}, along with newer discoveries reported by AAVSO members and confirmed by AAVSO staff.  The VSX catalog is updated weekly and provides a classification of the variable stars of all types. As of  {{2013 June 30}}, the catalog contains entries for  {{243,610}} variable stars. We note that although the name of the catalog implies that the entries are restricted to stars, there are a very small number of objects (0.08\%) in the catalog classified as AGN, BL Lacs, or QSOs. 

We cross matched our {{4}}$\sigma$ variables with objects in the VSX that are listed as known or suspected variables (\texttt{variability flag} in VSX equal to 0 or 1) by selecting the closest positional match within {{1}}\arcsec\ (a one-to-one match).  There are {{9,843}} objects that are cross matched between the catalogs, meaning that more than  {{96}}\% of candidate variables in our catalog are not in the VSX catalog.  
The VSX catalog is not complete, and suffers from other biases (e.g., sky coverage). Therefore we cannot claim to have found $\approx$  {{230,000}} previously unknown variable stars and quasars.  However, the fact that there is such little overlap between the catalogs suggests that many of the variable candidates in our catalog could be new discoveries.

One example of a potential new discovery is shown in Figure \ref{fig: thumb var mira}.
This object (SDSS J171120.07-124003.8) is a {{4}}$\sigma$ variable that a) has one of the largest values of $|m^{\prime\prime}_{USNO} - m_{SDSS}|$, b) is not in the VSX catalog, and c) appears to be a genuine variable upon visual inspection (e.g. the variability is not attributable to any of the enumerated causes in \S\ref{sec: limitations}). 
The object has a very red optical color ($g - i = 5.9$ at epoch 2005.43) and it is within 0.5\arcsec\ of a very bright infrared source ($K = 6.8$ mag) in the Two Micron All Sky Survey Point Source Catalog \citep[2MASS;][]{Skrutskie+06}. The large amplitude of its variability, red color, and high IR brightness is consistent with the object being a Mira variable \citep{Smak66}.  This object is not among the $\approx$  {{18,000}} Mira stars classified by the VSX catalog, and it cannot be cross matched to objects in any other variability catalog hosted by the Strasbourg Astronomical Data Center\footnote{\href{http://vizier.u-strasbg.fr/viz-bin/VizieR}{http://vizier.u-strasbg.fr/viz-bin/VizieR}} (CDC). This suggests that our catalog could contain many previously unknown Miras. The long time period between USNO-B and SDSS observations could be used to identify any changes in the periodicity of Miras, and hence to trace the poorly understood process of mass loss among AGB stars \citep[e.g.,][]{Zijlstra+02}.

\section{Summary}
\label{sec: summary}

We have combined the USNO-B and SDSS DR{{9}} catalogs to produce a multi-epoch photometric catalog of {{43,647,887}} optical point sources covering more than 14,500 deg$^2$ of the northern sky.  
We carefully cross matched the catalogs and incorporated the proper motions in a self-consistent way.
We robustly identified blended objects and objects affected by optical artifacts from bright stars by taking into account the different sensitivities and resolutions of the originating catalogs. Through careful recalibration of the USNO-B data, the catalog has a photometric accuracy of $\approx$ 0.1 mag and has a limiting sensitivity of $\approx 20$ mag. 
The improvements to the accuracy of USNO-B magnitudes are significantly better than those described by \citet{Sesar+06} and \citet{DR9}. 
{{For each object,}} up to ten {{magnitude values}} are reported  ({{for each USNO band, one is derived from the original USNO-B catalog and one is derived from SDSS). The photometry of individual objects }}  span a time period of up to 60 years.
We compared our recalibrated photometry with photometry of other catalogs based on the same photographic plates in order to identify objects with abnormal magnitudes. 
The catalog provides a number of flags that indicate the quality of the photometry for each object; this enables users to select data of varying quality depending on the application.  

We identified a number of causes for the presence of spurious objects or inaccurate magnitudes in the catalog. We employed a number of methods to remove or identify such objects. However, there are some objects in the catalog that may not be genuine. A visual inspection of plate and SDSS images reveal that the most common source of  {{contamination}} among large amplitude variable candidates is {{the presence of artifacts on the photographic plates}}. 

We used the catalog to identify $\approx$  {{250,000}} stars and quasars with magnitudes that have changed significantly (ie., at the {{4}}$\sigma$ level) between the USNO-B and SDSS observations. Almost all of these objects do not appear in catalogs of known variable stars and quasars.  The single-epoch optical colors of $\approx$ 20\% of these variable candidates are consistent with low-redshift quasars.  
The analysis and classification of these candidates could be improved through multi-wavelength imaging and spectroscopy, along with light curves from large area surveys such as the Catalina Sky Surveys\footnote{\href{http://nesssi.cacr.caltech.edu/DataRelease/}{http://nesssi.cacr.caltech.edu/DataRelease/}} \citep{Drake+09}.  The catalog could also be used to study the ensemble properties of different types of variable and  to constrain models for the population and discovery rate of variables in upcoming surveys such as those carried by Large Synoptic Survey Telescope \citep[e.g.,][]{Prsa+11} or the ESA Gaia mission \citep{Robin+12}.

By design, the sky coverage of our catalog is restricted to the coverage of the SDSS survey.  Future versions of our catalog will incorporate other deep, accurate photometric surveys over large areas of the sky such as those from PanSTARRS in the north \citep{Chambers13} or SkyMapper in the south \citep{Keller+07} in order to provide a recalibration of USNO-B over the entire sky.

\section{ACKNOWLEDGEMENTS}

This work was supported by the Centre for All-sky Astrophysics (CAASTRO), an Australian Research Council Centre of Excellence (grant CE110001020) and through the Science Leveraging Fund of the New South Wales Department of Trade and Investment.  We thank Jay Banyer for sharing his computational expertise during the project. We also thank Tara Murphy{{, Sergey Koposov, and Josh Grindlay}} for {{helpful}} discussions  {{in various}} stages of this work.
{{We are indebted to the anonymous referee whose comments led to substantial improvements to the paper.}}

Funding for SDSS-III has been provided by the Alfred P. Sloan Foundation, the Participating Institutions, the National Science Foundation, and the U.S. Department of Energy Office of Science.
SDSS-III is managed by the Astrophysical Research Consortium for the Participating Institutions of the SDSS-III Collaboration including the University of Arizona, the Brazilian Participation Group, Brookhaven National Laboratory, University of Cambridge, Carnegie Mellon University, University of Florida, the French Participation Group, the German Participation Group, Harvard University, the Instituto de Astrofisica de Canarias, the Michigan State/Notre Dame/JINA Participation Group, Johns Hopkins University, Lawrence Berkeley National Laboratory, Max Planck Institute for Astrophysics, Max Planck Institute for Extraterrestrial Physics, New Mexico State University, New York University, Ohio State University, Pennsylvania State University, University of Portsmouth, Princeton University, the Spanish Participation Group, University of Tokyo, University of Utah, Vanderbilt University, University of Virginia, University of Washington, and Yale University. 
This research has made use of the VizieR catalogue access tool, CDS, Strasbourg, France. 
This research has made use of data obtained from the SuperCOSMOS Science Archive, prepared and hosted by the Wide Field Astronomy Unit, Institute for Astronomy, University of Edinburgh, which is funded by the UK Science and Technology Facilities Council.
The Guide Star Catalogue-II is a joint project of the Space Telescope Science Institute and the Osservatorio Astronomico di Torino. Space Telescope Science Institute is operated by the Association of Universities for Research in Astronomy, for the National Aeronautics and Space Administration under contract NAS5-26555. The participation of the Osservatorio Astronomico di Torino is supported by the Italian Council for Research in Astronomy. Additional support is provided by European Southern Observatory, Space Telescope European Coordinating Facility, the International GEMINI project and the European Space Agency Astrophysics Division. 
The Digitized Sky Surveys were produced at the Space Telescope Science Institute under U.S. Government grant NAG W-2166. The images of these surveys are based on photographic data obtained using the Oschin Schmidt Telescope on Palomar Mountain and the UK Schmidt Telescope. The plates were processed into the present compressed digital form with the permission of these institutions.
The National Geographic Society - Palomar Observatory Sky Atlas (POSS-I) was made by the California Institute of Technology with grants from the National Geographic Society.
The Second Palomar Observatory Sky Survey (POSS-II) was made by the California Institute of Technology with funds from the National Science Foundation, the National Geographic Society, the Sloan Foundation, the Samuel Oschin Foundation, and the Eastman Kodak Corporation.
The Oschin Schmidt Telescope is operated by the California Institute of Technology and Palomar Observatory.
The UK Schmidt Telescope was operated by the Royal Observatory Edinburgh, with funding from the UK Science and Engineering Research Council (later the UK Particle Physics and Astronomy Research Council), until 1988 June, and thereafter by the Anglo-Australian Observatory (now the Australian Astronomical Observatory). The blue plates of the southern Sky Atlas and its Equatorial Extension (together known as the SERC-J), as well as the Equatorial Red (ER), and the Second Epoch [red] Survey (SES) were all taken with the UK Schmidt. 
All data are subject to the copyright given in the copyright summary. Copyright information specific to individual plates is provided in the downloaded FITS headers.
Supplemental funding for sky-survey work at the STScI is provided by the European Southern Observatory. 
We acknowledge the use of NASA's {\it{SkyView}} facility (\href{http://skyview.gsfc.nasa.gov}{http://skyview.gsfc.nasa.gov)} located at NASA Goddard Space Flight Center.
This research made use of data provided by \href{Astrometry.net}{Astrometry.net}.

\appendix
\section{Selection criteria for SDSS DR9 objects in \S\ref{sec: dr8}}
\label{sec: app}

We select only objects from the \texttt{Star} table in  {{DR9}},  {{i.e.\ primary}} objects classified by  {{DR9}} as unresolved point sources \citep{DR9}.
{{We impose a lower limit on the brightness of these objects by requiring the PSF magnitude to be less than 22 in at least one of the $g,r,$ or $i$ bands.  In order to avoid objects with abnormally faint magnitudes or those with inaccurate photometry, we also impose the joint condition that \texttt{psfmag\_g < 23}, \texttt{psfmag\_r < 22}, and \texttt{psfmag\_i < 22} for each object. 
}}   
We require that the \texttt{CalibStatus} flag is set to \texttt{PHOTOMETRIC} or \texttt{INCREMENT\_CALIB}  {{in each of the $g$, $r$, and $i$ bands}}.

 We  {{require}} the \texttt{PhotoFlags} in each of the $g$, $r$, and $i$ bands {{to satisfy the following conditions}}:
\begin{itemize}
\item{{{the bit corresponding to \texttt{BINNED1} is set}}, and}
\item{{{none of the bits corresponding to \texttt{PSF\_FLUX\_INTERP, BAD\_COUNTS\_ERROR, BADSKY, NOTCHECKED, SATURATED, NOPROFILE, PEAKCENTER,} and \texttt{EDGE} are set}}, and }
\item{{{the bit corresponding to \texttt{DEBLEND\_NOPEAK} is not set}} or \texttt{psfmagerr $\le$ 0.2}, and}
\item{{{the bit corresponding to \texttt{INTERP\_CENTER} is not set or the bit corresponding to \texttt{COSMIC\_RAY} is not set}}
}
\end{itemize}

{{We avoid moving objects by requiring that the bit in \texttt{PhotoFlags} corresponding to \texttt{DEBLEND\_AS\_MOVING} is not set in any of the $ugriz$ bands. We require that the \texttt{ImageStatus} flag is set to \texttt{CLEAR} for each object in each of the $g$, $r$, and $i$ bands. We use the \texttt{InsideMask} field to ensure that an object is not inside a mask of any kind. 
}}

In order to exclude objects with inaccurate proper motions, we {{join}} the {{\texttt{ProperMotions}}} table in  {{DR9}} {{with the \texttt{Stars} table}} and require each object to satisfy \texttt{Match = 1}, \texttt{NFit $\ge$ 5}, \texttt{SigRA < 350} and \texttt{SigDec < 350}.

{{After we apply the above criteria, a visual inspection of SDSS images revealed a number of objects that have highly inaccurate magnitudes.  The images implied that the reported magnitudes were underestimated by more than 1 magnitude. These kinds of objects tend to be in the vicinity bright stars or were deblended as part of very large SDSS objects. Another indication that the photometry is incorrect is that the fiber magnitude of these objects are very different from the PSF magnitudes.  The mean value of \texttt{fibermag - psfmag} among DR9 point sources is about 0.35 mag with a standard deviation of 0.05 mag.  In order to remove objects with inaccurate magnitudes, we require that the value of \texttt{fibermag - psfmag} is between 0.0 mag and 0.7 mag in each of the $g,r,$ and $i$ bands, i.e.\ we remove the 7$\sigma$ outliers.}} 

\section{Selection criteria for SDSS DR9 objects in \S\ref{sec: blended}}
\label{sec: app b}

We select point sources and extended sources from the \texttt{Star} and \texttt{Galaxy} tables, respectively in SDSS  {{DR9}}.   
{{We require the composite model magnitude to be less than 22 in at least one of the $g,r,$ or $i$ bands.}}
 We require that the \texttt{CalibStatus} flag is set to \texttt{PHOTOMETRIC} or \texttt{INCREMENT\_CALIB} {{in each of the $g$, $r$, and $i$ bands}}.

{{We require the \texttt{PhotoFlags} in each of the $g$, $r$, and $i$ bands to satisfy the following conditions}}:

\begin{itemize}
\item{{{the bit corresponding to \texttt{BINNED1} is set}}, and}

\item{{{none of the bits corresponding to \texttt{BADSKY, NOTCHECKED,} or \texttt{NOPROFILE} are set}}, and }

\item{{{the bit corresponding to \texttt{DEBLEND\_NOPEAK} is not set}} or \texttt{psfmagerr $\le$ 0.2} }

\end{itemize}

For objects in the \texttt{Star} table, we impose the additional requirement that the bit in \texttt{PhotoFlag} corresponding to \texttt{EDGE} is not set.
{{We avoid moving objects by requiring that the bit in \texttt{PhotoFlags} corresponding to \texttt{DEBLEND\_AS\_MOVING} is not set in any of the $ugriz$ bands.}}


\begin{thebibliography}{51}
\expandafter\ifx\csname natexlab\endcsname\relax\def\natexlab#1{#1}\fi

\bibitem[{{Adelman-McCarthy} {et~al.}(2006){Adelman-McCarthy}, {Ag{\"u}eros},
  {Allam}, {Anderson}, {Anderson}, {Annis}, {Bahcall}, {Baldry}, {Barentine},
  \& {Berlind}}]{DR4}
{Adelman-McCarthy}, J.~K., {Ag{\"u}eros}, M.~A., {Allam}, S.~S., {et~al.} 2006,
  \apjs, 162, 38, 38

\bibitem[{{Ahn} {et~al.}(2012){Ahn}, {Alexandroff}, {Allende Prieto},
  {Anderson}, {Anderton}, {Andrews}, {Aubourg}, {Bailey}, {Balbinot}, {Barnes},
  \& et~al.}]{DR9}
{Ahn}, C.~P., {Alexandroff}, R., {Allende Prieto}, C., {et~al.} 2012, \apjs,
  203, 21, 21

\bibitem[{{Barron} {et~al.}(2008){Barron}, {Stumm}, {Hogg}, {Lang}, \&
  {Roweis}}]{Barron+08}
{Barron}, J.~T., {Stumm}, C., {Hogg}, D.~W., {Lang}, D., \& {Roweis}, S. 2008,
  \aj, 135, 414, 414

\bibitem[{{Bigg}(2000)}]{Bigg00}
{Bigg}, C. 2000, Acta Historica Astronomiae, 9, 90, 90

\bibitem[{{Cannon}(1984)}]{Cannon84}
{Cannon}, R.~D. 1984, in Astrophysics and Space Science Library, Vol. 110, IAU
  Colloq. 78: Astronomy with Schmidt-Type Telescopes, ed. M.~{Capaccioli}, 25

\bibitem[{{Chambers}(2013)}]{Chambers13}
{Chambers}, K.~C. 2013, in American Astronomical Society Meeting Abstracts,
  Vol. 221, American Astronomical Society Meeting Abstracts, 215.08

\bibitem[{{Clayton}(2012)}]{Clayton12}
{Clayton}, G.~C. 2012, Journal of the American Association of Variable Star
  Observers (JAAVSO), 40, 539, 539

\bibitem[{{Djorgovski} {et~al.}(1998){Djorgovski}, {Gal}, {Odewahn}, {de
  Carvalho}, {Brunner}, {Longo}, \& {Scaramella}}]{Djorgovski+98}
{Djorgovski}, S.~G., {Gal}, R.~R., {Odewahn}, S.~C., {et~al.} 1998, in Wide
  Field Surveys in Cosmology, ed. S.~{Colombi}, Y.~{Mellier}, \& B.~{Raban}, 89

\bibitem[{{Drake} {et~al.}(2009){Drake}, {Djorgovski}, {Mahabal}, {Beshore},
  {Larson}, {Graham}, {Williams}, {Christensen}, {Catelan}, {Boattini},
  {Gibbs}, {Hill}, \& {Kowalski}}]{Drake+09}
{Drake}, A.~J., {Djorgovski}, S.~G., {Mahabal}, A., {et~al.} 2009, \apj, 696,
  870, 870

\bibitem[{{Gandhi} {et~al.}(2012){Gandhi}, {Yamamura}, \& {Takita}}]{GYT12}
{Gandhi}, P., {Yamamura}, I., \& {Takita}, S. 2012, \apjl, 751, L1, L1

\bibitem[{{Giveon} {et~al.}(1999){Giveon}, {Maoz}, {Kaspi}, {Netzer}, \&
  {Smith}}]{Giveon+99}
{Giveon}, U., {Maoz}, D., {Kaspi}, S., {Netzer}, H., \& {Smith}, P.~S. 1999,
  \mnras, 306, 637, 637

\bibitem[{{Grindlay} {et~al.}(2012){Grindlay}, {Tang}, {Los}, \&
  {Servillat}}]{Grindlay+12}
{Grindlay}, J., {Tang}, S., {Los}, E., \& {Servillat}, M. 2012, in IAU
  Symposium, Vol. 285, IAU Symposium, ed. E.~{Griffin}, R.~{Hanisch}, \&
  R.~{Seaman}, 29--34

\bibitem[{{Hambly} {et~al.}(2001{\natexlab{a}}){Hambly}, {Davenhall}, {Irwin},
  \& {MacGillivray}}]{SSS-Astrometry}
{Hambly}, N.~C., {Davenhall}, A.~C., {Irwin}, M.~J., \& {MacGillivray}, H.~T.
  2001{\natexlab{a}}, \mnras, 326, 1315, 1315

\bibitem[{{Hambly} {et~al.}(2001{\natexlab{b}}){Hambly}, {Irwin}, \&
  {MacGillivray}}]{SSS-Images}
{Hambly}, N.~C., {Irwin}, M.~J., \& {MacGillivray}, H.~T. 2001{\natexlab{b}},
  \mnras, 326, 1295, 1295

\bibitem[{{Hambly} {et~al.}(2001{\natexlab{c}}){Hambly}, {MacGillivray},
  {Read}, {Tritton}, {Thomson}, {Kelly}, {Morgan}, {Smith}, {Driver},
  {Williamson}, \& et~al.}]{SSS-Overview}
{Hambly}, N.~C., {MacGillivray}, H.~T., {Read}, M.~A., {et~al.}
  2001{\natexlab{c}}, \mnras, 326, 1279, 1279

\bibitem[{{Hartley} \& {Dawe}(1981)}]{HD81}
{Hartley}, M., \& {Dawe}, J.~A. 1981, Proceedings of the Astronomical Society
  of Australia, 4, 251, 251

\bibitem[{{Hartmann} \& {Kenyon}(1996)}]{HK96}
{Hartmann}, L., \& {Kenyon}, S.~J. 1996, \araa, 34, 207, 207

\bibitem[{{Herbig} \& {Boyarchuk}(1968)}]{HB68}
{Herbig}, G.~H., \& {Boyarchuk}, A.~A. 1968, \apj, 153, 397, 397

\bibitem[{{H{\o}g} {et~al.}(2000){H{\o}g}, {Fabricius}, {Makarov}, {Urban},
  {Corbin}, {Wycoff}, {Bastian}, {Schwekendiek}, \& {Wicenec}}]{Hog+00}
{H{\o}g}, E., {Fabricius}, C., {Makarov}, V.~V., {et~al.} 2000, \aap, 355, L27,
  L27

\bibitem[{{Hudec}(2011)}]{Hudec11}
{Hudec}, R. 2011, Journal of Astrophysics and Astronomy, 32, 91, 91

\bibitem[{{Ivezi{\'c}} {et~al.}(2005){Ivezi{\'c}}, {Vivas}, {Lupton}, \&
  {Zinn}}]{Ivezic+05}
{Ivezi{\'c}}, {\v Z}., {Vivas}, A.~K., {Lupton}, R.~H., \& {Zinn}, R. 2005,
  \aj, 129, 1096, 1096

\bibitem[{{Keller} {et~al.}(2007){Keller}, {Schmidt}, {Bessell}, {Conroy},
  {Francis}, {Granlund}, {Kowald}, {Oates}, {Martin-Jones}, {Preston},
  {Tisserand}, {Vaccarella}, \& {Waterson}}]{Keller+07}
{Keller}, S.~C., {Schmidt}, B.~P., {Bessell}, M.~S., {et~al.} 2007, \pasa, 24,
  1, 1

\bibitem[{{Koposov} \& {Bartunov}(2006)}]{KB06}
{Koposov}, S., \& {Bartunov}, O. 2006, in Astronomical Society of the Pacific
  Conference Series, Vol. 351, Astronomical Data Analysis Software and Systems
  XV, ed. C.~{Gabriel}, C.~{Arviset}, D.~{Ponz}, \& S.~{Enrique}, 735

\bibitem[{{Lasker} {et~al.}(2008){Lasker}, {Lattanzi}, {McLean}, {Bucciarelli},
  {Drimmel}, {Garcia}, {Greene}, {Guglielmetti}, {Hanley}, {Hawkins}, \&
  et~al.}]{Lasker+08}
{Lasker}, B.~M., {Lattanzi}, M.~G., {McLean}, B.~J., {et~al.} 2008, \aj, 136,
  735, 735

\bibitem[{{MacLeod} {et~al.}(2012){MacLeod}, {Ivezi{\'c}}, {Sesar}, {de Vries},
  {Kochanek}, {Kelly}, {Becker}, {Lupton}, {Hall}, {Richards}, {Anderson}, \&
  {Schneider}}]{Macleod+12}
{MacLeod}, C.~L., {Ivezi{\'c}}, {\v Z}., {Sesar}, B., {et~al.} 2012, \apj, 753,
  106, 106

\bibitem[{{Markwardt}(2009)}]{Markwardt09}
{Markwardt}, C.~B. 2009, in Astronomical Society of the Pacific Conference
  Series, Vol. 411, Astronomical Society of the Pacific Conference Series, ed.
  {D.~A.~Bohlender, D.~Durand, \& P.~Dowler}, 251

\bibitem[{{McLean} {et~al.}(2000){McLean}, {Greene}, {Lattanzi}, \&
  {Pirenne}}]{McLean+00}
{McLean}, B.~J., {Greene}, G.~R., {Lattanzi}, M.~G., \& {Pirenne}, B. 2000, in
  Astronomical Society of the Pacific Conference Series, Vol. 216, Astronomical
  Data Analysis Software and Systems IX, ed. N.~{Manset}, C.~{Veillet}, \&
  D.~{Crabtree}, 145

\bibitem[{{Mickaelian} {et~al.}(2011){Mickaelian}, {Mikayelyan}, \&
  {Sinamyan}}]{MMS11}
{Mickaelian}, A.~M., {Mikayelyan}, G.~A., \& {Sinamyan}, P.~K. 2011, \mnras,
  415, 1061, 1061

\bibitem[{{Minkowski} \& {Abell}(1963)}]{MA63}
{Minkowski}, R.~L., \& {Abell}, G.~O. 1963, {The National Geographic
  Society-Palomar Observatory Sky Survey}, ed. K.~A. {Strand} (the University
  of Chicago Press), 481

\bibitem[{{Monet} {et~al.}(2003){Monet}, {Levine}, {Canzian}, {Ables}, {Bird},
  {Dahn}, {Guetter}, {Harris}, {Henden}, {Leggett}, \& et~al.}]{usnob}
{Monet}, D.~G., {Levine}, S.~E., {Canzian}, B., {et~al.} 2003, \aj, 125, 984,
  984

\bibitem[{{Munn} {et~al.}(2004){Munn}, {Monet}, {Levine}, {Canzian}, {Pier},
  {Harris}, {Lupton}, {Ivezi{\'c}}, {Hindsley}, {Hennessy}, \&
  et~al.}]{Munn+04}
{Munn}, J.~A., {Monet}, D.~G., {Levine}, S.~E., {et~al.} 2004, \aj, 127, 3034,
  3034

\bibitem[{{Munn} {et~al.}(2008){Munn}, {Monet}, {Levine}, {Canzian}, {Pier},
  {Harris}, {Lupton}, {Ivezi{\'c}}, {Hindsley}, {Hennessy}, \&
  et~al.}]{Munn+08}
---. 2008, \aj, 136, 895, 895

\bibitem[{{Pr{\v s}a} {et~al.}(2011){Pr{\v s}a}, {Pepper}, \&
  {Stassun}}]{Prsa+11}
{Pr{\v s}a}, A., {Pepper}, J., \& {Stassun}, K.~G. 2011, \aj, 142, 52, 52

\bibitem[{{Reid} {et~al.}(1991){Reid}, {Brewer}, {Brucato}, {McKinley},
  {Maury}, {Mendenhall}, {Mould}, {Mueller}, {Neugebauer}, {Phinney},
  {Sargent}, {Schombert}, \& {Thicksten}}]{Reid+91}
{Reid}, I.~N., {Brewer}, C., {Brucato}, R.~J., {et~al.} 1991, \pasp, 103, 661,
  661

\bibitem[{{Richards} {et~al.}(2002){Richards}, {Fan}, {Newberg}, {Strauss},
  {Vanden Berk}, {Schneider}, {Yanny}, {Boucher}, {Burles}, {Frieman}, {Gunn},
  {Hall}, {Ivezi{\'c}}, {Kent}, {Loveday}, {Lupton}, {Rockosi}, {Schlegel},
  {Stoughton}, {SubbaRao}, \& {York}}]{Richards+02}
{Richards}, G.~T., {Fan}, X., {Newberg}, H.~J., {et~al.} 2002, \aj, 123, 2945,
  2945

\bibitem[{{Robin} {et~al.}(2012){Robin}, {Luri}, {Reyl{\'e}}, {Isasi}, {Grux},
  {Blanco-Cuaresma}, {Arenou}, {Babusiaux}, {Belcheva}, {Drimmel}, {Jordi},
  {Krone-Martins}, {Masana}, {Mauduit}, {Mignard}, {Mowlavi},
  {Rocca-Volmerange}, {Sartoretti}, {Slezak}, \& {Sozzetti}}]{Robin+12}
{Robin}, A.~C., {Luri}, X., {Reyl{\'e}}, C., {et~al.} 2012, \aap, 543, A100,
  A100

\bibitem[{{Samus} {et~al.}(2009){Samus}, {Durlevich}, \& {et al.}}]{GCVS}
{Samus}, N.~N., {Durlevich}, O.~V., \& {et al.} 2009, VizieR Online Data
  Catalog, 1, 2025, 2025

\bibitem[{{Schlegel} {et~al.}(1998){Schlegel}, {Finkbeiner}, \&
  {Davis}}]{SFD98}
{Schlegel}, D.~J., {Finkbeiner}, D.~P., \& {Davis}, M. 1998, \apj, 500, 525,
  525

\bibitem[{{Scranton} {et~al.}(2002){Scranton}, {Johnston}, {Dodelson},
  {Frieman}, {Connolly}, {Eisenstein}, {Gunn}, {Hui}, {Jain}, {Kent},
  {Loveday}, {Narayanan}, {Nichol}, {O'Connell}, {Scoccimarro}, {Sheth},
  {Stebbins}, {Strauss}, {Szalay}, {Szapudi}, {Tegmark}, {Vogeley}, {Zehavi},
  {Annis}, {Bahcall}, {Brinkman}, {Csabai}, {Hindsley}, {Ivezic}, {Kim},
  {Knapp}, {Lamb}, {Lee}, {Lupton}, {McKay}, {Munn}, {Peoples}, {Pier},
  {Richards}, {Rockosi}, {Schlegel}, {Schneider}, {Stoughton}, {Tucker},
  {Yanny}, \& {York}}]{Scranton+02}
{Scranton}, R., {Johnston}, D., {Dodelson}, S., {et~al.} 2002, \apj, 579, 48,
  48

\bibitem[{{Sesar} {et~al.}(2006){Sesar}, {Svilkovi{\'c}}, {Ivezi{\'c}},
  {Lupton}, {Munn}, {Finkbeiner}, {Steinhardt}, {Siverd}, {Johnston}, {Knapp},
  \& et~al.}]{Sesar+06}
{Sesar}, B., {Svilkovi{\'c}}, D., {Ivezi{\'c}}, {\v Z}., {et~al.} 2006, \aj,
  131, 2801, 2801

\bibitem[{{Skrutskie} {et~al.}(2006){Skrutskie}, {Cutri}, {Stiening},
  {Weinberg}, {Schneider}, {Carpenter}, {Beichman}, {Capps}, {Chester},
  {Elias}, {Huchra}, {Liebert}, {Lonsdale}, {Monet}, {Price}, {Seitzer},
  {Jarrett}, {Kirkpatrick}, {Gizis}, {Howard}, {Evans}, {Fowler}, {Fullmer},
  {Hurt}, {Light}, {Kopan}, {Marsh}, {McCallon}, {Tam}, {Van Dyk}, \&
  {Wheelock}}]{Skrutskie+06}
{Skrutskie}, M.~F., {Cutri}, R.~M., {Stiening}, R., {et~al.} 2006, \aj, 131,
  1163, 1163

\bibitem[{{Smak}(1966)}]{Smak66}
{Smak}, J.~I. 1966, \araa, 4, 19, 19

\bibitem[{{Storkey} {et~al.}(2004){Storkey}, {Hambly}, {Williams}, \&
  {Mann}}]{Storkey+04}
{Storkey}, A.~J., {Hambly}, N.~C., {Williams}, C.~K.~I., \& {Mann}, R.~G. 2004,
  \mnras, 347, 36, 36

\bibitem[{{Udalski} {et~al.}(2008){Udalski}, {Szymanski}, {Soszynski}, \&
  {Poleski}}]{Udalski+08}
{Udalski}, A., {Szymanski}, M.~K., {Soszynski}, I., \& {Poleski}, R. 2008,
  \actaa, 58, 69, 69

\bibitem[{{Watson}(2006)}]{VSX}
{Watson}, C.~L. 2006, Journal of the American Association of Variable Star
  Observers (JAAVSO), 35, 318, 318

\bibitem[{{West}(1984)}]{West84}
{West}, R.~M. 1984, in Astrophysics and Space Science Library, Vol. 110, IAU
  Colloq. 78: Astronomy with Schmidt-Type Telescopes, ed. M.~{Capaccioli}, 13

\bibitem[{{Wils} {et~al.}(2010){Wils}, {G{\"a}nsicke}, {Drake}, \&
  {Southworth}}]{Wils+10}
{Wils}, P., {G{\"a}nsicke}, B.~T., {Drake}, A.~J., \& {Southworth}, J. 2010,
  \mnras, 402, 436, 436

\bibitem[{{Wo{\'z}niak} {et~al.}(2004){Wo{\'z}niak}, {Williams}, {Vestrand}, \&
  {Gupta}}]{Wozniak+04}
{Wo{\'z}niak}, P.~R., {Williams}, S.~J., {Vestrand}, W.~T., \& {Gupta}, V.
  2004, \aj, 128, 2965, 2965

\bibitem[{{York} {et~al.}(2000){York}, {Adelman}, {Anderson}, {Anderson},
  {Annis}, {Bahcall}, {Bakken}, {Barkhouser}, {Bastian}, {Berman}, \&
  et~al.}]{York+00}
{York}, D.~G., {Adelman}, J., {Anderson}, Jr., J.~E., {et~al.} 2000, \aj, 120,
  1579, 1579

\bibitem[{{Zijlstra} {et~al.}(2002){Zijlstra}, {Bedding}, \&
  {Mattei}}]{Zijlstra+02}
{Zijlstra}, A.~A., {Bedding}, T.~R., \& {Mattei}, J.~A. 2002, \mnras, 334, 498,
  498

\end{thebibliography}

\clearpage

\clearpage

\clearpage

\clearpage

\clearpage

\clearpage

\clearpage

\clearpage

\clearpage

\clearpage

\clearpage

\clearpage

\clearpage

\clearpage

\clearpage

\clearpage

\begin{figure}
\includegraphics[scale=0.65,angle=90]{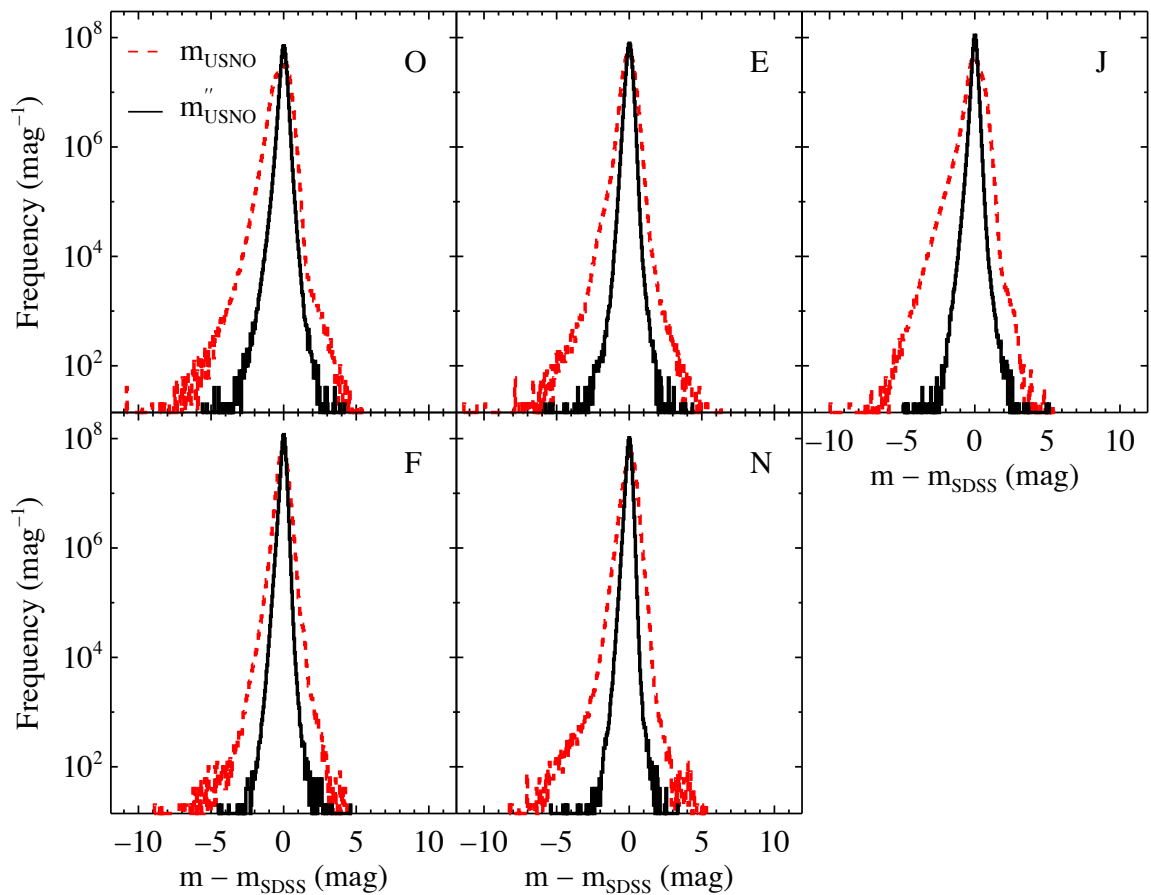}
\caption{Same as Figure \ref{fig: histo alt}, but the solid black line only shows recalibrated objects that are unblended and consistent with other catalogs as described in \S\ref{sec: pre-flag} and \S\ref{sec: consistent}, respectively (the areas under the curves in each panel are not the same). 
Most of the outlier objects in the red histogram (with large $|m_{USNO} - m_{SDSS}|$) are blended and/or inconsistent.   
\label{fig: histo culled}
}
\end{figure}

\clearpage

\begin{figure}
\includegraphics[scale=0.5,angle=90]{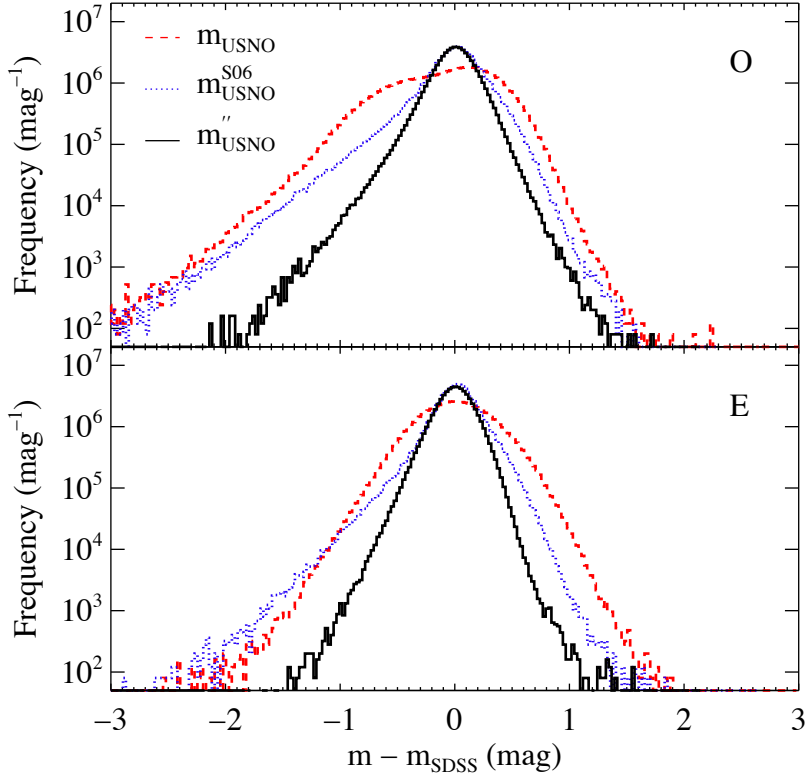}
\caption{Comparison of recalibrated USNO-B magnitudes between the publicly available subset of the results of \citet{Sesar+06} ($m^{S06}_{USNO}$; dotted blue) and our data (solid black).  For ease of comparison with Figures \ref{fig: histo summary} - \ref{fig: histo culled}, the original USNO-B magnitudes are shown as dashed red lines. 
For the \citet{Sesar+06} data, only those objects classified as having `good' photometry are shown (i.e. \texttt{goodphotoO/E} = 1). 
For our data, only unblended objects are shown; for band{{s}} $O$ {{and $E$}}, we impose the additional requirement that the magnitudes are consistent with GSC-II {{and SSS, respectively}}. 
The number of objects represented by the dashed red and dotted blue lines are identical within each panel; the number of objects represented by the black curve is smaller. 
In the central region of the histogram where $|m^{\prime\prime}_{USNO} - m_{SDSS}| < 0.3$ mag, our results are similar to \citet{Sesar+06}. However, at larger values of $|m^{\prime\prime}_{USNO} - m_{SDSS}|$,
the distribution of our data is much narrower and is more symmetric.
\label{fig: histo sesar}
}
\end{figure}

\clearpage
\begin{figure}
\includegraphics[scale=0.6,angle=90]{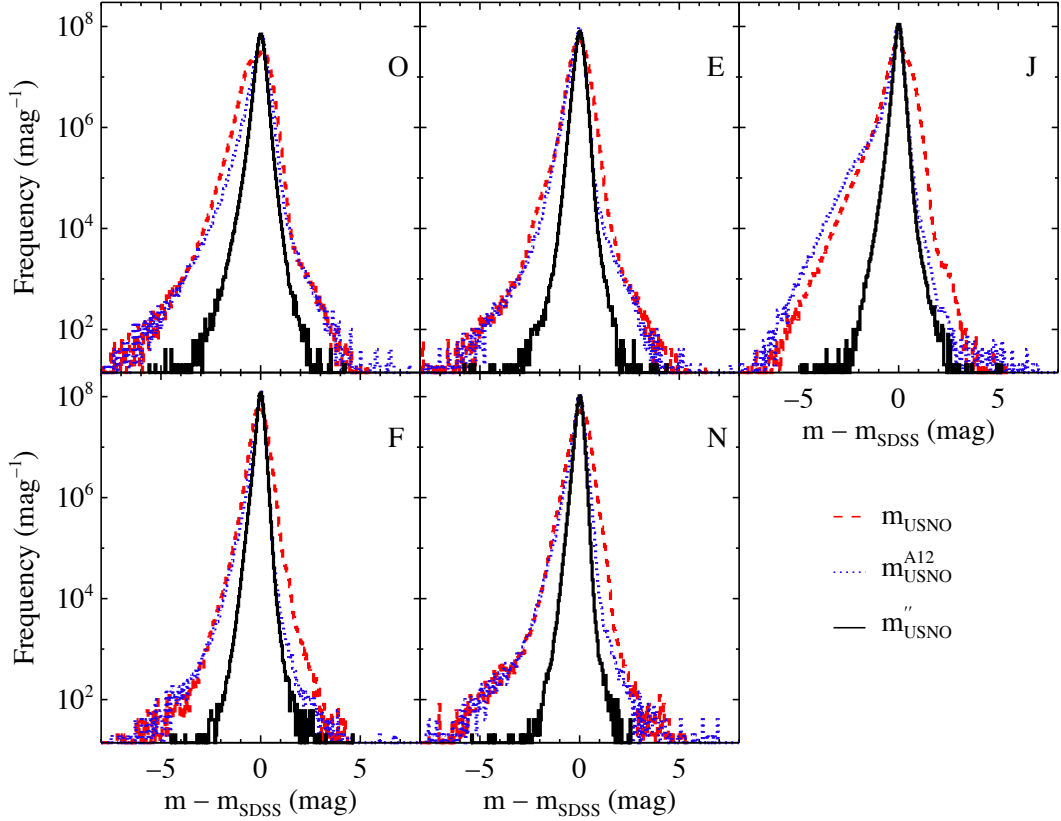}
\caption{Similar to Figure \ref{fig: histo culled}, but also showing the recalibrated USNO-B magnitudes from \citet{DR9} ($m^{A12}_{USNO}$; dotted blue).  For our data (solid black), only data that are unblended and consistent with other catalogs are shown.  
The areas under the dashed red and dotted blue lines are identical within each panel; the area under the black curve is smaller. 
The Figure shows that for objects with large values of $|m^{\prime\prime}_{USNO} - m_{SDSS}|$, the recalibrated magnitudes reported by \citet{DR9} are inaccurate and in many cases share the same distribution (or the same values) as the original USNO-B magnitudes. 
\label{fig: histo aihara}
}
\end{figure}

\clearpage

\begin{figure}
\includegraphics[scale=0.35]{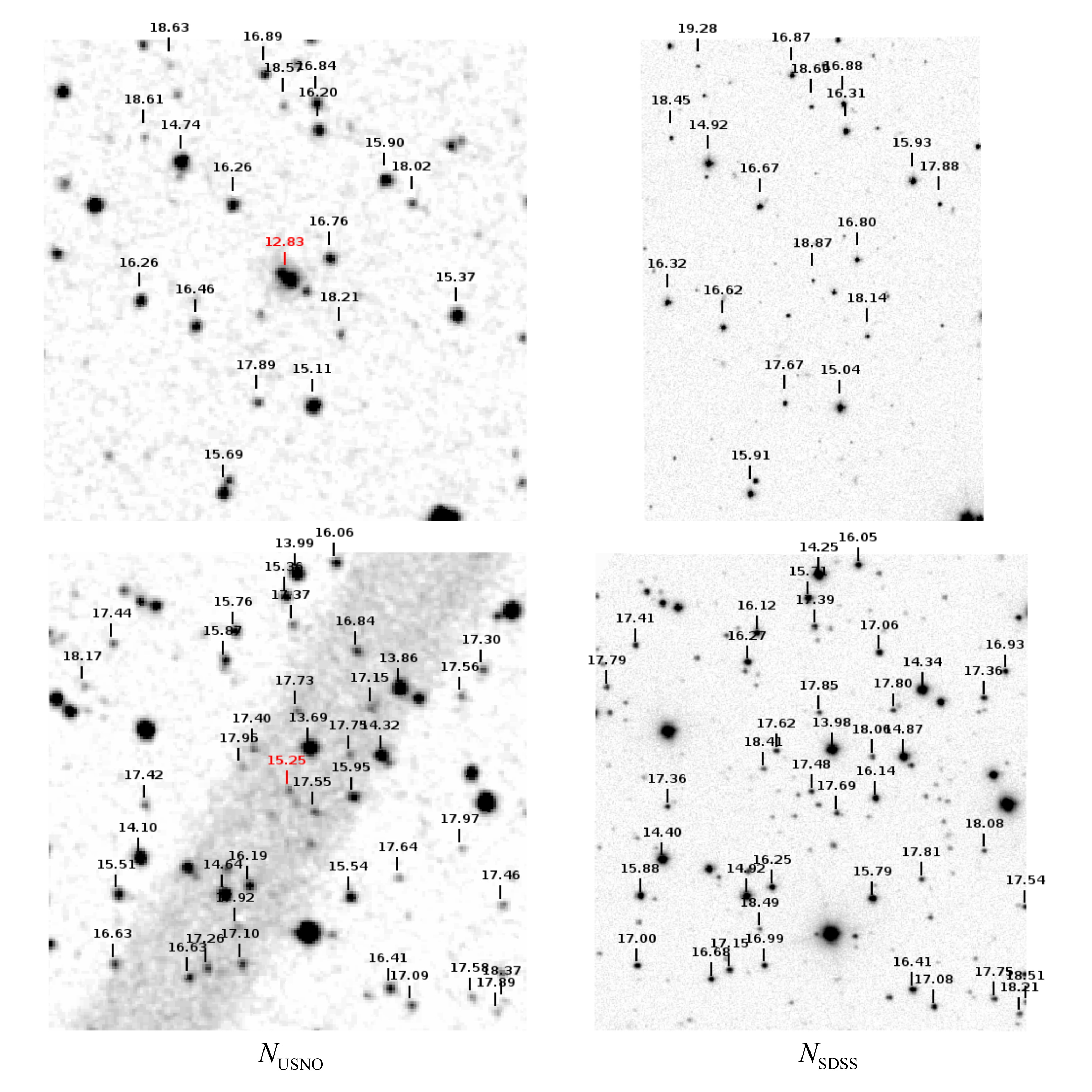}
\caption{Four images, each 3\arcmin\ on a side, that show examples of objects in our catalog with USNO-B photometry affected adversely by plate artifacts. North is up and east is to the left. 
The top row shows images from the DSS POSS-II {{$N$}} survey (left) and SDSS DR{{9}} {{$i$}} band (right), both toward {{$(\alpha, \delta) = (40.9521\arcdeg,+44.7367\arcdeg)$}}.
The bottom row shows images from {{the same surveys, but toward $(\alpha, \delta) = (93.056\arcdeg,+25.4443\arcdeg)$}}.
The tick marks and labels show the locations and magnitudes of objects from our cross matched catalog. 
Original USNO-B magnitudes ($m_{USNO}$) and DR{{9}} magnitudes converted to the USNO-B system ($m_{SDSS}$) are shown in the left and right columns, respectively.
The objects in red at the centers of the images have very large values of $|m^{\prime\prime}_{USNO} - m_{SDSS}| \gtrsim 3.0$. However, the images show that these are not genuinely variable objects and that the USNO-B magnitudes are underestimated (e.g., their brightnesses are overestimated) due to plate artifacts. {{Objects that overlap with artifacts are the single largest contributor to the very small level of contamination in our catalog.}}
\label{fig: thumbnail artefacts}
}
\end{figure}

\begin{figure}
\includegraphics[scale=0.6]{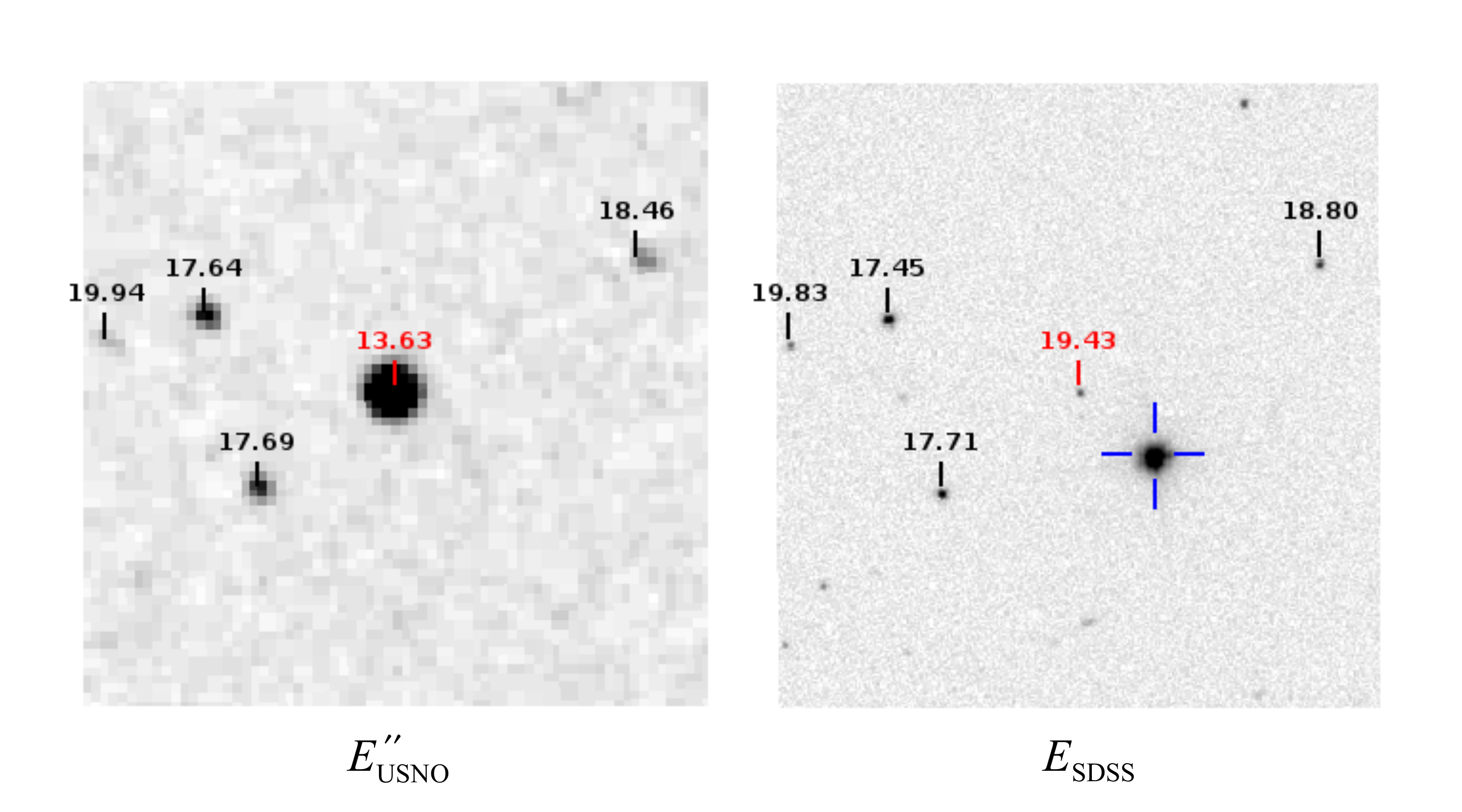}
\caption{Two images, 2\arcmin\ on a side, both toward {{$(\alpha, \delta) = (140.3766\arcdeg, +14.8777\arcdeg)$}} from the DSS POSS-I $E$ survey (left), and SDSS $r$ band (right), illustrating the effect of inaccurate proper motions. 
North is up, east is to the left, and the tick marks and labels indicate the locations and magnitudes of objects in our cross matched catalog.
Recalibrated USNO-B magnitudes ($E^{\prime\prime}_{USNO}$) and DR{{9}} magnitudes converted to the $E$ band ($E_{SDSS}$) are shown in the left and right panels, respectively.
The objects labelled in red are cross matched to each other have a very large difference in magnitude, $|m^{\prime\prime}_{USNO} - m_{SDSS}|$ = {{5.8}} mag. However, the images show that the object is not authentically variable but rather is the consequence of the chance alignment of two stars at the USNO-B epoch.  There is relatively bright star ($r$ = {{13.4}} mag), in the blue cross hairs on the panel on the right, that has moved $\approx$ {{19}}\arcsec\ to the south{{west}} during the $\approx$ {{55}} years between the two images ($\approx$ 0\farcs{{3}} per year). However, this bright star is reported by DR{{9}} as {{not having reliable photometry}}, and hence it was not cross matched to the correct USNO-B object (see \S\ref{sec: limitations}).
\label{fig: thumbnail proper motion}
}
\end{figure}

\clearpage
\begin{figure}
\includegraphics[scale=0.6,angle=90]{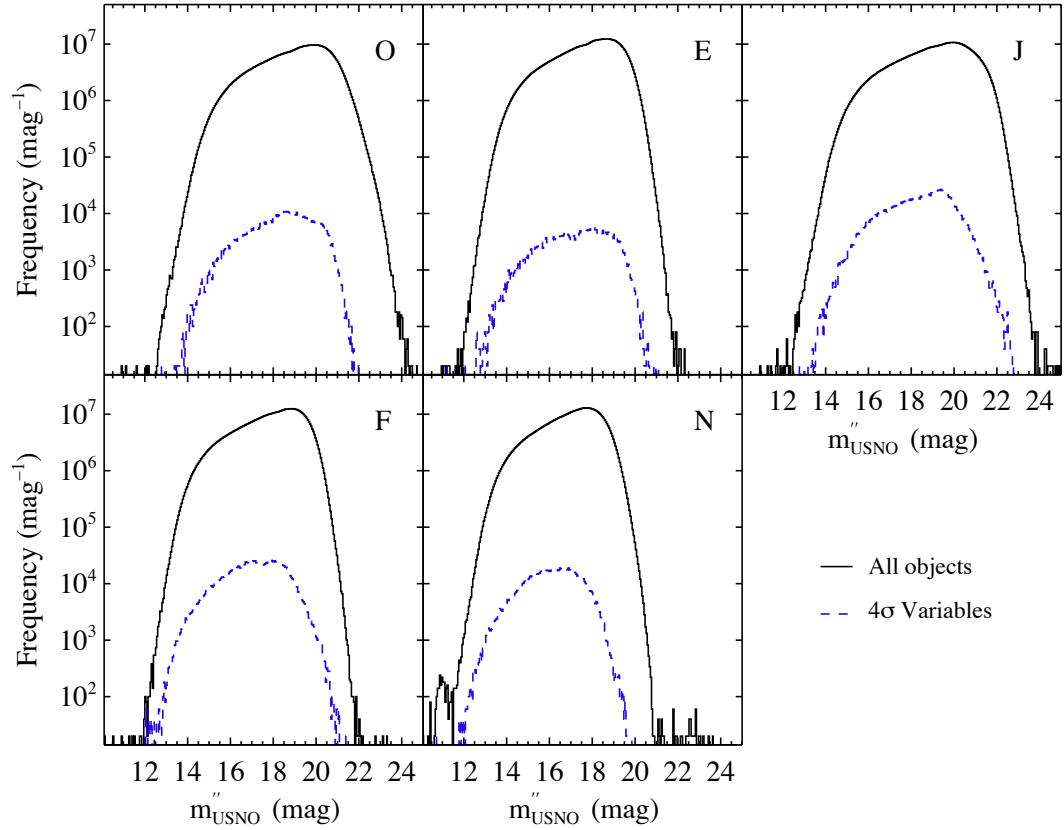}
\caption{Histograms of recalibrated magnitudes. The solid dark lines show all of the objects in the catalog; the dashed blue line shows objects that {{are candidate 4$\sigma$ variables. The y-axis is chosen to show bins with as few as one object in them.}}  The Figure shows the approximate limiting magnitude of the catalog for each of the USNO-B bands. The USNO-B band is labelled in each panel. 
\label{fig: histo mags}
}
\end{figure}

\clearpage
\begin{figure}
\includegraphics[scale=0.6,angle=90]{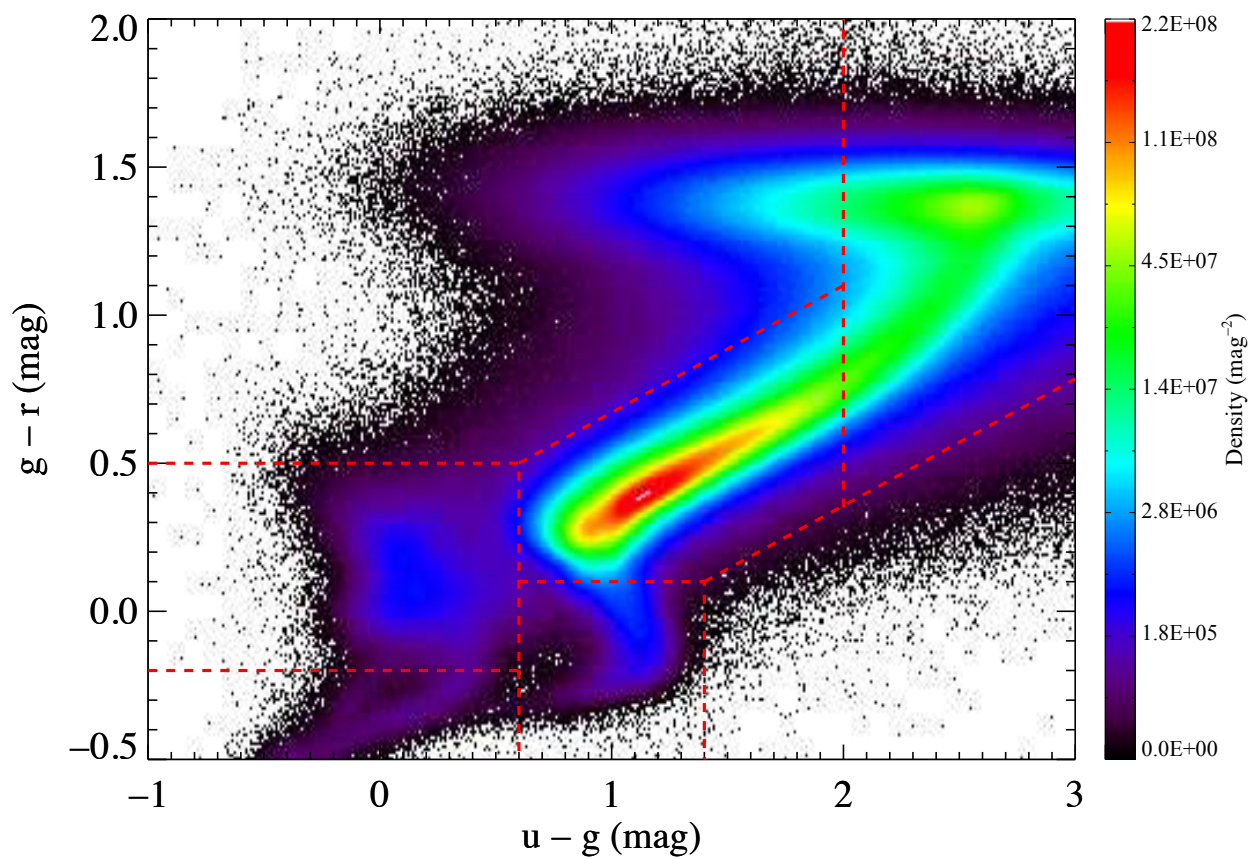}
\caption{Extinction corrected color-color diagram for  {{37,164,783}} object{{s}} in our catalog.  The color {{bar}} indicates the number of objects at each location in color-color space (with bin widths of 0.01 mag for each color). The dashed red lines outline the boundaries from \citet{Richards+02}  that are used to classify quasars and stars. {{The locus of points near $(u-g = 0.1,~g-r = 0)$ are consistent with low-redshift quasars, and the locus of points near $(u-g = 1.0,~g-r = -0.2)$ are consistent with RR Lyrae stars. }}
\label{fig: all stars cmd}
}
\end{figure}

\clearpage
\begin{figure}
\includegraphics[scale=0.6,angle=90]{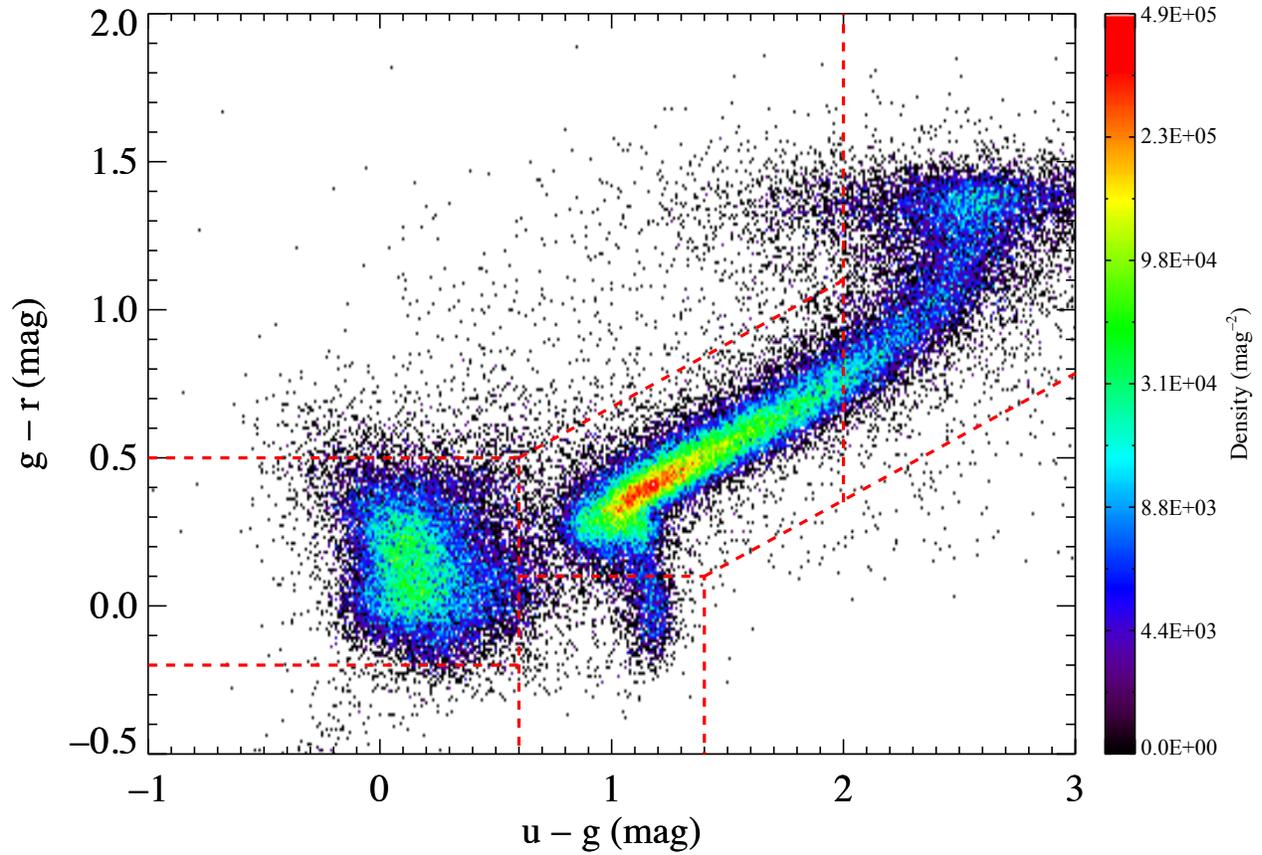}
\caption{Same as Figure \ref{fig: all stars cmd}, but for {{74,062}} objects with large amplitude variability between USNO-B and SDSS in the USNO-B $J$ band.  A {{quantitative}} comparison between this Figure and Figure \ref{fig: all stars cmd} shows that the fraction of object with colors consistent with low redshift quasars is much higher for the candidate variables than for the catalog as a whole (see \S\ref{sec: vars}).
\label{fig: var stars cmd}
}
\end{figure}

\clearpage
\begin{figure}
\includegraphics[scale=0.6,angle=-90]{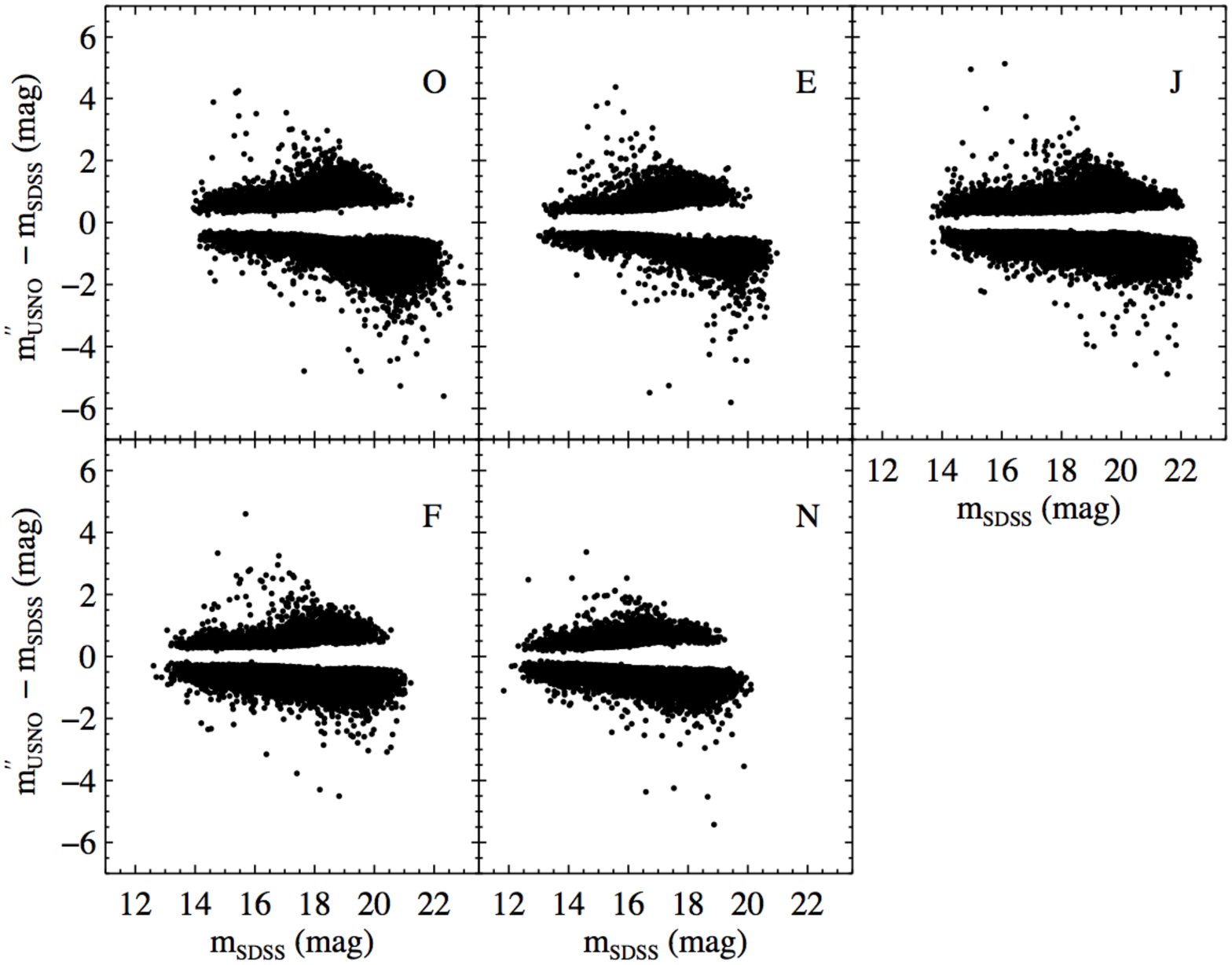}
\caption{{{
Distribution of $m^{\prime\prime}_{USNO} - m_{SDSS}$ as a function of $m_{SDSS}$ for our 4$\sigma$ candidate variables.  
The USNO-B band is labelled in each panel.
}} 
\label{fig: delta mag vars}
}
\end{figure}

\clearpage

\begin{figure}
\includegraphics[scale=0.6]{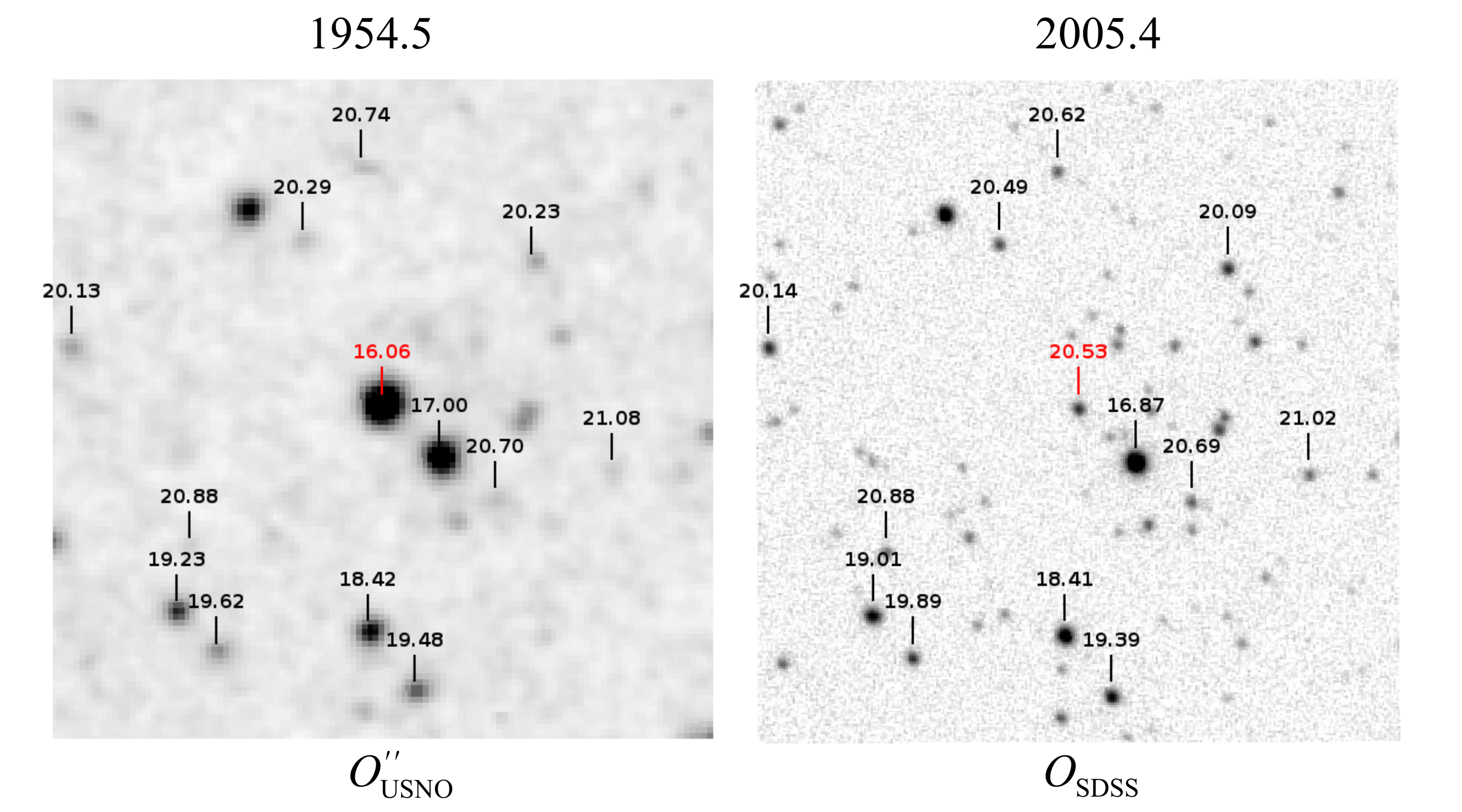}
\caption{Two images, 2\arcmin\ on a side, both toward $(\alpha, \delta) = (257.8337\arcdeg, -12.6677\arcdeg)$ from the DSS POSS-I $O$ survey (left), and SDSS $g$ band (right), illustrating a genuine, large amplitude variable star. 
North is up, east is to the left, the tick marks and labels indicate the locations and magnitudes of objects in our cross matched catalog.  
Recalibrated USNO-B magnitudes ($O^{\prime\prime}_{USNO}$) and DR{{9}} magnitudes converted to the $O$ band ($O_{SDSS}$) are shown in the left and right panels, respectively. The epoch of each observation, in units of years, is shown above the panels.
The object labelled in red  {{(SDSS J171120.07-124003.8)}} has decreased its brightness by 4.5 magnitudes between observations. 
The magnitude of the variability, the extremely red color ($g - i = 5.9$ at the DR{{9}} epoch 2005.43), and high infrared brightness ($K = 6.8$ mag from 2MASS) suggests that the object is a Mira variable (see \S\ref{sec: vars}).
\label{fig: thumb var mira}
}
\end{figure}

\clearpage

\begin{deluxetable}{ccclccclrrr}
\tabletypesize{\scriptsize}
\tablecaption{Recalibration coefficients from small-scale correction (\S\ref{sec: recal 1}) \label{tab: coeffs}}
\tablewidth{0pt}
\tablehead{
\multicolumn{3}{c}{SDSS DR9} & & \multicolumn{2}{c}{USNO-B\tablenotemark{a}} \\
\cline{1-3} \cline{5-6}
\colhead{Run} & \colhead{Field} & \colhead{Camcol} & &
\colhead{Survey} & \colhead{Field} &
\colhead{$A$} & \colhead{$B$} & \colhead{$C$} & \colhead{$\sigma_m$}
}
\startdata
109&37&5&&2&831&0.80152&0.02327&3.52059&0.10898\\
109&37&5&&3&831&0.91455&-0.16782&1.84326&0.08480\\
109&37&5&&8&831&0.80325&0.22527&2.97896&0.12718\\
109&38&5&&1&591&0.83905&-0.04885&3.01893&0.21691\\
109&38&5&&2&831&0.77061&-0.02857&4.15806&0.16747\\
109&38&5&&3&831&0.88260&0.07456&2.10249&0.15485\\
109&38&5&&8&831&0.78817&0.14289&3.19963&0.13145\\
109&39&4&&0&591&0.83039&-0.21656&3.51583&0.14375\\
109&39&4&&1&591&0.83085&0.26772&2.90034&0.18247\\
109&39&4&&2&831&0.79432&0.17378&3.45662&0.12339\\
109&39&4&&3&831&0.83876&0.22968&2.69934&0.19561\\
109&64&6&&0&592&0.95783&0.04863&0.46269&0.16118\\
109&64&6&&1&592&0.92757&0.15528&1.24565&0.20639\\
109&64&6&&2&832&0.96665&0.12493&0.02146&0.18579\\
109&64&6&&3&832&0.78084&0.07070&3.61143&0.18879\\
109&64&6&&8&832&0.83402&0.15937&2.41671&0.14567\\
109&65&6&&0&592&0.98637&0.04982&-0.15571&0.13329\\
109&65&6&&1&592&0.87617&0.16879&2.01625&0.20248\\
109&65&6&&2&832&1.01364&0.01284&-0.72019&0.11190\\
109&65&6&&3&832&0.80166&0.05639&3.18888&0.14805\\
109&65&6&&8&832&0.84832&0.07956&2.25547&0.15250\\
109&108&6&&0&593&1.01925&-0.09219&-0.66425&0.20051\\
109&108&6&&1&593&0.81592&0.08879&3.05121&0.19019\\
109&108&6&&2&833&0.94078&0.02257&1.03253&0.21214\\
109&108&6&&3&833&0.86891&-0.00146&2.20399&0.15810
\enddata
\tablecomments{This table is published in its entirety in the electronic edition of ApJS.  A portion is shown here for guidance regarding its form and content. 
}
\tablenotetext{a}{The survey and field number shown here are from the original USNO-B catalog. They indicate the survey from which the photometry was derived (e.g., POSS-I $O$-band, ESO/SERC $J$-band, etc.) and the plate identification number in the corresponding survey, respectively. 
}
\end{deluxetable}

\begin{deluxetable}{llll}
\tabletypesize{\scriptsize}
\tablecaption{Description of columns in the catalog \label{tab: columns}}
\tablewidth{0pt}
\tablehead{
\colhead{Column} & \colhead{Name} & \colhead{Units} & \colhead{Notes}
}
\startdata
1 & USNO ID &  & Object identification in USNO-B\\
2 & SDSS OBJID   &  & Object identification in DR{{9}} of SDSS-III\\
3 & RA & deg & Right ascension of object in DR{{9}} (J2000; epoch 2000.0) \\
4 & Dec & deg &  Declination of object in DR{{9}} (J2000; epoch 2000.0)\\
5 & $t_{SDSS}$ & yrs & Epoch of DR{{9}} observation (decimal years) \\
6 & Artifact Flag &  & Bit mask indicating object is likely an artifact in USNO-B\tablenotemark{a}\\
\\
7 & $m^{\prime}_{USNO}$ & mag & Small-scale recalibrated $O$-band USNO-B magnitude \\
8 & $m^{\prime\prime}_{USNO}$ & mag & Final recalibrated $O$-band USNO-B magnitude \\
9 & $\sigma_{USNO}$ & mag & Uncertainty of $O$-band $m^{\prime\prime}_{USNO}$ \\
10 & $t_{USNO}$ & yrs & Epoch of $O$-band observation (decimal years)\\
11 & $m_{SDSS}$ & mag & Magnitude of DR{{9}} object in $O$-band \\
12 & $\sigma_{SDSS}$ & mag & Uncertainty of $O$-band $m_{SDSS}$ \\
13 & Consistent & & Boolean flag indicating consistency of $O$-band $m_{USNO}$ with other catalogs\tablenotemark{b}\\
14 & Quality Flag & & Bit mask indicating $O$-band data quality\tablenotemark{c}\\
15 & Variability Flag & & Boolean flag indicating 4$\sigma$ variability\tablenotemark{d} \\
\\
16-24 & & &  Columns 7-15, but for $E$-band\\
25-33 & & &  Columns 7-15, but for $J$-band\\
34-42 & & &  Columns 7-15, but for $F$-band\\
43-51 & & &  Columns 7-15, but for $N$-band\tablenotemark{e}
\enddata
\tablecomments{Columns with null values indicate there are no data for a given USNO-B band.} 
\tablenotetext{a}{See \S\ref{sec: spurious}; the zeroth bit indicates the object is classified as a diffraction spike; the first bit indicates the object is classified as reflection halo artifact as determined by \citet{Barron+08}. }
\tablenotetext{b}{See \S\ref{sec: consistent}; this flag is set to a null value if the consistency could not be determined.}
\tablenotetext{c}{See Table \ref{tab: bitmask quality} for details.} 
\tablenotetext{d}{This flag is set to true if the object satisfies the criteria described in \S\ref{sec: vars} for 4$\sigma$ variables.} 
\tablenotetext{e}{See \S\ref{sec: usno}; rows with null values for $t_{USNO}$ are from $N$ band plates with unidentifiable epochs}
\end{deluxetable}

\begin{deluxetable}{llrrlcccccccccc}
\tabletypesize{\tiny}
\rotate
\tablecaption{Sample of the catalog \label{tab: sample}}
\tablewidth{0pt}
\tablehead{
\multicolumn{6}{c}{} & \multicolumn{8}{c}{$O$-band} \\
\cline{7-15}
\colhead{USNO ID} & \colhead{SDSS OBJID} & \colhead{RA} & \colhead{Dec} & \colhead{$t_{SDSS}$} &
\colhead{Art.\tablenotemark{a}} & \colhead{$m^{\prime}_{USNO}$} & \colhead{$m^{\prime\prime}_{USNO}$} & \colhead{$\sigma_{USNO}$} & \colhead{$t_{USNO}$} &
\colhead{$m_{SDSS}$} & \colhead{$\sigma_{SDSS}$} & \colhead{Cons.\tablenotemark{b}} & \colhead{Qual.\tablenotemark{c}} & \colhead{Var.\tablenotemark{d}} \\
\colhead{} & \colhead{} & \colhead{(deg)} & \colhead{(deg)} & \colhead{(yrs)} &
\colhead{} & \colhead{(mag)} & \colhead{(mag)} & \colhead{(mag)} & \colhead{(yrs)} &
\colhead{(mag)} & \colhead{(mag)} & \colhead{} & \colhead{}
}
\startdata
0875-0822160&1237672795031994468&0.000036&-2.417371&2006.7123&0&19.98&19.94&0.23&1954.6762&20.08&0.03&t&0&f\\
1222-0000001&1237663307453563363&0.000059&32.289956&2003.7452&0&20.72&21.47&0.43&1952.7132&21.73&0.05&t&0&f\\
1029-0000001&1237678920204681228&0.000087&12.998416&2008.8388&0&16.50&16.42&0.10&1949.8823&16.40&0.03&t&0&f\\
0847-0000001&1237672793422495967&0.000097&-5.271091&2006.7123&0&21.24&21.73&0.41&1954.6762&21.80&0.06&...&0&f\\
1000-0000001&1237678906782515326&0.000106&10.023592&2008.8361&0&...&...&...&...&...&...&...&...&...\\
0933-0000001&1237678620102164731&0.000149&3.352021&2008.7568&0&17.84&17.90&0.14&1951.6071&18.08&0.03&t&4&f\\
1161-0000001&1237666310706430148&0.000153&26.120653&2004.7295&0&18.24&18.34&0.14&1953.6085&18.34&0.03&t&0&f\\
1063-0000001&1237679454925095064&0.000153&16.358324&2009.0466&0&18.19&18.27&0.12&1954.6735&18.12&0.04&t&0&f\\
0891-0000001&1237663783123681350&0.000198&-0.838722&2003.8877&0&19.39&19.36&0.18&1951.6071&19.08&0.03&t&0&f\\
1064-0000001&1237678600231518582&0.000200&16.411836&2008.7541&0&20.90&21.49&0.17&1954.6735&21.55&0.07&...&0&f\\
0900-0000001&1237663784197423365&0.000225&0.077152&2003.8877&0&...&...&...&...&...&...&...&...&...\\
1235-0000002&1237663307989910044&0.000231&33.596348&2003.7452&0&...&...&...&...&...&...&...&...&...\\
1129-0000002&1237680298891805154&0.000250&22.956606&2009.7945&0&...&...&...&...&...&...&...&...&...\\
0889-0000001&1237656906355048600&0.000268&-1.034285&2001.7890&0&17.91&17.97&0.15&1951.6071&18.15&0.03&t&0&f\\
0969-0000001&1237669680114106516&0.000269&6.962031&2005.7342&0&17.09&17.12&0.09&1955.8590&17.07&0.03&t&0&f\\
 &  &  &  &  &  &  &  &  &  &  &  &  \\ 
0838-0625831&1237652600099242573&314.895431&-6.107006&2000.6749&1&19.28&19.26&0.21&1953.6277&19.62&0.02&f&0&f\\
0952-0255833&1237662263254515957&241.184508&5.257190&2003.3233&0&18.02&18.09&0.11&1954.3943&18.08&0.02&t&4&f\\
0950-0255852&1237655745099858328&241.596907&5.052574&2001.4575&0&20.22&20.51&0.15&1954.3943&20.65&0.05&t&2&f\\
0944-0313291&1237668572552039405&269.411447&4.442867&2005.4247&0&17.95&18.00&0.14&1950.5202&18.09&0.01&t&3&f\\
1255-0231699&1237661465986465996&236.674151&35.507975&2003.1945&0&19.35&19.43&0.19&1950.2793&20.27&0.04&t&0&t
\enddata
\tablecomments{This table is published in its entirety in the electronic edition of ApJS.  A portion is shown here for guidance regarding its form and content. The top 15 rows show columns 1-15 of the first rows of the catalog.  The next 5 rows show selected entries that demonstrate a range of values for the artefact flag (column 6), consistency flag (column 13), quality flag (column 14){{, and variability flag (column 15)}}. 
}
\tablenotetext{a}{Artifact flag described in Table \ref{tab: columns}.}
\tablenotetext{b}{Flag indicating if $m^{\prime\prime}_{USNO}$ is consistent with the magnitude in other catalogs, as described in \S\ref{sec: consistent} and in Table \ref{tab: columns}.}
\tablenotetext{c}{Quality flag described in Table \ref{tab: bitmask quality}. For example, a quality flag with a value of 5 ($=2^0+2^2$) indicates that bit numbers 0 and 2 are set, e.g. the object is within the exclusion radius of a bright star and it is blended with one or more DR{{9}} objects.}
\tablenotetext{d}{{{Flag indicating if the object is a candidate variable at the 4$\sigma$ level, as described in \S\ref{sec: vars} and in Table \ref{tab: columns}.}}}
\end{deluxetable}

\clearpage

\begin{deluxetable}{cl}
\tablecaption{Bit mask values for quality flags described in Table {\ref{tab: columns}}
\label{tab: bitmask quality}}
\tabletypesize{\scriptsize}
\tablewidth{0pt}
\tablehead{
\colhead{Bit number} & \colhead{Description}
}
\startdata
0 & Object is within exclusion radius of a bright star (see \S\ref{sec: blend bright}) \\
1 & Object is blended with one or more USNO-B objects (see \S\ref{sec: blended}) \\
2 & Object is blended with one or more DR{{9}} objects (see \S\ref{sec: blended})\\
3 & Uncertainty in recalibrated magnitude ($\sigma_{USNO}$) is derived from sparse data (see \S\ref{sec: recal 2})\\
\enddata
\end{deluxetable}

\end{document}